\documentclass[12pt]{article}
\usepackage{tabularx}
\usepackage{amsmath} 
\usepackage{authblk}
\usepackage{graphicx}
\usepackage{xcolor}
\usepackage{appendix}
\usepackage{comment}
\usepackage[margin=1in,letterpaper]{geometry}
\usepackage{colortbl}
\usepackage[hyphens]{url}
\usepackage{natbib}
\setcitestyle{aysep={}} 
\usepackage{subcaption}
\usepackage[final]{hyperref} 

\usepackage[symbol]{footmisc}
\hypersetup{
	colorlinks=true,       
	linkcolor=blue,        
	citecolor=blue,        
	filecolor=magenta,     
	urlcolor=blue         
}
\definecolor{darkGreen}{rgb}{0,0.6,0}

\graphicspath{{Figures/}}

\begin{document}

\begin{center}
{\Large Impacts of California Proposition 47 on \\ Crime in Santa Monica, CA}\\
\vspace{0.5cm}
\noindent Jennifer Crodelle$^{1,*}$, Celeste Vallejo$^{2,*}$, Markus Schmidtchen$^{3}$,\\
Chad M. Topaz$^{4,5}$, Maria R. D'Orsogna$^{6,7,\dagger}$\\
\end{center}
\footnotesize
$^1$Courant Institute of Mathematical Sciences, New York University, New York, NY 10012\\
$^2$Mathematical Biosciences Institute, The Ohio State University, Columbus, OH 43210\\
$^3$Laboratoire Jacques-Louis Lions,  Sorbonne Universit\'e, 75005 Paris, France\\
$^4$Department of Mathematics and Statistics, Williams College, Williamstown, MA 01267\\
$^5$Institute for the Quantitative Study of Inclusion, Diversity, and Equity, Williamstown, MA 01267\\
$^6$Department of Computational Medicine, UCLA, Los Angeles, CA 90095\\
$^7$Department of Mathematics, CSUN, Los Angeles, CA 91330\\
$^*$C. Vallejo and J. Crodelle contributed equally to this work.\\
$^\dagger$Corresponding author
\vfill
\normalsize
\begin{center}
April 28, 2020
\end{center}
\vfill
\noindent
\textbf{Objectives:} 
We examine crime patterns in Santa Monica, California 
before and after passage of Proposition 47, a 2014 initiative that reclassified some 
non-violent felonies to misdemeanors.
We also study how the 2016 opening of four new light rail stations, and how
more community-based policing starting in late 2018, impacted crime.
\\
\noindent
\textbf{Methods:}
A series of statistical analyses are performed on reclassified (larceny, fraud, possession of narcotics, 
forgery, receiving/possessing stolen property) and non-reclas\-sified crimes by probing publicly 
available databases from 2006 to 2019. 
We compare data before and after passage of Proposition 47, city-wide and within eight neighborhoods. 
Similar analyses are conducted within a 450 meter radius of the new transit stations. 
\\
\noindent
\textbf{Results:}
Reports of monthly reclassified crimes increased city-wide by approximately 15$\%$ after enactment of Proposition 47, 
with a significant drop observed in late 2018. Downtown exhibited the largest overall surge.
The reported incidence of larceny intensified throughout the city.
Two new train stations, including Downtown, 
reported significant crime increases in their vicinity after service began. 
\\
\noindent
\textbf{Conclusions:}
While the number of reported reclassified crimes increased after passage of Proposition 47,  
those not affected by the new law decreased or stayed constant, suggesting that Proposition 47 
strongly impacted crime in Santa Monica. Reported crimes decreased in late 2018 concurrent with the adoption of 
new policing measures that enhanced outreach and patrolling. These findings may be relevant 
to law enforcement and policy-makers. 
Follow-up studies needed to confirm long-term trends may be affected by the
COVID-19 pandemic that drastically changed societal conditions. 
\vfill
\noindent
\textbf{Keywords:}
California Proposition 47  $\cdot$ Welch's t-test $\cdot$ change-point analysis $\cdot$
segmented regression $\cdot$ light rail $\cdot$ community policing

\newpage

\section{Introduction and background}

\noindent On November 4, 2014, voters of the state of California passed Proposition 47 (hereafter Prop.\,47), also known as the ``Criminal Sentences. 
Misdemeanor Penalties. Initiative Statute.'' or ``The Safe Neighborhoods and Schools'' Act.  The referendum, which was approved with 59.6$\%$ of the vote, 
 went into effect the following day, November 5, 2014  \citep{BAL14}. 
 Prop.\,47 imparted three broad changes to felony sentencing laws in the state of California: i) certain non-violent theft and drug possession offenses would be
reclassified from felonies to misdemeanors; ii) those serving sentences for the reclassified offenses would be allowed to petition courts for re-sentencing; 
iii) those who had completed felony sentences now classified as misdemeanors would be able to petition courts to amend their criminal records. 
Felonies reclassified as misdemeanors under Prop.\,47 include shoplifting, attempted shoplifting, grand theft auto, receiving stolen property, forgery, fraud, 
writing bad checks; each up to a maximum monetary value of 950 USD. Possession of most illegal drugs for personal use, including methamphetamine, 
heroin and cocaine, was also reclassified as a misdemeanor. The law allows for some exceptions, for instance reclassification may not apply if
perpetrators have a criminal record 
including violence or sexual offenses.

Prop.\,47 was  part of a series of initiatives designed to lessen California's incarcerated population in response to allegations of inadequate inmate medical and 
mental health care, amounting to cruel and unusual punishment.  In 2009, federal courts required the state to reduce prison overcrowding and set an occupancy threshold of 137.5$\%$ of design capacity to guarantee inmates' Eighth Amendment rights.  On May 23, 2011 and upon appeal by the
State of California, the US Supreme Court 
upheld this decision in \textit{Brown vs.\,Plata}: California's prison population would have to decrease from approximately 156,000 to 110,000 individuals \citep{NEW12}. 
To comply with federal orders, state lawmakers enacted significant legislative reforms over the years, 
including Prop.\,47. Assembly Bill (AB) 109, 
also known as the Public Safety Realignment Bill, and 
Assembly Bill 117, also known as the Criminal Justice Realignment Bill, were approved and went into effect on October 1, 2011 \citep{OWE12}. 
These laws allowed 
those convicted of certain non-violent crimes to serve their sentences in county facilities, under house arrest, or in alternative sentencing schemes, 
rather than in state prisons. Overall, 500 criminal statutes were amended and penalties for parole violations were reduced. 
On November 6, 2012, voters also approved Prop.\,36, which revised California's 1994 Three Strikes Law mandating a sentence of 25 years to life 
for those convicted of a third felony. Under Prop.\,36, to be considered a strike, the third offense must be a serious or violent felony, or the perpetrator must have  
been previously convicted of murder, rape, or child molestation.

Although the state prison population fell after enactment of AB 109 and AB 117, it was only after passage of Prop.\,47 that the incarcerated population
dropped below the 2009 court-mandated target \citep{ROM15, GRA18, MOO19}. One study found a 50$\%$ decline in the number of individuals 
being held or serving sentences for the reclassified crimes \citep{BIR16}.
Prop.\,47 also stipulated that any resulting monetary savings 
should be diverted to crime prevention programs targeting youth and recidivists. The Safe Neighborhoods and Schools Fund was 
specifically created to manage these savings, estimated to be between 150 and 250 million USD per year. To date, $65\%$ of payments 
have been distributed to the Board of State and Community Correction,  with 
the Department of Education and the Victim Compensation and Government Claims Board receiving minor percentages \citep{TAY16}.
Reducing penalties for drug possession may have also lessened racial and ethnic disparities in the California criminal justice system \citep{MOO18}.
  
While Prop.\,47 helped reduce incarceration, determining its effects on the reclassified crime rates has proven more controversial. 
Several parties, including law enforcement officials, district attorneys, and mayors point to the law 
for rising crime \citep{CPC16, CAS18, EGE18, WEI19}.  Some studies link 
moderate \citep{BAR18} or sustained \citep{BIR18, FIS18} crime increases to the enactment of Prop.\,47, while other groups maintain 
that current data is inconclusive and that a longer term perspective is necessary \citep{MAL16}. 
Aside from the disputed effects of Prop.\,47 on crime rates, it has also been claimed that the new law 
brought unintended consequences such as the elimination of DNA collection for the reclassified crimes, 
restrictions in arresting repeat offenders, declines in the reporting of crimes as victims learned that police 
would not be able to apprehend and punish perpetrators, and making habitual drug users less likely to seek treatment \citep{HAN18}. 
A preliminary analysis conducted state-wide by the California Police Chiefs Association finds that
the consequences of Prop.\,47 are not homogeneous among cities of comparable size and that 
county specific factors, such as efficacy of monitoring and treatment programs, 
and how probation and/or incarceration are handled locally, may affect crime rates \citep{CPC16}. 
Judging the outcomes of Prop.\,47 has led to a contentious debate within academic, political, and community settings, 
culminating in a growing movement to reverse some of its reforms through a proposed 2020 ballot initiative \citep{KCS20}. 
Public safety agencies, caught between opposite viewpoints on the overall positive or negative
societal effects of Prop.\,47, have often expressed the need to better understand its consequences 
to optimize operations and budgets, to improve procedures, and to share impartial findings with stakeholder groups \citep{HUN17}. 

The goal of this work is to investigate the impacts of Prop.\,47 on crime rates in the coastal city of Santa Monica, California, population 91,411 (2019).
Located in Los Angeles County, the city is bordered by Los Angeles proper and the Pacific Ocean. 
Its downtown core has recently undergone intense revitalization, fueled by high-tech start-ups, increased tourism,
and the 2016 opening of the Metro Expo Line light rail extension. The latter now connects
the beach with nearby Culver City and inner Los Angeles neighborhoods
through seven new stations of which four are located within municipal borders. 
The city has also experienced rising housing costs,  the displacement of long-term tenants, 
and increasing levels of homelessness \citep{KAM12, LAT17}. In recent years both the Santa Monica Police Department 
(SMPD) and the local press have reported increases in crime,  
including robbery, burglary, aggravated assault, and homicide, with large numbers of repeat offenders \citep{LAC18, CAG18, CAT19, PAU19}. 
Dedicated social media accounts and resident neighborhood groups \citep{RES19, SMN19, SMC19, SMP19} have been 
awash with images, anecdotal evidence and speculations on the root cause of these trends. Passage of Prop.\,47 is among the
theories offered to explain the rise in crime; another is the opening of the 
Expo Line allowing for easier transportation to and from the city \citep{CER16, NEW17, HAR18}. 
After a change in leadership in May 2018, the SMPD launched a series of new
public safety initiatives. These included hiring twenty new police officers, increasing patrolling and outreach efforts, 
establishing a dedicated unit to analyze crime trends, 
deploying nightly security guards, adding lighting and CCTV cameras to public garages, and even limiting
the hours of operations of some businesses that attracted large amounts of crime \cite{REN19}.
The SMPD also increased its engagement with people experiencing homelessness and former inmates, helping them
connect to services. Among the newly established programs are the Neighborhood Resource Officers to facilitate
community-oriented policing, the Homeless Liaison Program, to assist the unhoused, and the Downtown Business Services Unit
to improve communication with business owners. These initiatives are reported to have mitigated crime in the city, especially those
affecting quality of life \cite{PAU19b, PAU20}.

As part of its pledge towards greater transparency, the SMPD maintains a publicly-available crime database,
listing dates, types, and locations of crimes within its jurisdiction starting from January 2006 until the present day \citep{SMPD18}.
Motivated by the many changes to the city, and to better quantify 
how violence and crime have changed over the past thirteen years, we performed several statistical analyses on this large body of data
with a specific focus on identifying possible effects of Prop.\,47. Our data analysis is meant as a first step in quantifying long-term
crime trends in Santa Monica, and as a way to go beyond casual information and/or personal opinion. 
Throughout our work, in every instance where we discuss changes to crime trends
it is important to note that any increase or decrease we present applies only to the reported crimes
listed by the SMPD. This qualifier is crucial, as inferring true crime trends
would require perfect knowledge, and the SMPD data may not be an unbiased representative sample
of the actual crimes committed. For example, the data may be affected by biases in collection methods,
changes to police routines, changes in the public's habits, and more.

Given the above caveat, we find that the average number of reclassified monthly crimes  
increased overall by about 15$\%$ after enactment of Prop.\,47. The sharp increase clearly emerges in the latter part of 2014, concurrent
with passage and implementation of Prop.\,47. A decrease in crime is observed towards the end of 2018
and persists through 2019, concurrent with the new police initiatives illustrated above. Longer term studies
would be needed to determine whether this decrease will stabilize in the future.
A geographical analysis reveals that overall the incidence of 
reclassified crimes after passage of Prop.\,47 increases or stays constant in all but one of the eight Santa Monica neighborhoods, with 
significant rises in monthly counts in Downtown ($+37.2\%$) and 
the North of Montana and Ocean Park neighborhoods ($+13.2\%$ and $+12.0\%$, respectively). Non-reclassified crimes instead appear
to decrease in all districts, except for Downtown which saw an $8.9\%$ increase.

Finally, we analyze monthly average crime rates within 450 meters of the four new Expo Line 
train stations opened in Santa Monica in May 2016. Of these, the Downtown Santa Monica and 17$^{th}$ Street/Santa Monica College stops
exhibit a statistically significant increase for all crimes after May 2016;  the difference 
is not significant for the 26$^{th}$ Street/Bergamot and Expo/Bundy stops.
The first two stations are marked by a larger number of crime attractors and foot traffic than the other two. 
Crime percent increases are similar for both reclassified (+30.6$\%$) and non-reclassified (+34.6$\%$) crimes
at Downtown Santa Monica. At 17$^{th}$ Street/Santa Monica College instead
the percent increase of reclassified crimes (+38.6$\%$) is much larger than that of non-reclassified crimes (+20.6$\%$).
Finally, reclassified crimes increase at 26$^{th}$ Street/Bergamot (+30.0$\%$) but 
non-reclassified ones do not vary appreciably. These results suggest that Prop.\,47 
led to differential crime increases at the Expo Line train stations. 

Our conclusions stem from multiple statistical 
analyses performed on the data, which we present in Sect.\,\ref{sect:data}. 
In Sect.\,\ref{subsec:meanCrimes}, we use a Welch's t-test to show that the average monthly number of crimes
affected by Prop.\,47  after November 2014 is significantly larger than prior to that date. We further decompose
the full 2006-2019 crime time series into three components: a trend, a periodic seasonality, and a random residual, 
in Sect.\,\ref{subsec:std}. The resulting trend increases around the end of 2014, 
as discussed in Sects.\,\ref{subsec:changePoint} and \ref{subsec:breakPoint}, where change-point analysis and segmented
regression analysis are used to determine trend change loci.  In Sect.\,\ref{subsec:eightneigh} and \ref{subsec:train}
we analyze crime trends in each of the eight neighborhoods that comprise the city of Santa Monica 
and the more circumscribed ones associated with the opening of the Expo Line.
Finally, we present a discussion and conclusion illustrating limitations and 
possible extensions of this work in Sect.\,\ref{subsec:concl}.

Understanding city-wide and neighborhood effects of Prop.\,47 as well as how the opening
of new train stations affect the geography of crime, may help administrators
and law enforcement better plan intervention strategies, 
optimize resource allocation, and prioritize budget spending, especially in light of savings due to reduced incarceration.


\section{Data}
\label{sect:data}

\begin{table}[t!]
    \centering
    \begin{tabular}{|l|r|}
    \hline
    \cellcolor[gray]{0.8} \textbf{Prop.\,47 crimes} & \cellcolor[gray]{0.8} \textbf{Count} \\\hline \hline
    Larceny & 37,082\\ \hline
    Fraud & 7,121\\ \hline
    Narcotics possession & 3,549\\ \hline
    Forgery& 1,458\\ \hline
    Receiving/Possessing stolen property & 636\\ \hline
    Total & 49,846 \\
    \hline\hline
   \cellcolor[gray]{0.8}  \textbf{non-Prop.\,47 crimes}  & \cellcolor[gray]{0.8} \textbf{Count} \\\hline
        Assault    & 13,870\\ \hline
    Public intoxication & 12,926\\ \hline
    Vandalism & 9,699\\ \hline
    Burglary & 8,724 \\ \hline
    Grand theft auto  (GTA) & 4,812\\ \hline
    Contempt of court &  4,295\\ \hline
    DUI & 3,724 \\ \hline
    Robbery & 2,317\\ \hline
    Sex Offenses & 1,128 \\ \hline
    Narcotic sale & 456\\ \hline
      Total & 61,951 \\\hline\hline
    \cellcolor[gray]{0.8}  \textbf{All Crimes} & 111,797 \\ \hline
    \end{tabular}
    \caption{Crimes considered
    in this work and their cumulative count between 2006 and 2019. The Prop.\,47  (reclassified) 
    crimes are those subject to reclassification; 
    the non-Prop.\,47 (non-reclassified) crimes are those not affected by legislative change. }
    \label{tab:crimeCount}
\end{table}

\noindent 
The data used in this study was obtained from an open source file 
managed and updated by the Santa Monica Police Department (SMPD)
from January 2006 to the present day \citep{SMPD18}.  
The raw data includes the Uniform Crime Reporting (UCR) classification code
as determined by the FBI \citep{FBI04}, a description of the type of crime, the date on which it occurred, and the latitude and longitude of
its location. 
To better manage the information and perform statistical analyses on categories with a sufficient amount of data,  
we collapsed some original crime categories into coarser ones. For instance, 
we grouped ``aggravated assault with a firearm," ``general aggravated assault," ``aggravated assault 
with hands," ``aggravated assault with knife," and ``aggravated assault with other weapon," 
under the new general category  ``assault."  We also set a threshold of
450 counts per category, so that if this minimum number of crime occurrences 
was not met over the 2006-2019 period, the category was excluded from our analysis due to insufficient data.
Among the crimes that did not meet the threshold were arson, embezzlement, blackmail and homicide. 
Misappropriation of property 
appears in the database only from 2010 onwards so we also discarded this
category from our analysis.  Finally, all incomplete or corrupted entries were discarded. 

\begin{figure}[t]
    \centering
\includegraphics[width=0.6\textwidth]{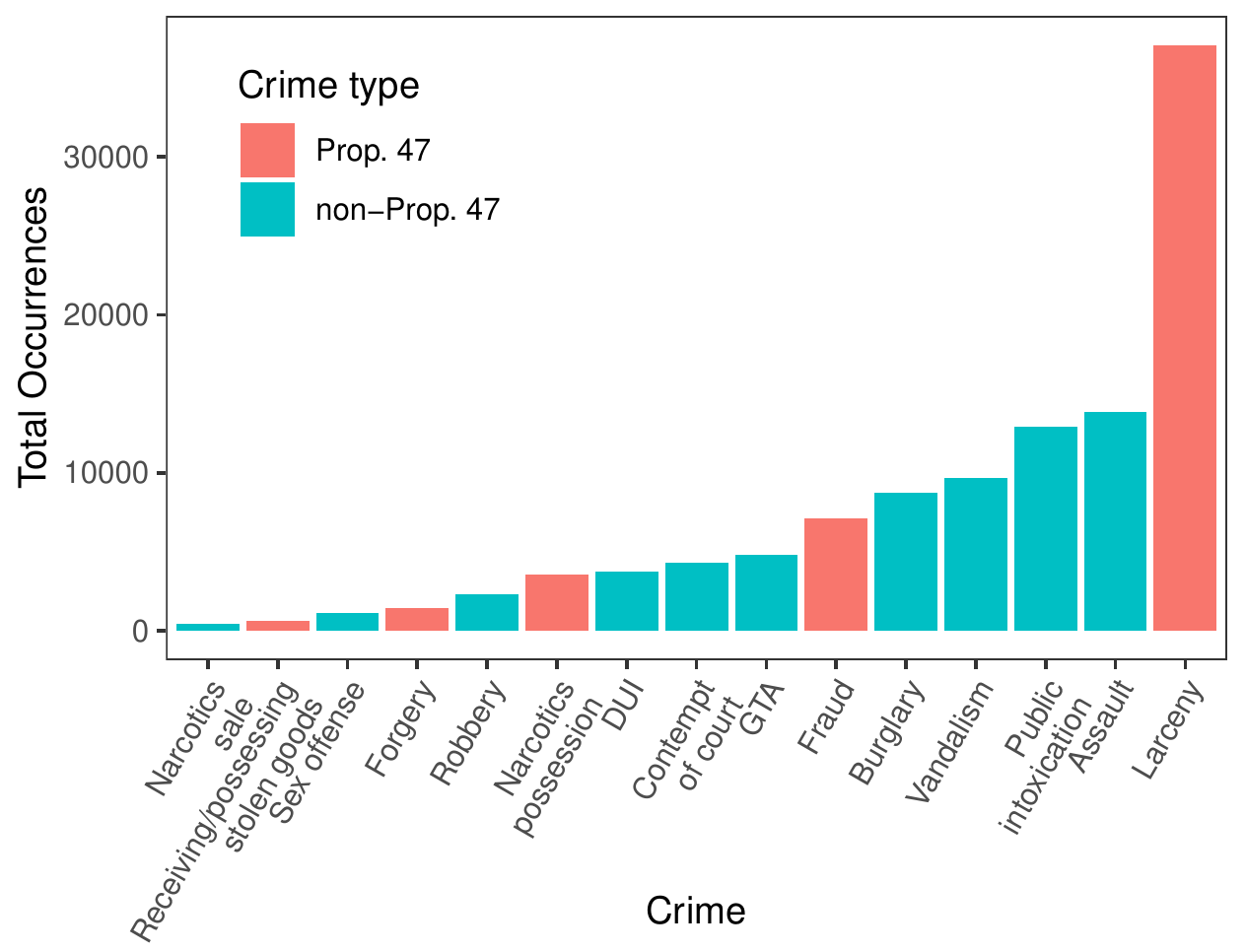}
    \caption{Visualization of the total number of crimes in Santa Monica from 2006 to 2019, as listed in Table \ref{tab:crimeCount}. 
    Receiving and possessing stolen goods, forgery, narcotics possession, fraud and larceny are all reclassified crimes.
    All others do not fall under the provisions of Prop.\,47.}
    \label{fig:barplot1}
\end{figure}

\begin{figure}[t]
    \centering
\includegraphics[width=0.6\textwidth]{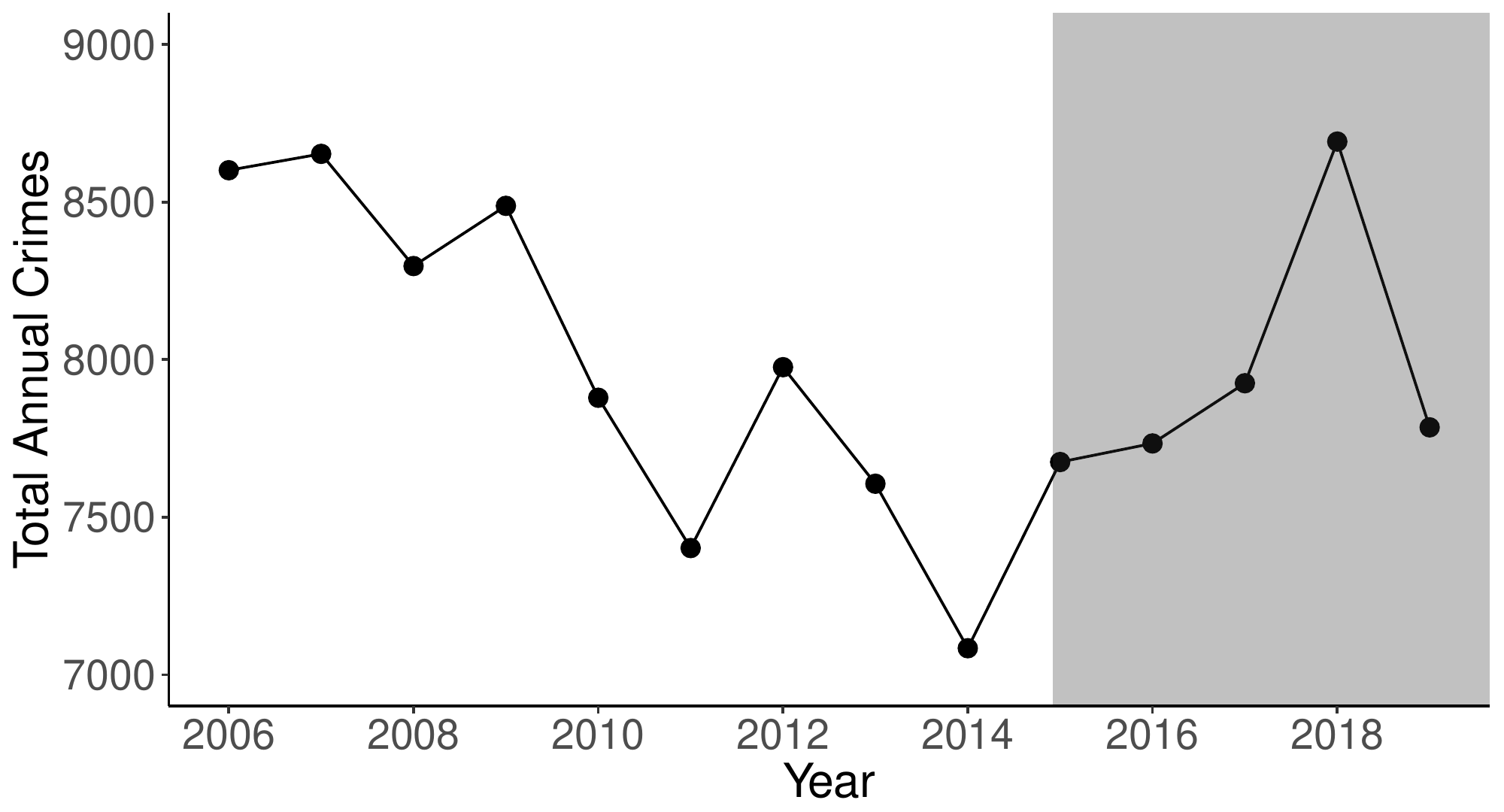}
    \caption{Total number of annual crimes from 2006 to 2019. The gray area indicates the years during which Prop.\,47 was enforced. 
    Note the lowest crime point in 2014, and the sharp decline in 2019.}
    \label{fig:barplot2}
\end{figure}

In Table \ref{tab:crimeCount} we list the general crime categories that were reclassified from felonies to misdemeanors under Prop.\,47
and their respective 2006-2019 city-wide counts. They are larceny, fraud, narcotics possession, forgery, and receiving/possessing stolen property.
We refer to these collectively as ``Prop.\,47 crimes" or ``reclassified crimes." 
Crimes not affected by Prop.\,47 are also listed in Table \ref{tab:crimeCount}  as ``non-Prop.\,47 crimes;'' 
we will also refer to them as ``non-reclassified crimes." Note that
their cumulative counts are of the same order of magnitude as the reclassified ones.
The crime database is updated by the Santa Monica Police Department 
daily; for most of our analysis we consider monthly, or in some cases
yearly, aggregates.  In the following sections we compare temporal trends between the two groups of crimes to determine the effects
of the 2014 initiative.  Fig.\,\ref{fig:barplot1} gives an overall view of the data. 
Larceny, one of the Prop.\,47 offenses, 
has the highest overall incidence followed by assault and public intoxication, both non-Prop.\,47 crimes.
Fig.\,\ref{fig:barplot2} displays the total (reclassified and non-reclassified) annual crime count 
from 2006 through 2019. 

Prop.\,47 imposed a maximum monetary value of 950 USD for crimes to be reclassified 
as misdemeanors; however, no dollar amount information is specified in the data set we examined. 
We used ancillary information to determine which categories should fall under the Prop.\,47 header,
depending on their typical economic value. For example, the  FBI's UCR 
Program for 2017 estimates the average value of property lost due to larceny to be roughly 1,007 USD per offense \citep{FBI17}.
In earlier years, the average value of losses due to larceny was lower, for example in 2006 it was 855 USD per offense, 
justifying its inclusion in the Prop.\,47 reclassified list for all years. Larceny is here broadly defined as the unlawful taking of property such 
as motor vehicle parts and accessories or bicycles, shoplifting and pick-pocketing.  
Since the average street value of cocaine, heroin or methamphetamine doses for personal use is well below the 950 USD threshold,
we also include possession of narcotics in the Prop.\,47 reclassified list. The Federal Reserve
estimates that for the year 2015 total fraud from bad checks, general-purpose transactions, 
and credit card accounts, resulted in 62 million single payments for a total of 8.3 billion USD, averaging
135 USD per transaction \citep{FDR18}.  We thus include fraud and forgery in the Prop.\,47 crime list. 
Finally, we do not include grand theft auto in the list of Prop.\,47 crimes, since as per conversations with the
SMPD, the typical value of stolen vehicles exceeds 950 USD, and thus grand theft auto incidents may
fall outside the scope of Prop.\,47.  The SMPD also confirmed that the monetary value associated with all the Prop.\,47 
crimes listed in Table \ref{tab:crimeCount} is usually under the 950 USD threshold
imposed for reclassification purposes.  

We do not adjust crime counts for population change since the number of Santa Monica inhabitants
has remained fairly stable in the thirteen year period under investigation. 
The city tallied approximately 87,000 residents in 2006, and after peaking at 93,000 in 2015, the population is
currently estimated to be 91,411 \citep{WPR19}.

\section{Effects of passage of Proposition 47} 
\label{sect:stats}

\begin{figure}[t!]
    \centering
 \includegraphics[width=0.85\textwidth]{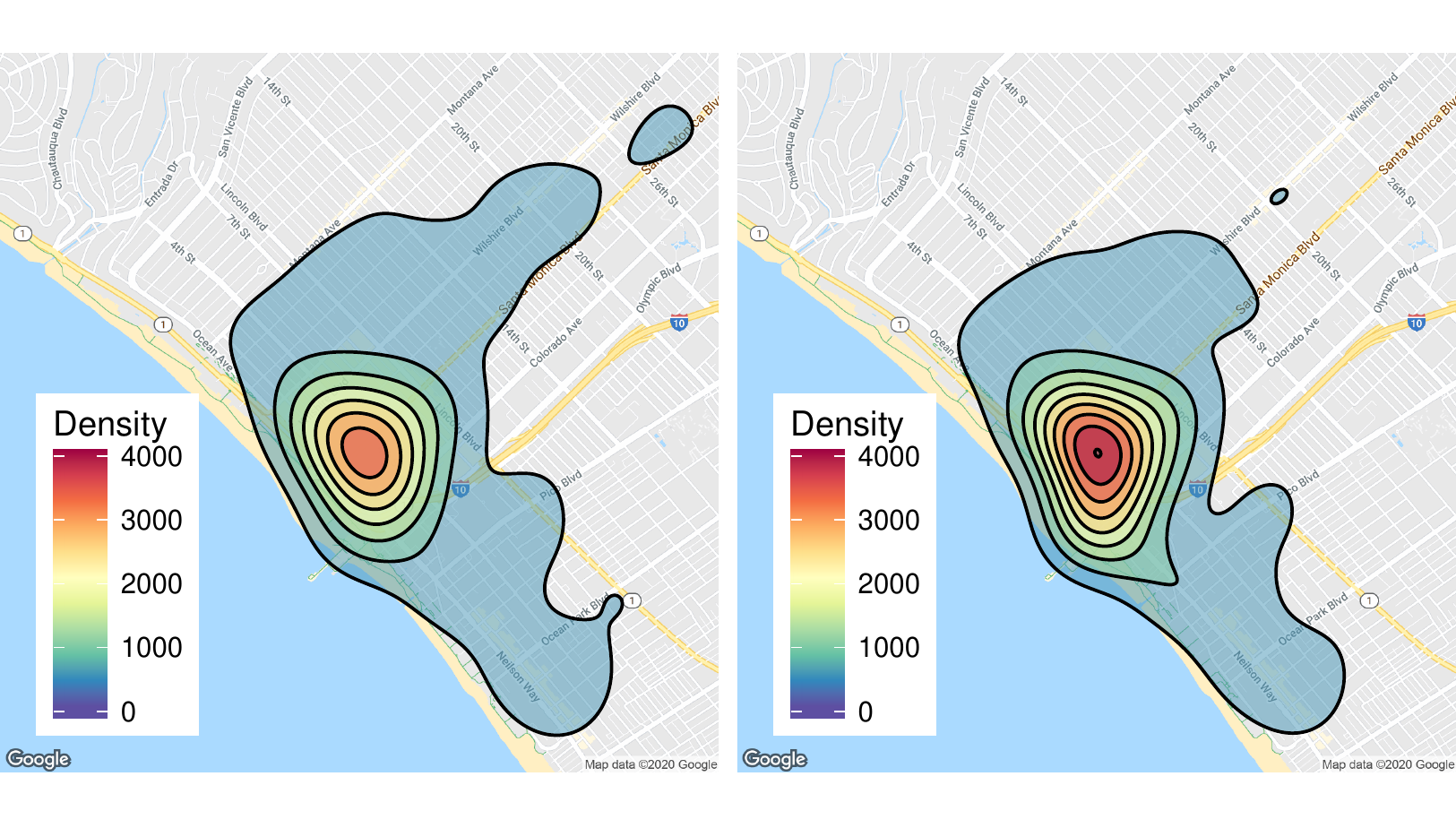}
    \caption{Contour map of annual larceny crimes before Prop.\,47 (left) and after Prop.\,47 (right).
    The right-hand panel displays more intense coloring in the central contours, reflecting higher incidence of larceny after implementation 
    of Prop.\,47. Although the specific visualization pertains to larceny, all crime types are more prevalent Downtown compared to residential areas.}
    \label{fig:density}
\end{figure}

\noindent
We now perform a series of statistical analyses on the two data sets identified 
in Table \ref{tab:crimeCount}: the Prop.\,47 crimes
(possession of narcotics, fraud, larceny, forgery, receiving and possessing stolen property), and the 
non-Prop.\,47 crimes (all others not affected by legislative change).
To illustrate the geographical variability of crime, 
Fig.\,\ref{fig:density} displays
a map of the average annual incidence of larceny (a reclassified crime) 
before (2006-2014) and after (2015-2019) implementation 
of Prop.\,47. Most events are located in downtown Santa Monica, 
with the average annual crime density increasing after 2014, 
as can be seen by the more intense coloring in the right-hand panel
of Fig.\,\ref{fig:density}.

\subsection{The monthly mean number of reclassified crimes increases after implementation of 
Prop.\,47} 
\label{subsec:meanCrimes}

\begin{figure}[t!]
    \centering
     \hspace{-0.4cm}
    \includegraphics[width=0.54\textwidth]{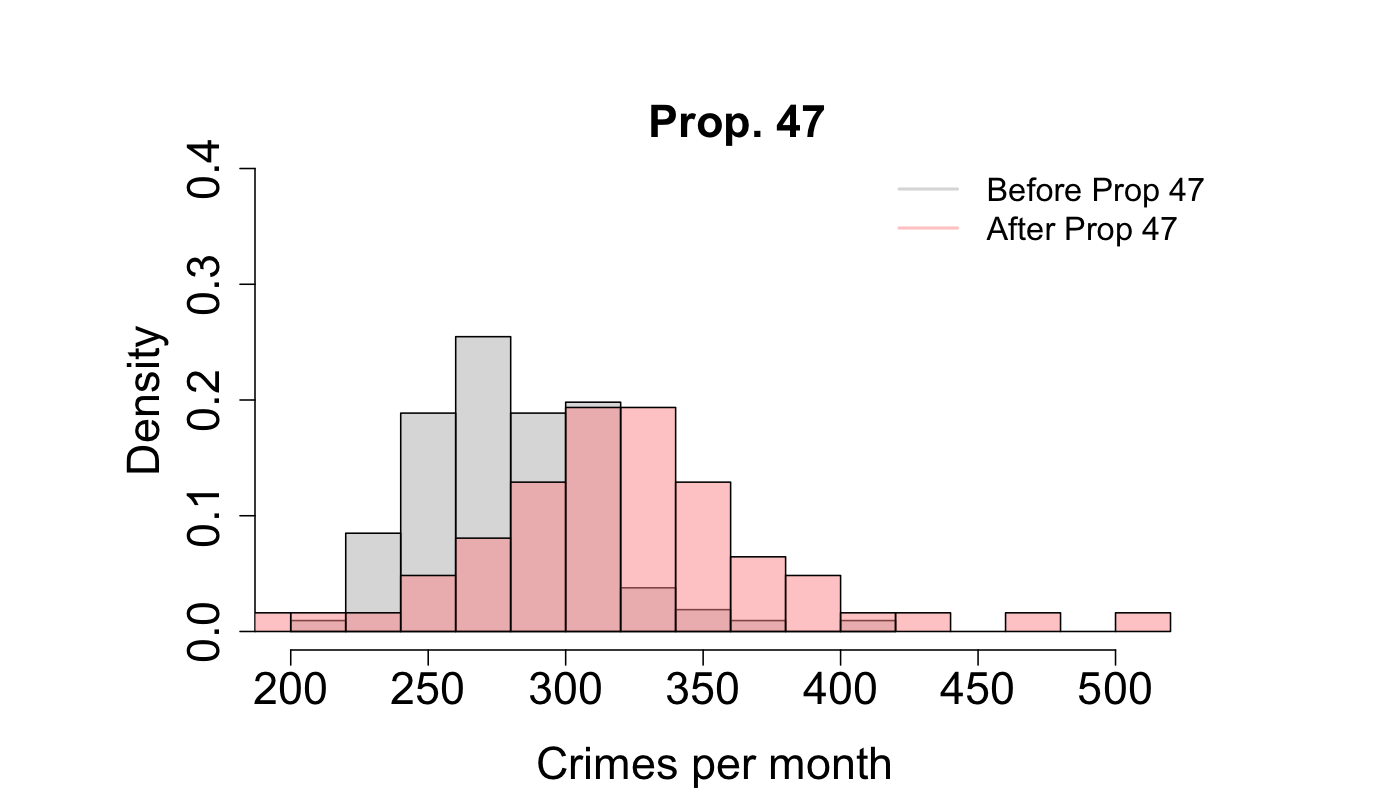}
    \hspace{-1.4cm}
      \includegraphics[width=0.54\textwidth]{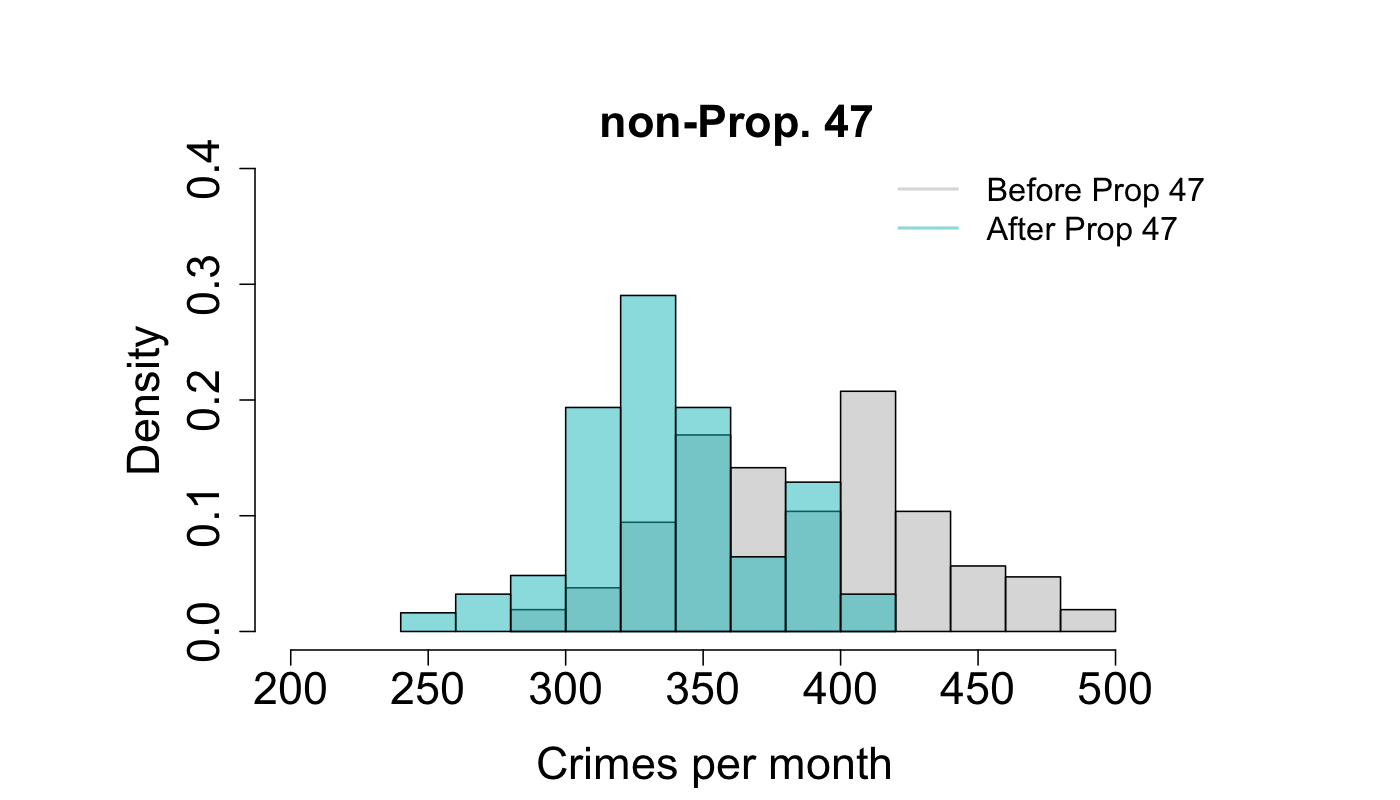}
    \caption{Histograms of crime counts per month before November 2014 for Prop.\,47 crimes, subject to reclassification (left),
    and non-Prop.\,47 crimes, not affected by legislative change (right). For Prop.\,47 crimes, 
    the monthly average before implementation of the new law was 281.4 crimes, whereas after 
    November 2014 the average was 322.9 crimes per month. Performing a Welch's t-test shows that this 
    14.7\% increase in the average number of crimes per month is statistically significant at the 0.05 level.
    The same computations applied to non-Prop.\,47 crimes show that, to the contrary, non-reclassified crimes decrease
    by 13\%, from 387.3 to 337.1 per month, after passage of the initiative.}
    \label{fig:histogram}
\end{figure}

To quantify the effects of Prop.\,47 on crime rates in Santa Monica, we
compute the average number of monthly offenses subject to reclassification
before and after passage of Prop.\,47.  For comparison we perform the same analysis on the non-Prop.\,47 crimes. 
Although the specific implementation of Prop.\,47 occurred on November 5, 2014,
we group all November 2014 events as occurring after passage of the new law
since we are binning data by the month.
Histograms for the four resulting subsets of data are shown in 
Fig.\,\ref{fig:histogram}. The before and after monthly 
crime distributions for Prop.\,47 offenses are presented in the left-hand panel of Fig.\,\ref{fig:histogram};  
before and after distributions for non-Prop.\,47 crimes are in the right-hand panel of Fig.\,\ref{fig:histogram}. 
As can be seen, the Prop.\,47 crime distribution is shifted to the right after November 2014
with respect to the pre-implementation data. On the other hand, the post-November 2014, 
non-Prop.\,47 crime distribution is shifted to the left.  
These histograms suggest that crimes affected by the reclassification process increased
after passage of Prop.\,47 whereas the incidence of crimes that were not subject to the new legislation decreased.
To determine whether these shifts are statistically significant we use 
Welch's unequal variances t-test (Welch's t-test) to compare the before and after mean monthly count of reclassified 
crimes. This is a two-sample test typically employed 
to compare two mean values when the respective samples have unequal size or variance \citep{WEL47}.
In our specific case, data is available over eight years (106 months) before November 2014 and only over
five years (62 months) after the same date, leading to very different sample sizes and variances.   
If the occurrence of Prop.\,47 crimes listed in Table \ref{tab:crimeCount}  
were not affected by the reclassification process, we would expect the 
difference between crime counts before and after passage of the law
as determined by Welch's t-test to be negligible. 

We denote by $\mu_{\text{b}}$ and $\mu_{\text{a}}$  the mean monthly number of Prop.\,47 crimes
before and after November 2014, respectively; 
$\sigma_{\text{b}}$ and $\sigma_{\text{a}}$ represent the associated standard deviations, and
$N_{\text{b}}$ and $N_{\text{a}}$ the respective number of months over which these averages were calculated.
The null hypothesis
is formulated as there being no difference in the mean values, 
$\mu_{\text{b}}$ = $\mu_{\text{a}}$,
while the alternative hypothesis 
posits that Prop.\,47 led to an increase in the reclassified offenses,  
$\mu_{\text{b}} < \mu_{\text{a}}$. 
Our data yields 
$\{\mu_{\text{b}}, \sigma_{\text{b}}, N_{\text{b}} \}_{\rm p47} = \{281.4, 33.2, 106\}$ and 
$\{\mu_{\text{a}}, \sigma_{\text{a}}, N_{\text{a}} \}_{\rm p 47} = \{322.9,53.9,62\}$. 
The `p47' subscript indicates that these statistical values are evaluated on Prop.\,47 offenses.
To verify whether the before-to-after crime increase is statistically significant we perform a one-tailed Welch's t-test by calculating
the following $t$-statistic

\begin{eqnarray}
\label{welch_tscore}
    t = \frac{\mu_{\rm{b}} -  \mu_{\rm{a}}}
    {\displaystyle{\sqrt{\frac{\sigma^2_{\rm{b}}}{N_{\text{b}}}
  + \frac{\sigma^2_{\rm{a}}}{N_{\text{a}}}}}}, 
    \end{eqnarray}

\noindent
yielding $t = 5.5$ for the values listed above. This quantity must be compared to the corresponding $t$-value 
from the Student's $t$-distribution \citep{WAL07}, once the number of degrees of freedom $\nu$ and the significance level are specified. We denote this reference $t$-value as $t_{\rm s}$. 
Since the before and after Prop.\,47 samples are associated to different data sets, 
each with their own degrees of freedom, we use the Welch-Satterthwaite equation to derive an effective $\nu$ \citep{SAT46}

\begin{eqnarray}
\label{ws-eqn}
\nu = \frac{\left(\displaystyle{\frac{\sigma_{\text{b}}^2}{N_{\text{b}}}} +\frac{\sigma_{\text{a}}^2}{N_{\text{a}}}\right)^2}
{\displaystyle{\frac{\sigma_{\text{b}}^4}{N_{\text{b}}^2 
  (N_{\text{b}} - 1) }+
\frac{\sigma_{\text{a}}^4}{N_{\text{a}}^2 
 (N_{\text{a}} - 1)}}}, 
\end{eqnarray}

\noindent
from which we obtain $\nu = 89$. Finally, we specify a significance level of 
$0.05$ to find the reference value  $t_{\rm s} = 1.66$ from the Student's $t$-distribution. Since this quantity
is much smaller than the $t=5.5$ statistic found from Eq.\,(\ref{welch_tscore}),
we reject the null hypothesis in favor of the alternative one:
the $15\%$ increase in the average monthly number of reclassified crimes after the introduction of Prop.\,47 is statistically significant. 

We perform a similar analysis for the non-reclassified crimes, using 
$\{\mu_{\text{b}}, \sigma_{\text{b}}, N_{\text{b}} \}_{\rm{non\,p47}} = \{387.3, 44.8, 106\}$
and $\{\mu_{\text{a}}, \sigma_{\text{a}}, N_{\text{a}} \}_{\rm{non\,p47}} = \{337.1,34.8,62\}$, 
where the subscript `non\,p47' refers to values being evaluated on non-reclassified crimes
before and after passage of Prop.\,47. 
We formulate the same null hypothesis as above, 
$\mu_{\text{b}}$ = $\mu_{\text{a}}$, 
with the alternative hypothesis 
set as there being a decrease in the mean monthly number of crimes 
after November 2014,
$\mu_{\text{b}} > \mu_{\text{a}}$.  
The $t$-statistic obtained from Eq.\,(\ref{welch_tscore}) and the non-Prop.\,47 values is $t=8.1$; 
Eq.\,(\ref{ws-eqn}) yields
$\nu = 153$, which results in  $t_{\rm s}$ = 1.66 at the 0.05 significance level. 
Since $t_{\rm s} = 1.66 < t = 8.1$, we reject the null hypothesis in favor of the alternative one: 
the $13 \%$ decrease in the average monthly number of non-reclassified crimes 
after the introduction of Prop.\,47 is statistically significant.

Thus far, our analysis suggests that the monthly occurrence of reclassified crimes increased
after passage of Prop.\,47, whereas crimes that were not affected by it, decreased.

\begin{figure}[t!]
\centering
\includegraphics[scale=1]{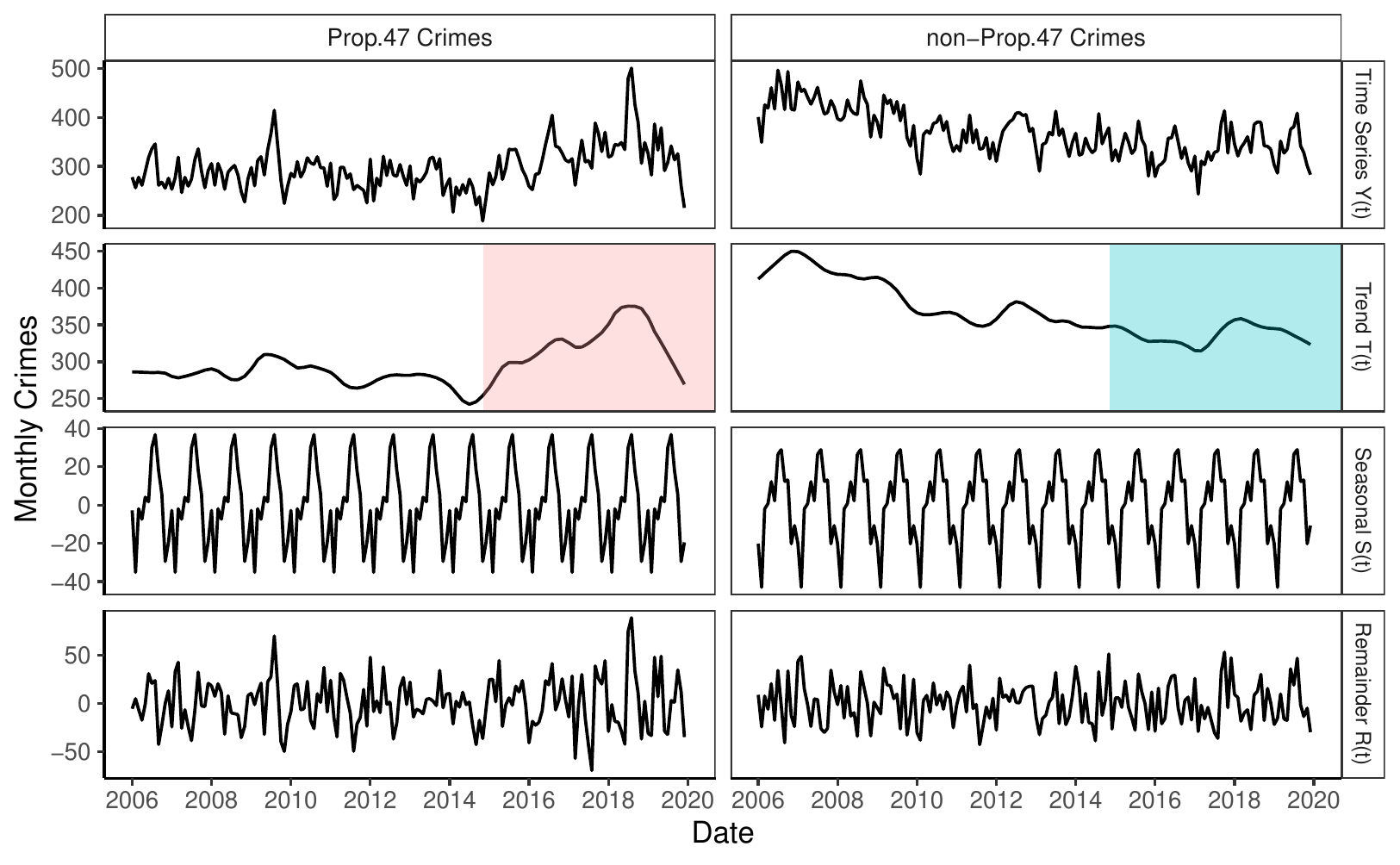}
\caption{Results from the Seasonal and Trend decomposition using Loess (STL) with an additive model for 
monthly crimes between 2006-2019.
The full time series of crimes per month $Y(t)$ is separated into trend $T(t)$, seasonality $S(t)$, and remainder $R(t)$
components. In each panel, the top row shows the monthly data $Y(t)$, 
the second row is the trend $T(t)$ as given by the loess smoothed data without seasonality, the third row contains the seasonal component $S(t)$. 
Finally, the remainder row contains the residual $R(t)$ with seasonality and trend removed from $Y(t)$. 
This decomposition is performed for Prop.\,47 crimes in the left panel and for non-Prop.\,47 crimes in the right panel. 
Note that an increase in the trend emerges for the Prop.\,47 crimes towards the end of 2014, 
but not for non-Prop.\,47 crimes. }
\label{fig:_decomp_prop47}
\end{figure}

\subsection{Reclassified crimes increase after implementation of Prop.\,47}
\label{subsec:std}

\noindent
To further identify differences in the temporal evolution of the reclassified and non-reclassified offenses
we analyze the entire 2006-2019 crime time series. 
Temperature variations, seasonal cycles, and
the increased criminal opportunities provided by travel and/or shopping
during holiday periods are well-known possible crime influencers \citep{FAL52, MCD12, LAU14}.
Although the climate in the coastal Los Angeles basin is typically mild-to-hot and dry throughout the year, 
heavy rainfall is concentrated in the months of February and March, 
potentially affecting crime rates. Similarly, large numbers of tourists visit Santa Monica during the summer.
It is thus important to remove seasonal effects from the time series to better understand underlying trends.  
As mentioned in Sect.\,\ref{sect:data}, the raw data lists the date of each 
crime; for convenience we aggregate all occurrences by month to produce a crime time series $Y(t)$ where $t$ is a discrete variable that labels each month
from January 2006 to December 2019.  In order to separate the main trend in crime progression from possible periodic perturbations, 
we use the Seasonal and Trend decomposition using Loess (STL decomposition) method on our data set
\citep{STL1990}. Here, the full crime time series $Y(t)$ is decomposed into a trend $T(t)$, a seasonality $S(t)$, and  a remainder $R(t)$ 
so that $Y(t) = T(t) + S(t) + R(t)$, where $S(t)$ is periodic and $R(t)$
represents any residual fluctuations of $Y(t)$.  We discard the multiplicative option where the
time series is expressed as a product of its components, $Y(t) = S(t) \, R(t) \,T(t) $, since 
we expect seasonality effects to remain relatively stable over the temporal arc of our data.  
We decompose the data using the `stl' function in the R statistical package \citep{R_stats}.
The algorithm requires several parameters to be be specified, including 
$w_{\rm trend}$, the time-frame over which the data is smoothed. Details are illustrated in 
Sect.\,\ref{sec:SItext1} of the Supplementary Information (SI).  

STL decomposition results are shown in Fig.\,\ref{fig:_decomp_prop47}, where $Y(t)$ is the 
number of crimes per month from January 2006 to December 2019 for both Prop.\,47 and non-Prop.\,47 crimes. 
The trend of the number of monthly Prop.\,47 crimes starts to increase
towards the end of 2014, but no corresponding rise is observed for the non-reclassified ones.
In fact, non-Prop.\,47 crimes appear to be declining from 2014 onwards apart from a slight increase around 2018.
This indicates that the rise in the observed reclassified crime trend should not be attributed to a general pattern of increasing crime rates in the city
of Santa Monica, rather it suggests that a specific event in late 2014 is responsible for the observed rise in Prop.\,47 crimes, without playing any role in 
the dynamics of the non-Prop.\,47 ones. We identify this event with the implementation of the new law. 
A significant drop in the $T(t)$ trend emerges for Prop.\,47 crimes towards the end of 2018, persisting
throughout 2019, and concurrent with the several new initiatives undertaken by the SMPD to improve public safety and 
towards more community-based operations.

Interesting observations can also be inferred from the seasonal component $S(t)$: for both 
Prop.\,47 and non-Prop.\,47 crimes the number of offenses increases over spring and summer,
reaching a peak in August, then declining through November. Crime rates increase again
throughout the end-of-the-year holiday season, in December and January, and decline
in February, during the rainy period. Although the main features of
the seasonality components $S(t)$  of the reclassified and non-reclassified crimes in Fig.\,\ref{fig:_decomp_prop47} are similar,
some differences arise, most notably behaviors in the spring and fall months. These slight discrepancies
might be ascribed to some crimes being more affected by seasonal changes than others.

\subsection{A change-point in the reclassified crime trend occurs in late 2014}
\label{subsec:changePoint}

\indent 
Having isolated the trend component $T(t)$ from the time series $Y(t)$, 
we determine whether any statistically significant changes in 
$T(t)$ arise. If so, we also aim to identify the times at which these changes occur, and the associated
confidence intervals.  To do this, we use change-point analysis, a well developed
method that has been applied to many disciplines, from economics to medicine \citep{PAG54, PAG57, CHE11}.
Once a time series is given, the basic foundation of change-point analysis is to evaluate a statistical quantity on 
a subsample of the data immediately prior and immediately after each time point.  
If the difference between the prior and after quantities surpasses a given threshold, the selected time point is the
locus of a change-point, given that some consistency requirements are met. 
This concept can be applied to the mean, variance, or any moment or derived property of the data  \citep{R_changepoint, R_changepoint_updated}.  
Operationally, the detection of a change-point is framed as a hypothesis test, where
the null hypothesis is that there are no change-points and the alternative hypothesis is that at least one exists \citep{PAG54, PAG57}. 

\begin{figure}[t!]
\includegraphics[width=\linewidth]{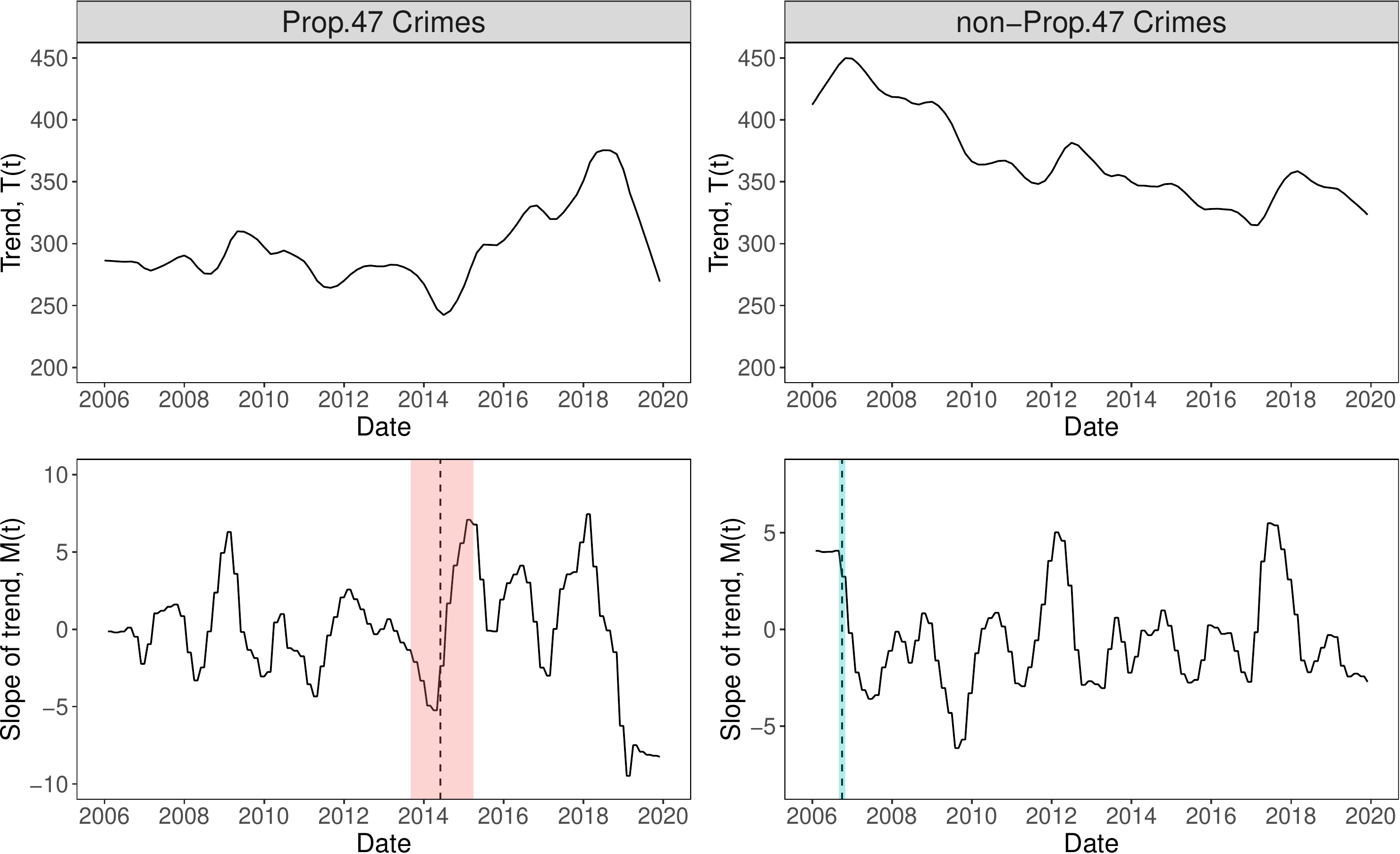}
\caption{Change point analysis on the $M(t)$ slope derived from the $T(t)$ trend for Prop.\,47 crimes (left) and 
    non-Prop.\,47 crimes (right). The top panels are the $T(t)$ trends derived from the $Y(t)$ monthly data using the smoothing 
    window $w_{\rm trend} = 19$ months  in the  decomposition. The lower panels are the derived $M(t)$ slopes 
    on which change-point analysis is performed after  the `mosum' test statistic parameters $G, \eta, \epsilon$ are specified 
    as detailed in the main text and in Sect.\,\ref{sec:SItext2} of the SI. For both Prop.\,47 and non-Prop.\,47 crimes
    $G=10$ months, $\eta = 12$, $\epsilon = 0.5$. The left hand panel for reclassified crimes shows a change-point date of July 2014 (dotted line) 
    with a 95\% confidence interval between September 2013 and April 2015, which includes November 2014.  
    Most of all other $\{w_{\rm trend}, G, \eta,  \epsilon\}$ combinations yield a change-point centered towards the latter part of 2014 with a 
    confidence interval that includes November 2014. For the non-reclassified crimes in the right panel, the date of the change-point is October 2006 (dotted line) 
    with a 95\% confidence interval between September 2006 and November 2006.  
    Note however, that the non-Prop.\,47 change-point is strongly dependent on the specific $\{w_{\rm trend}, G, \eta,  \epsilon\}$ values chosen and no 
    time-frame emerges that is robust to parameter changes.}
    \label{fig:prop_47_changepoint}
\end{figure}

In our case, since we expect Prop.\,47 to affect the crime trend, we seek to identify
the time when $T(t)$ exhibits the largest rate of change. We thus calculate the difference 
between subsequent $T(t)$ values and derive a new time series for the slope of $T(t)$, which we refer to as $M(t)$. 
The $M(t)$ time series is constructed by evaluating the backward difference on each time point:  
if $t_{i-1} < t_{i}$ are consecutive times, 
we define $M(t_{i}) =(T(t_{i}) - T(t_{i-1}))/(t_{i} - t_{i-1})$.
Since our data points are monthly values, $t_{i} - t_{i-1} = 1$ month and
$M(t_{i}) =(T(t_{i}) - T(t_{i-1}))$.  We then perform a change-point analysis on $M(t)$ 
to detect where changes to the slope are largest.
We compute $M(t)$ from the trend $T(t)$ rather than from the monthly time series $Y(t)$ because 
fluctuations in the latter would propagate to the slope, rendering a change-point analysis inconclusive. 
The trade-off in choosing to work with $T(t)$ rather than $Y(t)$ is that the  smoothing process
may affect our analysis; for example the change-points may depend on the smoothing 
window length $w_{\rm trend}$, as discussed in Sect.\,\ref{sec:SItext1} of the SI.   

Once $w _{\rm trend}$ is selected, the change-points of $M(t)$ are calculated
through the R package `mosum' \citep{R_mosum}; 
the procedure is described in Sect.\,\ref{sec:SItext2} of the SI.  As illustrated, several parameters must be specified,
including the window $G$ over which the prior and after subsamples are evaluated, the minimum allowed distance
$\eta G$ between change-points, and the minimum width $\epsilon G$
of a neighborhood of the change-point where the mosum test statistic surpasses the reference threshold for
all data points in the $\epsilon G$ neighborhood. 
These $G, \eta, \epsilon$ parameters affect the location of the change-point
and the associated confidence intervals, in addition to $w_{\rm trend}$ from the  decomposition.

Results for Prop.\,47 and non-Prop.\,47 crimes are shown in Fig.\,\ref{fig:prop_47_changepoint}.
The estimated change-point for  the Prop.\,47 crimes for $w_{\rm trend} = 19$ months, $G=10$ months, $\eta = 12$, $\epsilon = 0.5$
is detected to be July 2014 with a 95\% margin of error that includes November 2014. Other choices of $\{w_{\rm trend}, G, \eta, \epsilon\}$  yield different change-point estimates, 
most notably reducing $w_{\rm trend}$ will shift the change-point towards later dates.
For example $w_{\rm trend} = 5$ months, $G=10$ months, $\eta = 10$, $\epsilon = 0.5$
yields a change point of August 2014 with a 95\% margin of error of ten months, 
which also includes November 2014.  We performed change-point analysis for a large set of $\{w_{\rm trend}, G, \eta, \epsilon \}$ combinations.
For all of them changes in the rate of Prop.\,47 crimes emerged towards the second half of 2014, between June 2014 and August 2014,
The choice of $w_{\rm trend}=1$, which corresponds to building the slope $M(t)$ from the full time series $Y(t)$
without any smoothing procedure, typically yields no change-points due to the irregularity of the data, as mentioned above. 
Some parameter choices allow us to identify additional change-points at the end
of 2018. For example, $w_{\rm trend} = 19$ months, $G=5$ months, $\eta = 5$, $\epsilon = 0.5$ yield
October 2018 as a new change-point in addition to June 2014. This is true for other $\{w_{\rm trend}, G, \eta,  \epsilon\}$ combinations that 
allow for a smaller window size and a smaller distance between change-points. Thus, while the main change-point remains between June 2014 and August 2014, 
a minor one also arises towards the end of 2018 for Prop.\,47 crimes.
The change-point loci for the non-Prop.\,47 crimes are more heavily dependent on the chosen $\{w_{\rm trend}, G, \eta, \epsilon \}$ parameters, and typically 
do not extend into 2014. In conclusion, most of the parameter combinations
tested yield change-point loci for the Prop.\,47 crimes that remain within the June 2014  to August 2014 window, 
with November 2014 falling within the 95\% confidence interval in all cases.

 \subsection{A breakpoint for Prop.\,47 crimes is located at November 2014}
\label{subsec:breakPoint}

We now perform segmented regression on the monthly time series $Y(t)$ and on the trend $T(t)$
of both reclassified and non-reclassified crimes as an alternative method to identify the time at which changes occur 
in the respective data sets. Segmented regression is typically used when abrupt changes are expected
in the relationship between an explanatory and a response variable.
This relationship is assumed to be piece-wise linear, with segments separated by so-called breakpoints.  
Under the assumption of a single breakpoint, an initial guess of its location
is made and the response variable is fit to two lines, one before and one after the putative breakpoint
with the constraint that the overall fit is continuous at the breakpoint itself. 
The resulting curve is the first estimate on which nonlinear regression models are iterated through
least squares, or weighted least squares, methods until convergence is reached and a breakpoint $t^*$ identified.  
This procedure can be extended to multiple breakpoints.

 \begin{figure}[t!]
\renewcommand{\thesubfigure}{\Alph{subfigure}}
\captionsetup[subfigure]{justification=centering}
\begin{subfigure}{.5\linewidth}
\includegraphics[width=\linewidth]{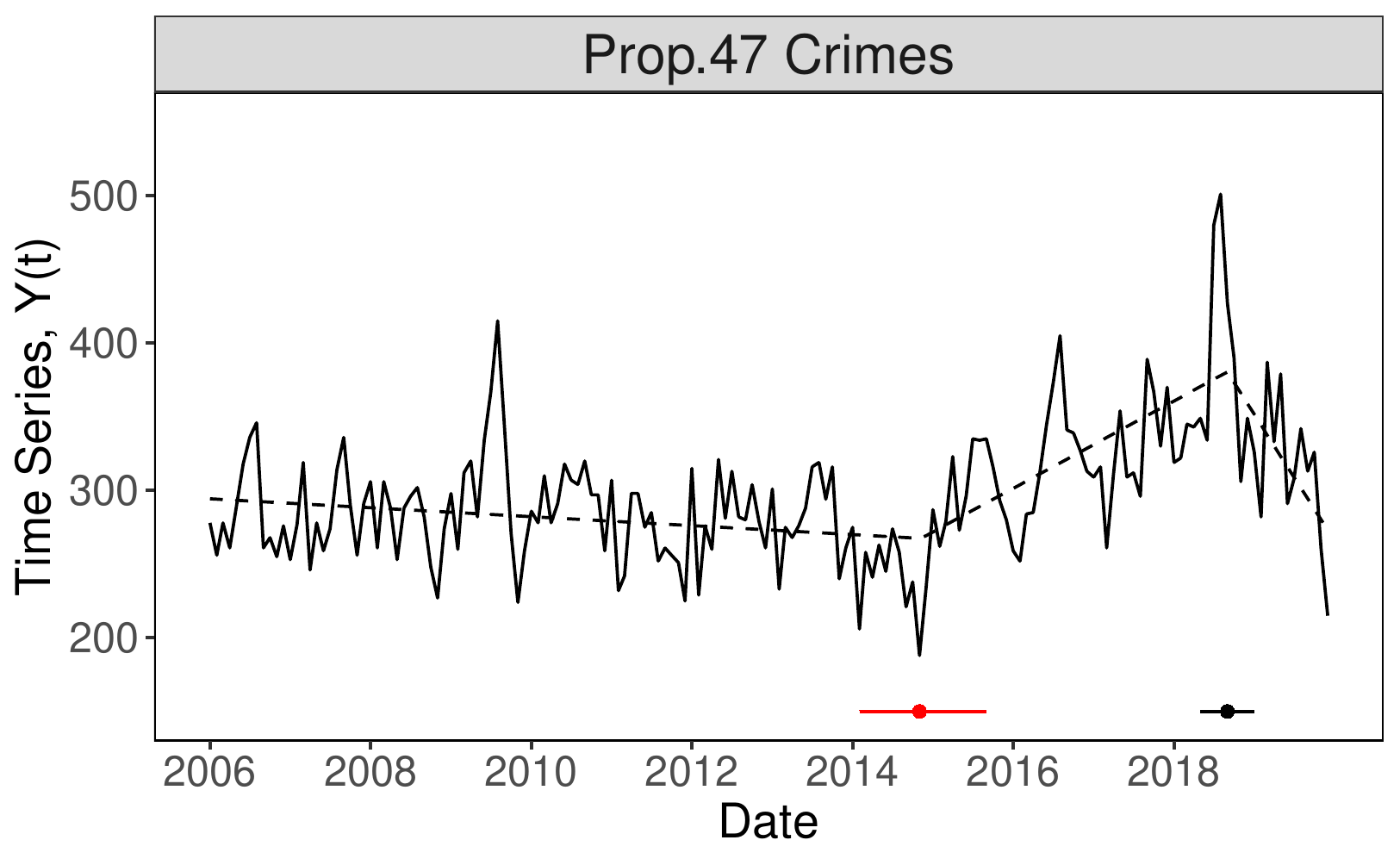}
\caption*{
BP 1: Nov 2014 (Feb 2014 - Sept 2015)\\
BP 2: Sept 2018 (May 2018 - Jan 2019)}
\label{fig:1a}
\end{subfigure}
\begin{subfigure}{.5\linewidth}
\includegraphics[width=\linewidth]{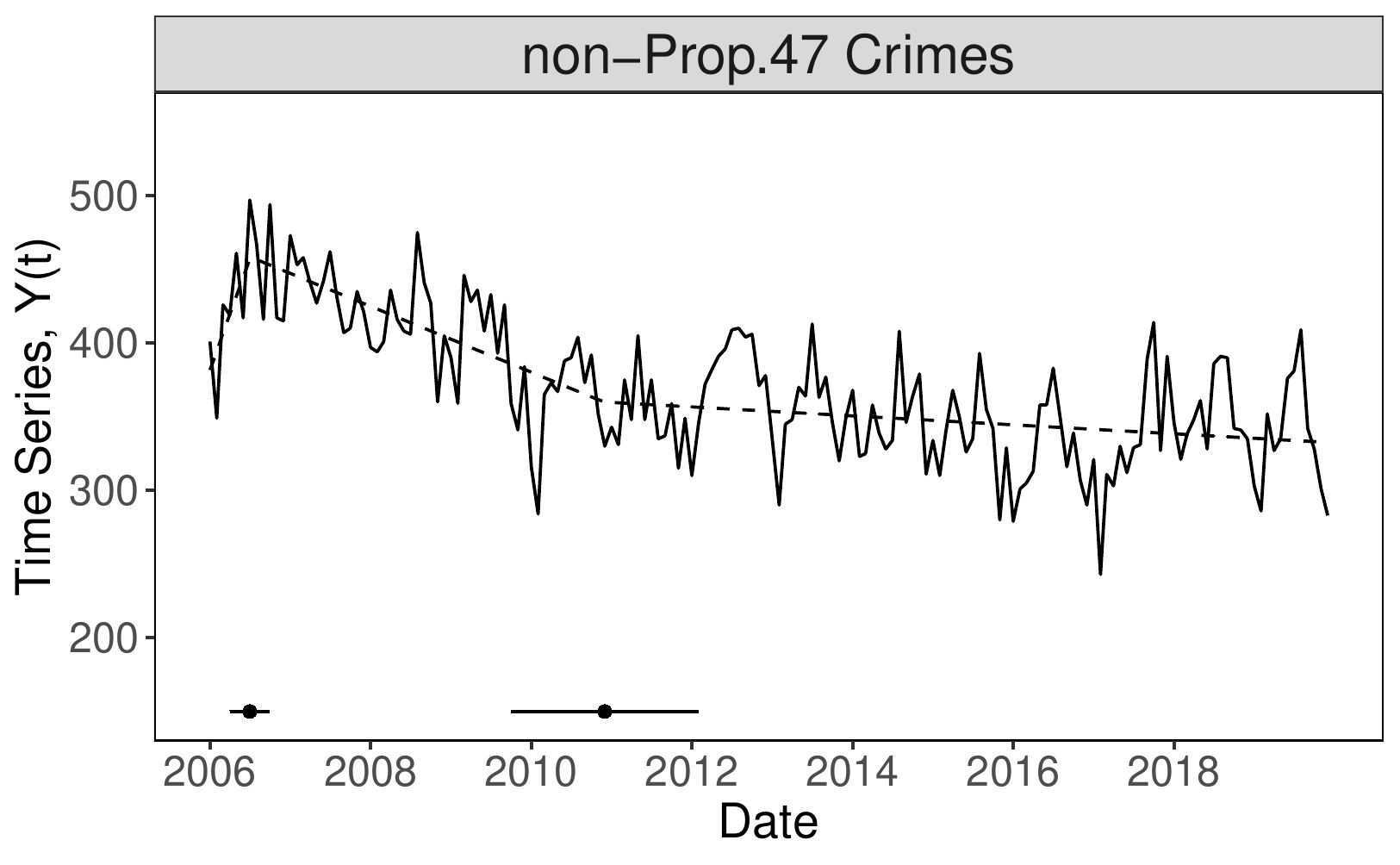}
\caption*{
BP 1: Jul 2006 (Apr 2006- Oct 2006)\\
BP 2: Dec 2010 (Oct 2009 - Feb 2012)}
\label{fig:1b}
\end{subfigure}
\begin{subfigure}{.5\linewidth}
\includegraphics[width=\linewidth]{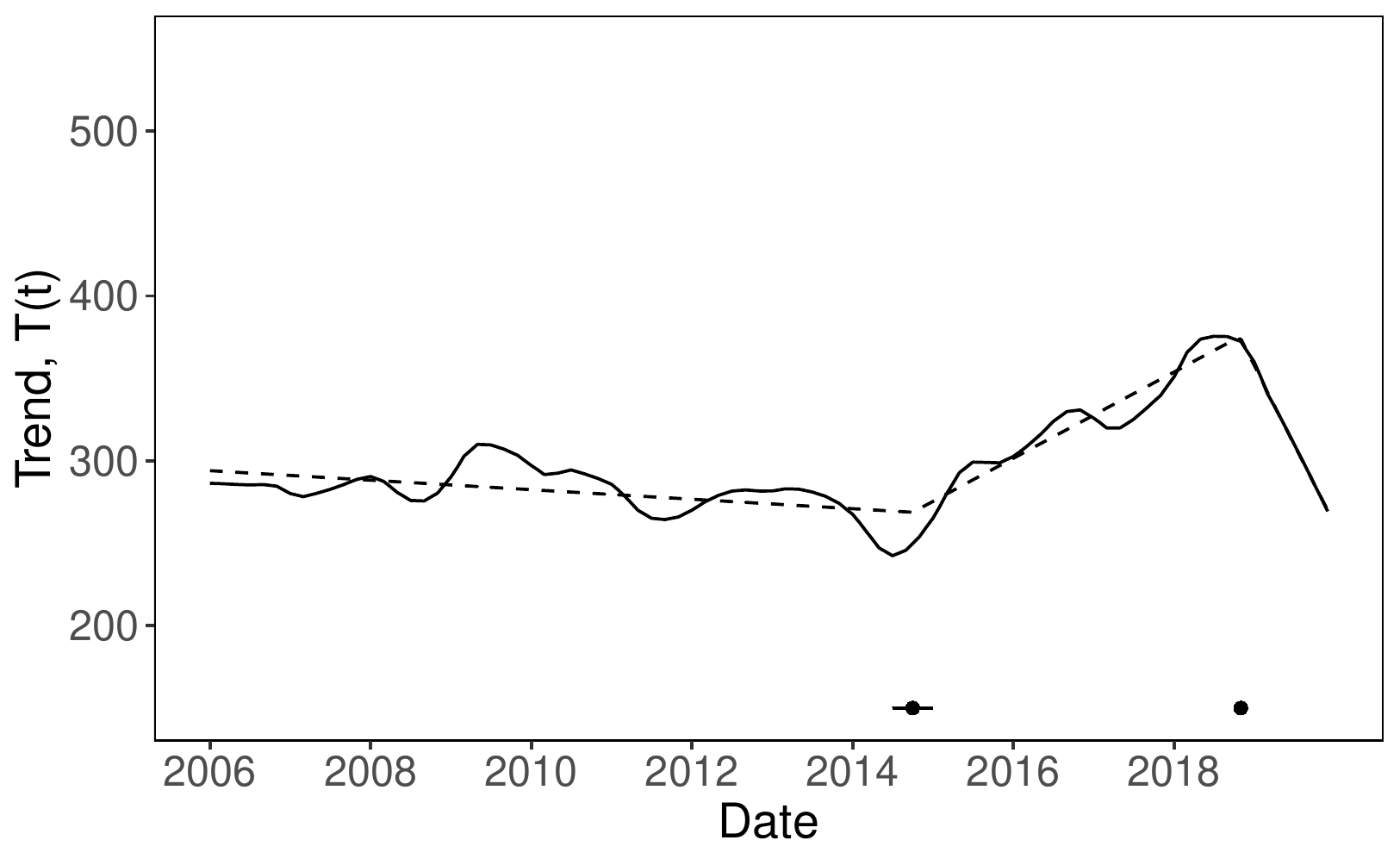}
\caption*{
BP 1: Oct 2014 (July 2014 - Jan 2015)\\
BP 2: Nov 2018 (Oct 2018 - Dec 2018)}
\label{fig:2a}
\end{subfigure}
\begin{subfigure}{.5\linewidth}
\includegraphics[width=\linewidth]{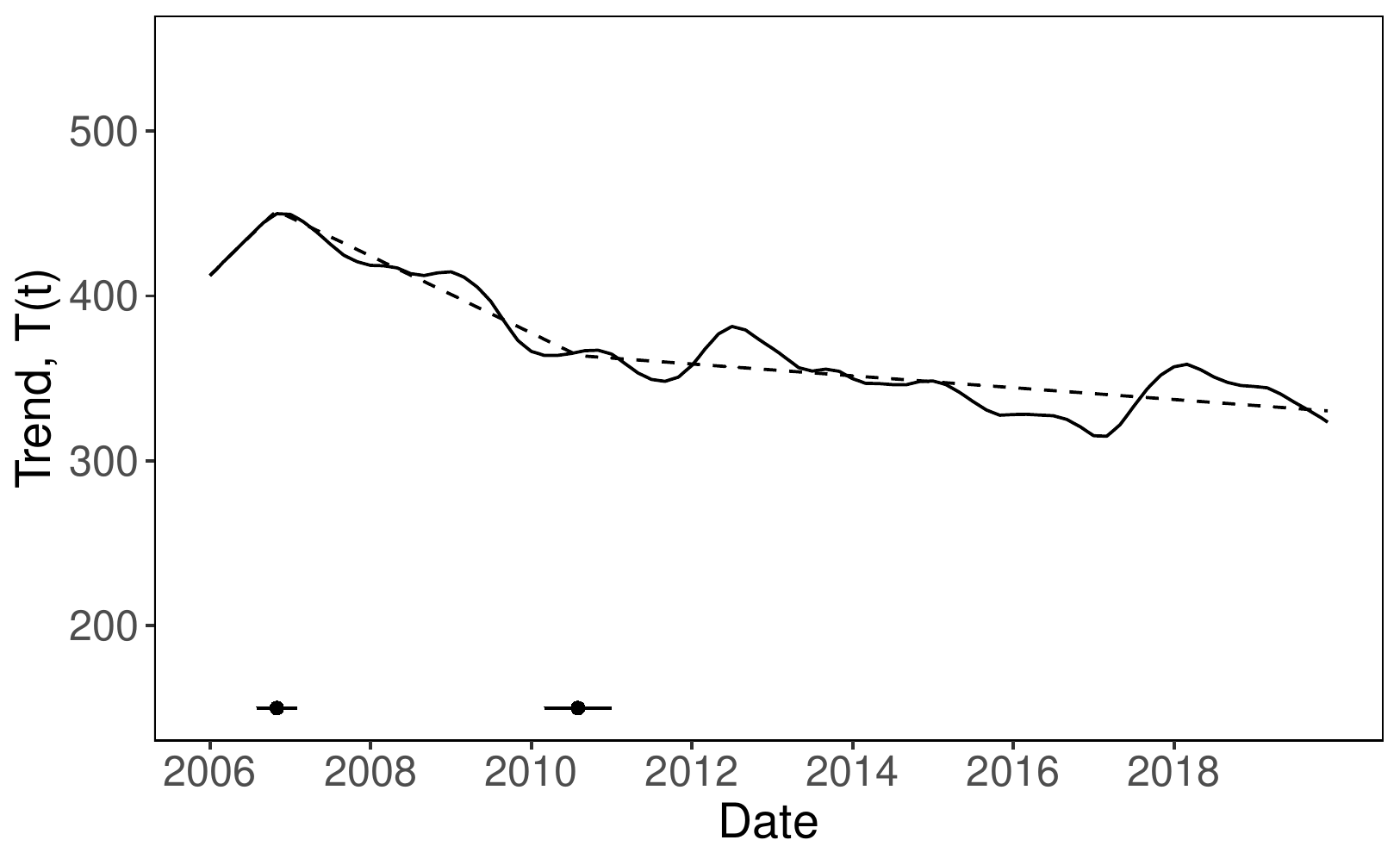}
\caption*{
BP 1: Nov 2006 (Aug 2006 - Feb 2007)\\
BP 2: Aug 2010 (Mar 2010 - Jan 2011)}
\label{fig:2b}
\end{subfigure}
\caption{
Segmented regression with two breakpoints on $Y(t)$ and $T(t)$ for the reclassified (left) and non-reclassified (right) crimes.
The  decomposition assumes $w_{\rm trend} = 19$ months. 
Breakpoints (BP) are represented by dots and the respective 95$\%$ confidence intervals
are marked by bars; they are both reported in the text below each panel.
For the Prop.\,47 crimes the first breakpoint is November 2014 for
$Y(t)$, and October 2014 for $T(t)$. Values of $1 < w_{\rm trend} < 19$ months
preserve the $T(t)$ breakpoint. The second breakpoint is September 2018 for
$Y(t)$, and November 2018 for $T(t)$.
The first breakpoint signals a transition to higher crime for both $Y(t)$ and $T(t)$,
the second one a decrease.
For the non-Prop.\,47 crimes, breakpoint dates are highly unstable and 
sensitive to changes in $w_{\rm trend}$.} 
\label{fig:seg_reg}
\end{figure}
  
We use the R package `segmented' \citep{R_seg_reg}
to perform segmented regression on $Y(t)$ for Prop.\,47 crimes, and on 
the corresponding $T(t)$ obtained by setting $w_{\rm trend} = 19$ months. Since a visual inspection of $T(t)$ 
reveals at least two possible major changes occurring between 2014 and 2015 and
between 2018 and 2019, we impose two breakpoints to the algorithm.
Results are shown in Figs.\,\ref{fig:seg_reg}. 
As can be seen, 
November 2014 emerges as one of the breakpoints for the monthly time series $Y(t)$; 
the corresponding trend $T(t)$ yields October 2014 as a breakpoint
with a 95$\%$ confidence interval that includes
November 2014. The location of the $T(t)$ breakpoint is
relatively stable: lower values of $1 < w_{\rm trend} < 19$ months still yield
October 2014 as breakpoint. The second breakpoint for $Y(t)$ is September 2018; for $T(t)$ and
 $w_{\rm trend} = 19$ months it is November 2018.  Lower values of $w_{\rm trend}$ preserve
 the November 2018 breakpoint, however $w_{\rm trend} = 3$ and 5 months yield September 2018.
If we allow for one or three breakpoints, results are highly unstable, with 
the breakpoint dates changing with each algorithm run. This behavior is indicative of poor \textit{a priori} assumptions on 
the number of breakpoints \citep{BAN06}.

Finally, we perform a segmented regression analysis on the non-Prop.\,47 crimes 
by similarly allowing for two breakpoints. As shown in Fig.\,\ref{fig:seg_reg}
when using the monthly time series $Y(t)$, breakpoints occur on December 2016 and August 2019 with 
 95$\%$ confidence intervals that do not contain November 2014.  Performing a segmented regression on
$T(t)$ obtained by setting $w_{\rm trend} = 19$ months yields one breakpoint on 
November 2006 and one on August 2010 with 95$\%$ confidence intervals that do not include November 2014
in either case. For the non-Prop.\,47 crimes,
breakpoints for both $Y(t)$ and $T(t)$ are highly unstable and highly sensitive to the choice of
$w_{\rm trend}$.  Furthermore, regardless of the value of $w_{\rm trend}$ or the number of breakpoints specified,
we found no 95$\%$ confidence interval that contain
November 2014, or even a proximal time frame, as a likely breakpoint for non-Prop.\,47 crimes.
Hence, the abrupt change in the reclassified time series observed in November 2014 
should be attributed not to an overall increase in crime, but rather to an event that specifically affected this 
category of crimes. As discussed above, we identify this event with the implementation of Prop.\,47 in November 2014.
Similarly, the second breakpoint evaluated on the Prop.\,47 time series and 
occurring in September 2018 may be attributed to the new SMPD policing strategies \cite{REN19, PAU19b, PAU20}
which may have had a stronger impact on Prop.\,47 crimes than on non-Prop.\,47 ones,
for which no corresponding breakpoint was observed. This is because reclassified crimes are 
generally low-level and quality of life offenses, more easily impacted by 
the community-based initiatives and the increased patrolling efforts undertaken by the SMPD,
such as engagement with vulnerable populations, increased illumination, and physical presence.

\section{Neighborhood effects}
\label{subsec:eightneigh}

\begin{figure}[t!]
\centering
\includegraphics[width = 0.6\textwidth]{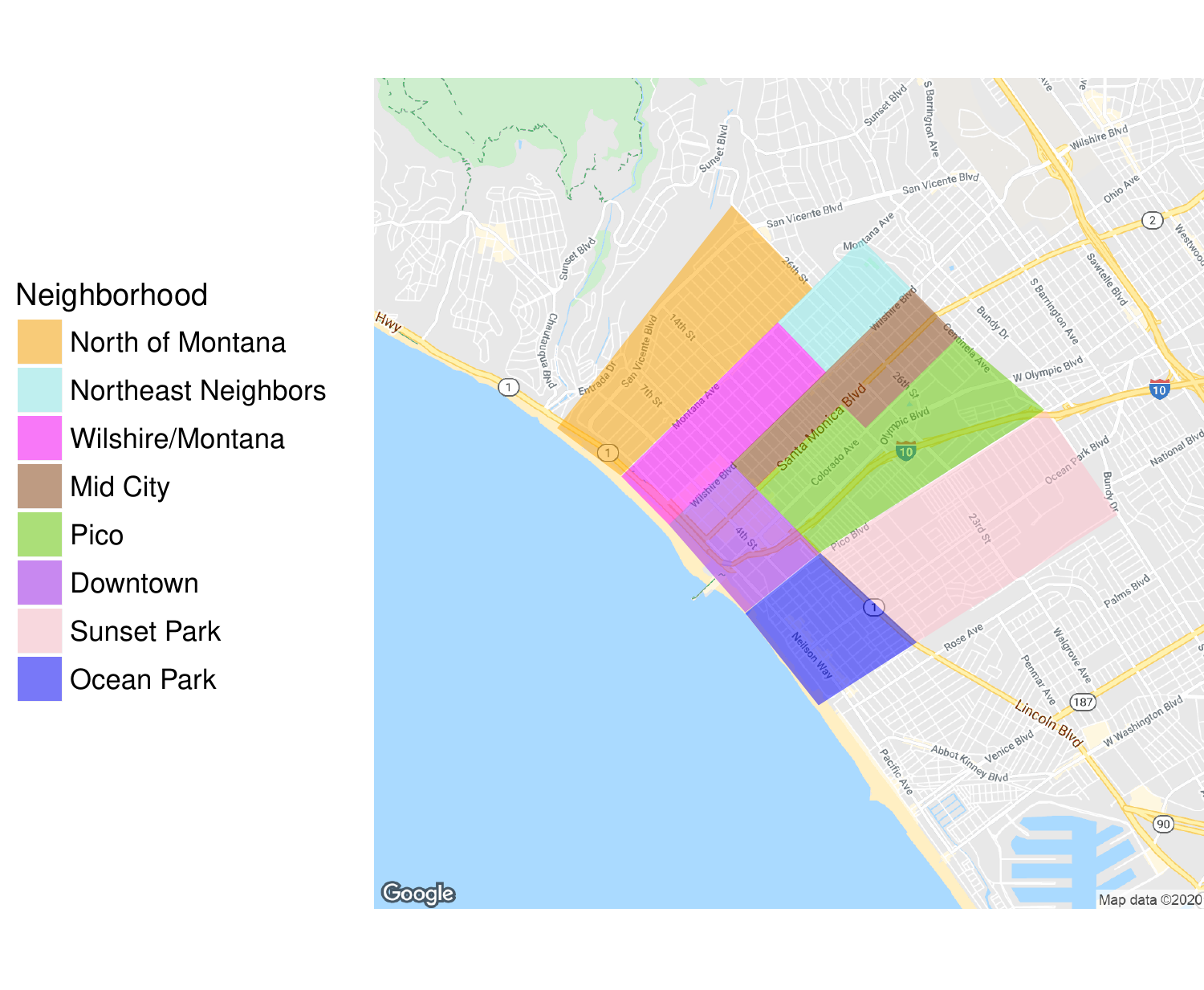}
\includegraphics[width = 0.6\textwidth]{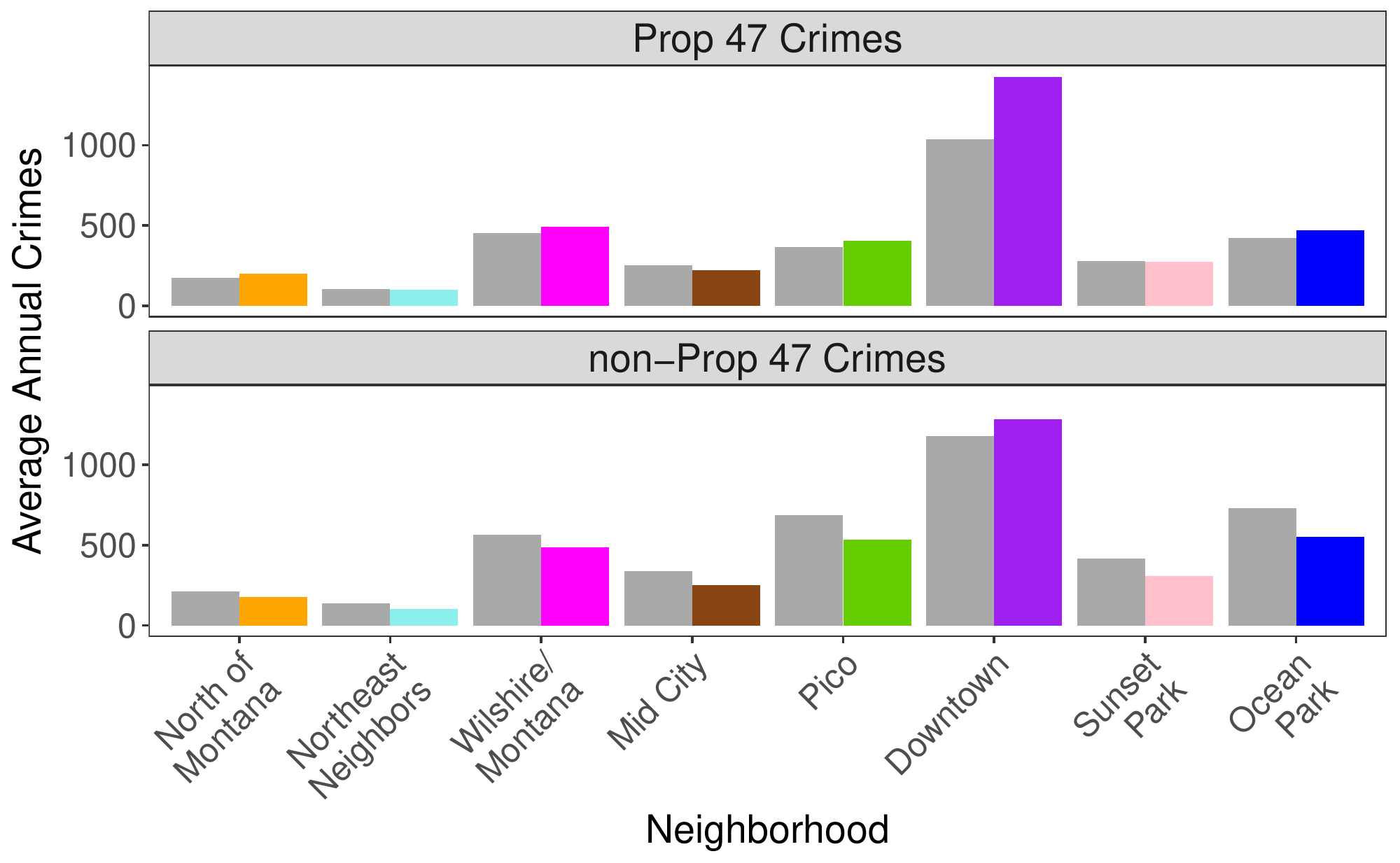}
\caption{(Top) Color-coding of the eight neighborhoods of the city of Santa Monica. 
To the north is Pacific Palisades,  to the south Venice, to the east West Los Angeles,
which are all part of the city of Los Angeles proper. To the west is the Pacific Ocean. 
(Bottom): Average yearly number of crimes in each neighborhood for Prop.\,47 crimes (top) and non-Prop.\,47 crimes (bottom). 
Gray bars indicate the average prior to implementation of the new law in November 2014;  the color-coded bars represent the averages after Prop.\,47 came into effect.}
\label{fig_neighborhood}
\end{figure}

The city of Santa Monica is divided into eight neighborhoods marked by specific boundaries as shown in the map in 
Fig.\,\ref{fig_neighborhood}. These are: North of Montana, Wilshire/Montana, Northeast Neighbors Association, Mid City, Pico, Downtown, Sunset Park, and Ocean Park. To investigate the geographic effects of Prop.\,47, we focus on the before and after Prop.\,47 
incidence of crime in the above districts and plot the average number of crimes per year for the reclassified and non-reclassified crimes in each neighborhood.
The gray bars represent the annual number of crimes before November 2014 and the  
colored ones those after November 2014, with the color-coding mirroring that of the neighborhood map.

\begin{figure}[t!]
\centering
\caption*{\textbf{Prop.\,47 crimes; Monthly averages}}
\includegraphics[width=0.27\textwidth]{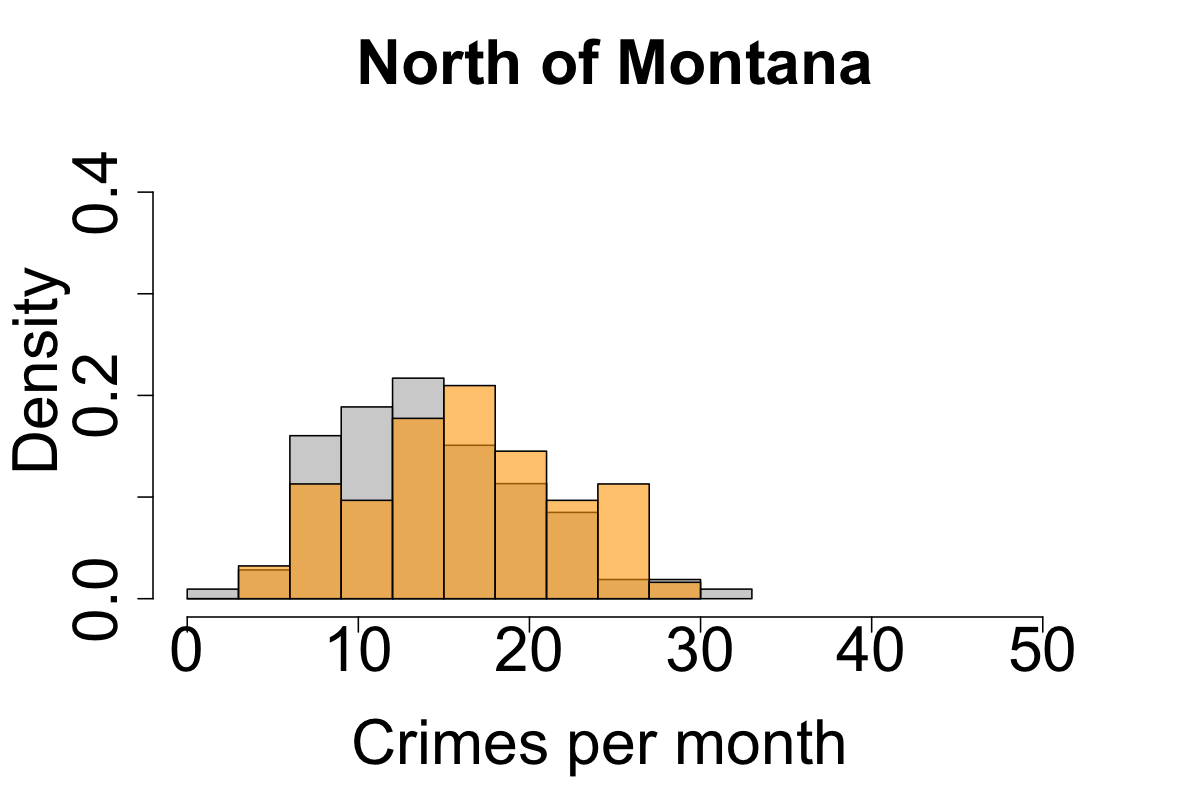}
\hspace{-0.7cm}
\includegraphics[width=0.27\textwidth]{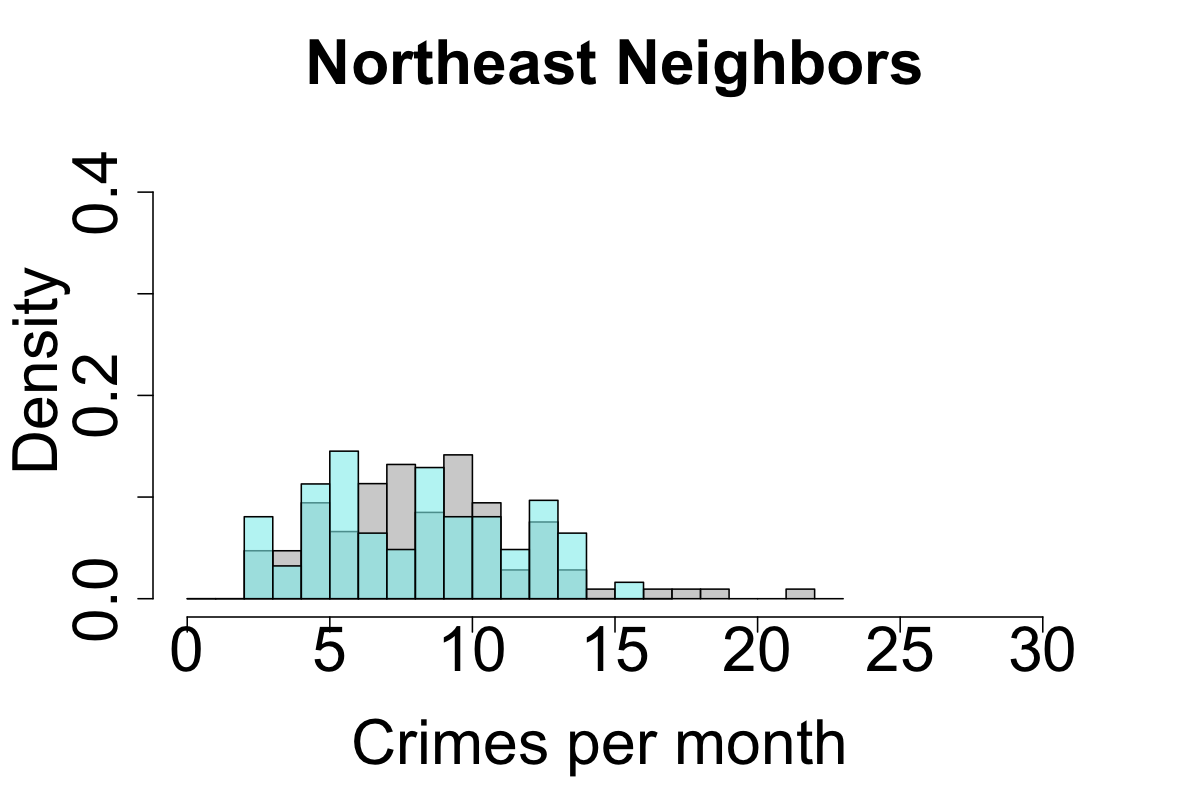}
\hspace{-0.7cm}
\includegraphics[width=0.27\textwidth]{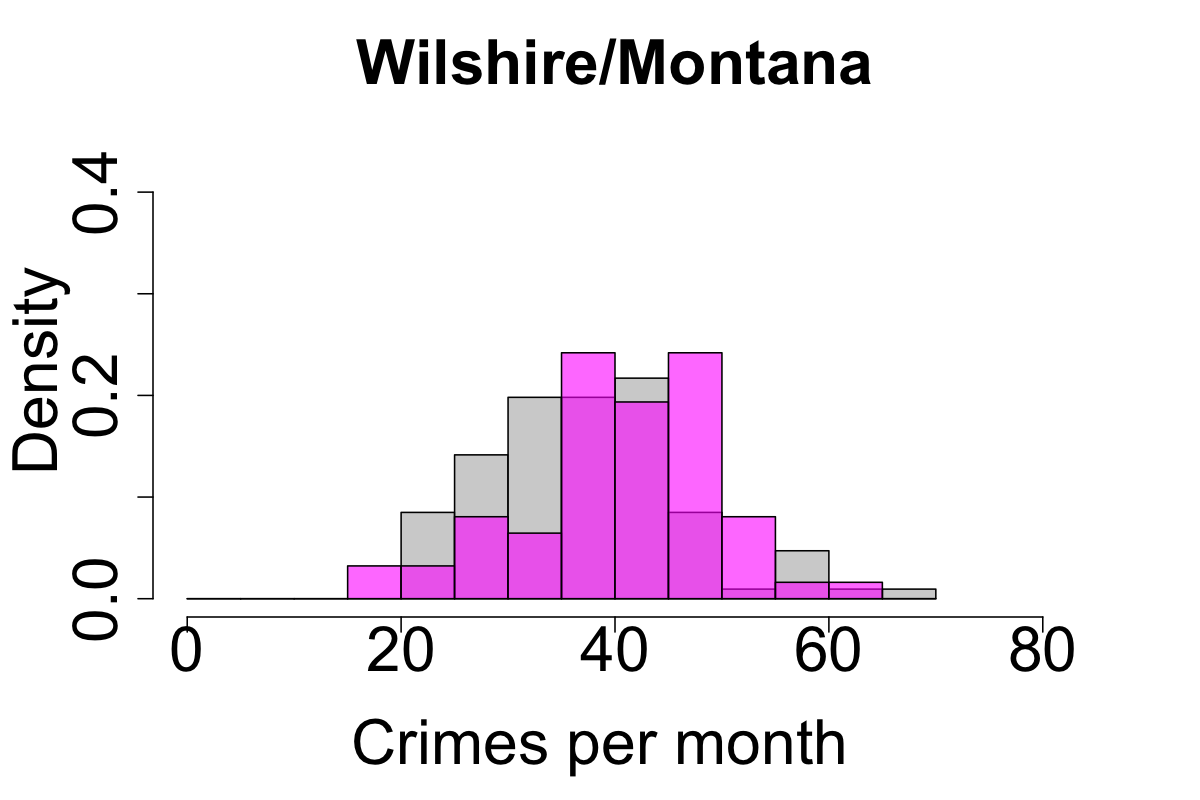}
\hspace{-0.7cm}
\includegraphics[width=0.27\textwidth]{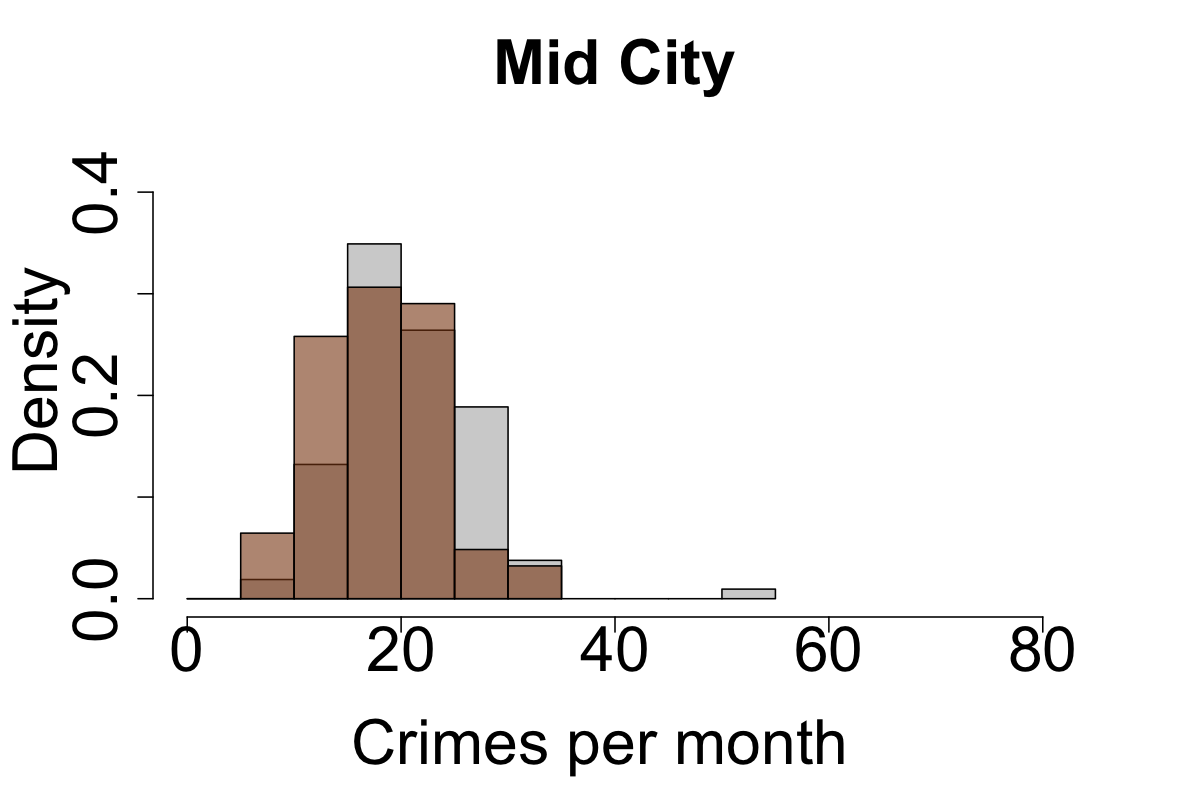}\\
\hspace{-0.7cm}
\includegraphics[width=0.27\textwidth]{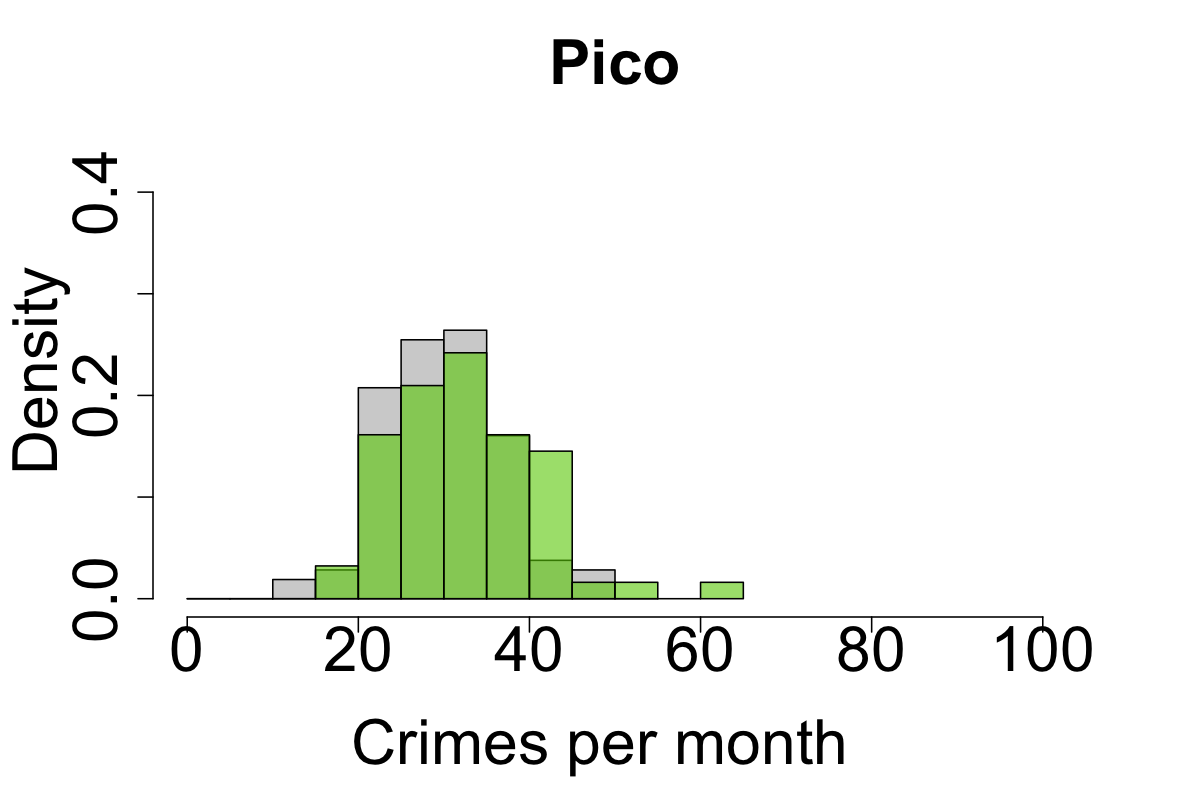}
\hspace{-0.7cm}
\includegraphics[width=0.27\textwidth]{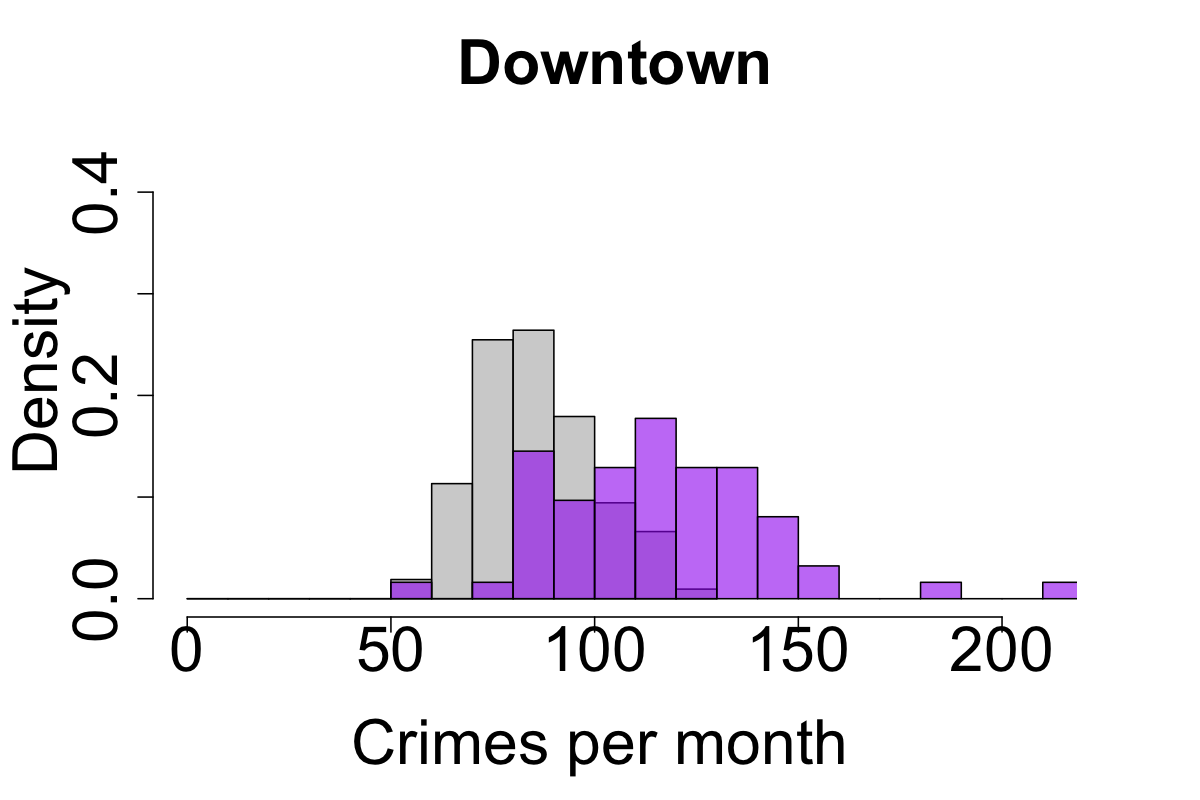}
\hspace{-0.7cm}
\includegraphics[width=0.27\textwidth]{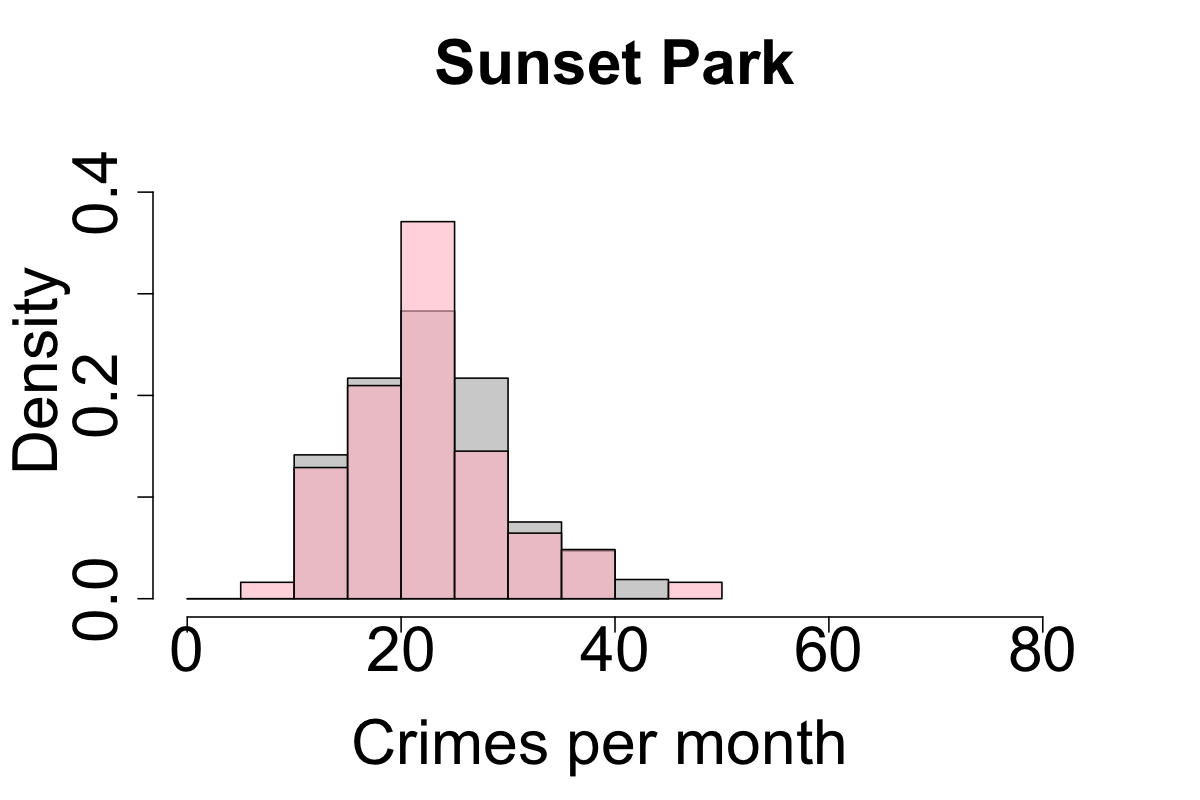}
\hspace{-0.7cm}
\includegraphics[width=0.27\textwidth]{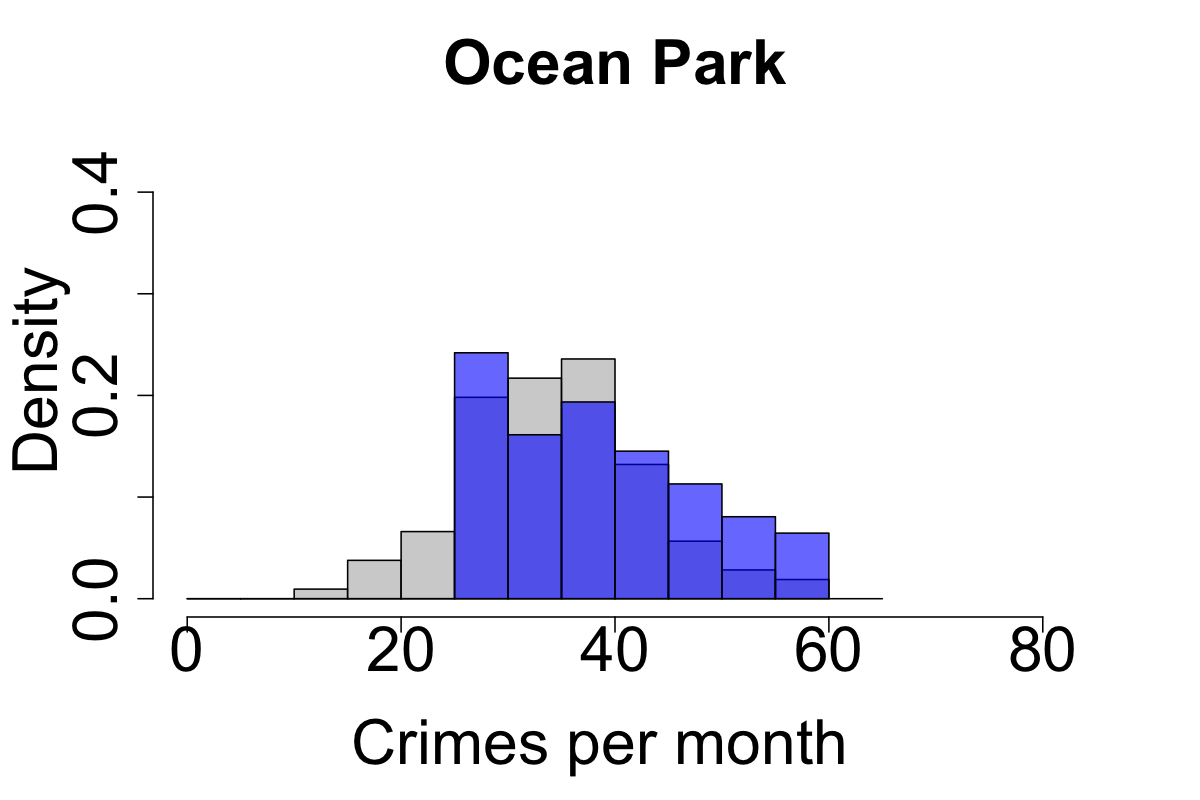}\\
\vspace{0.5cm}
 \caption*{\textbf{non-Prop.\,47 crimes; Monthly averages}}
\includegraphics[width=0.27\textwidth]{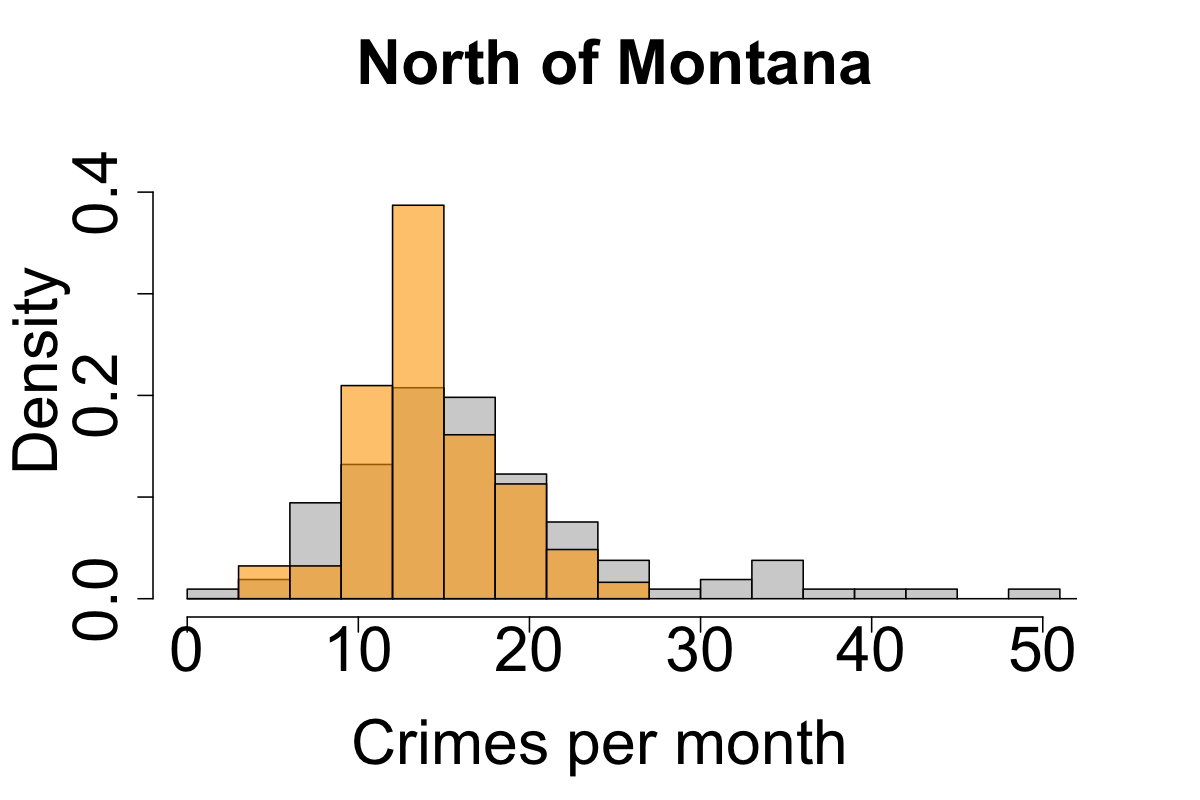}
\hspace{-0.7cm}
\includegraphics[width=0.27\textwidth]{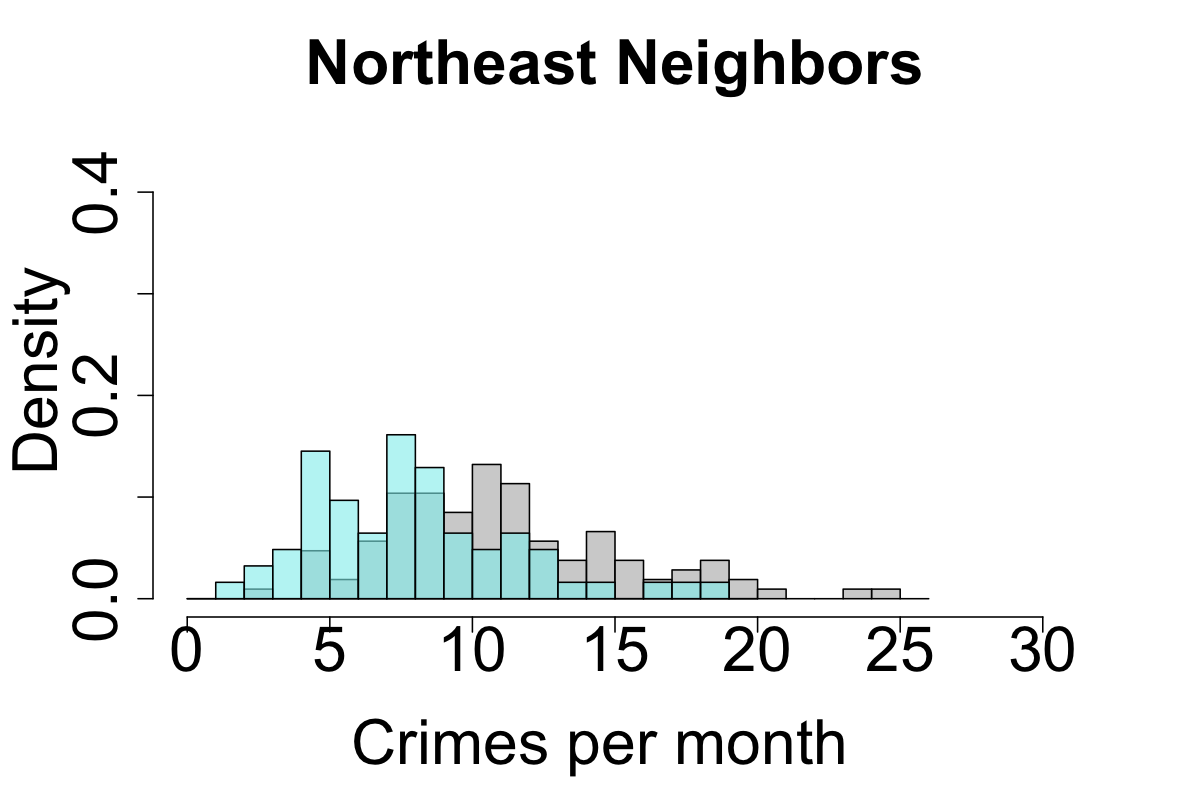}
\hspace{-0.7cm}
\includegraphics[width=0.27\textwidth]{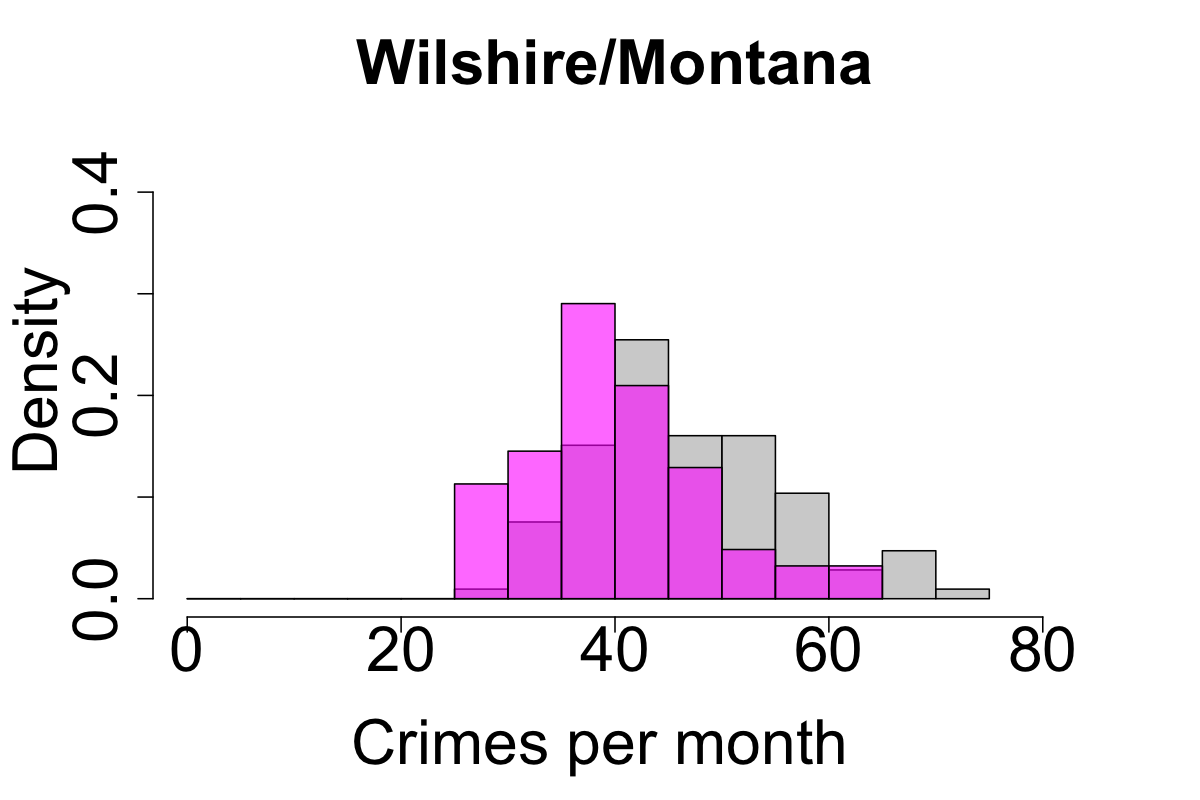}
\hspace{-0.7cm}
\includegraphics[width=0.27\textwidth]{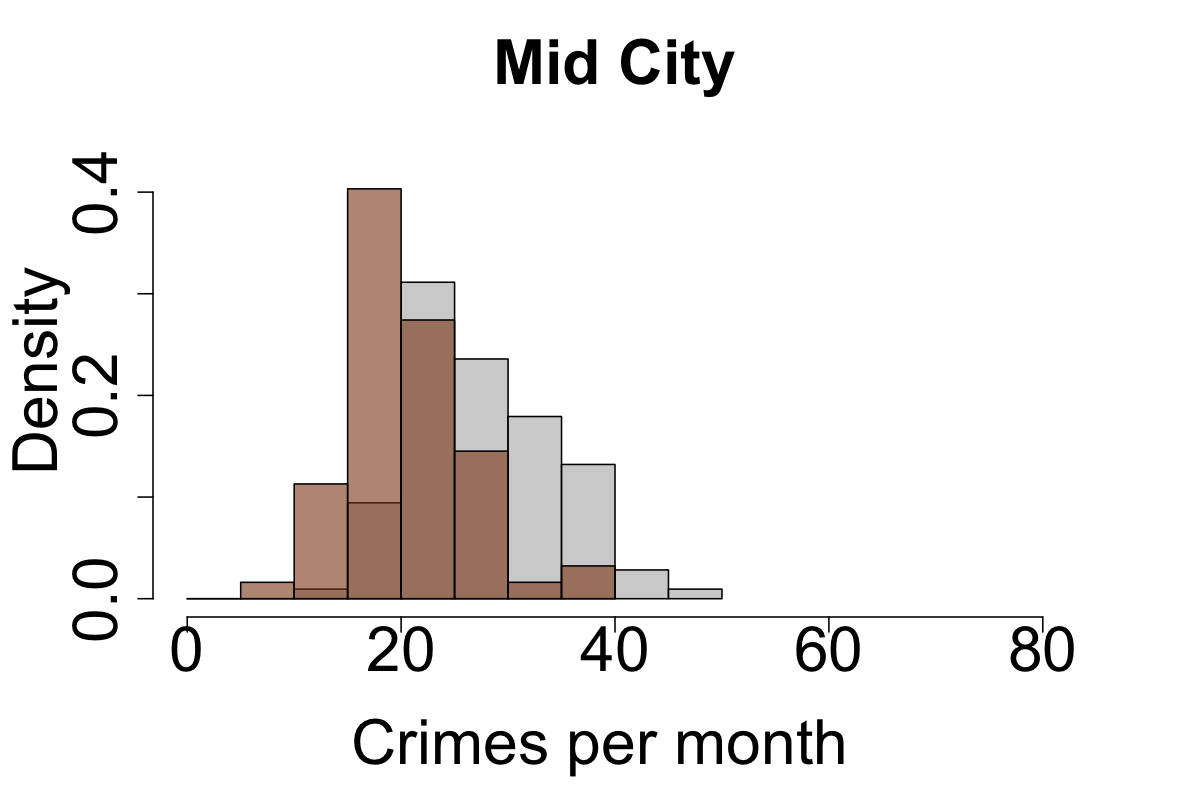}\\
\includegraphics[width=0.27\textwidth]{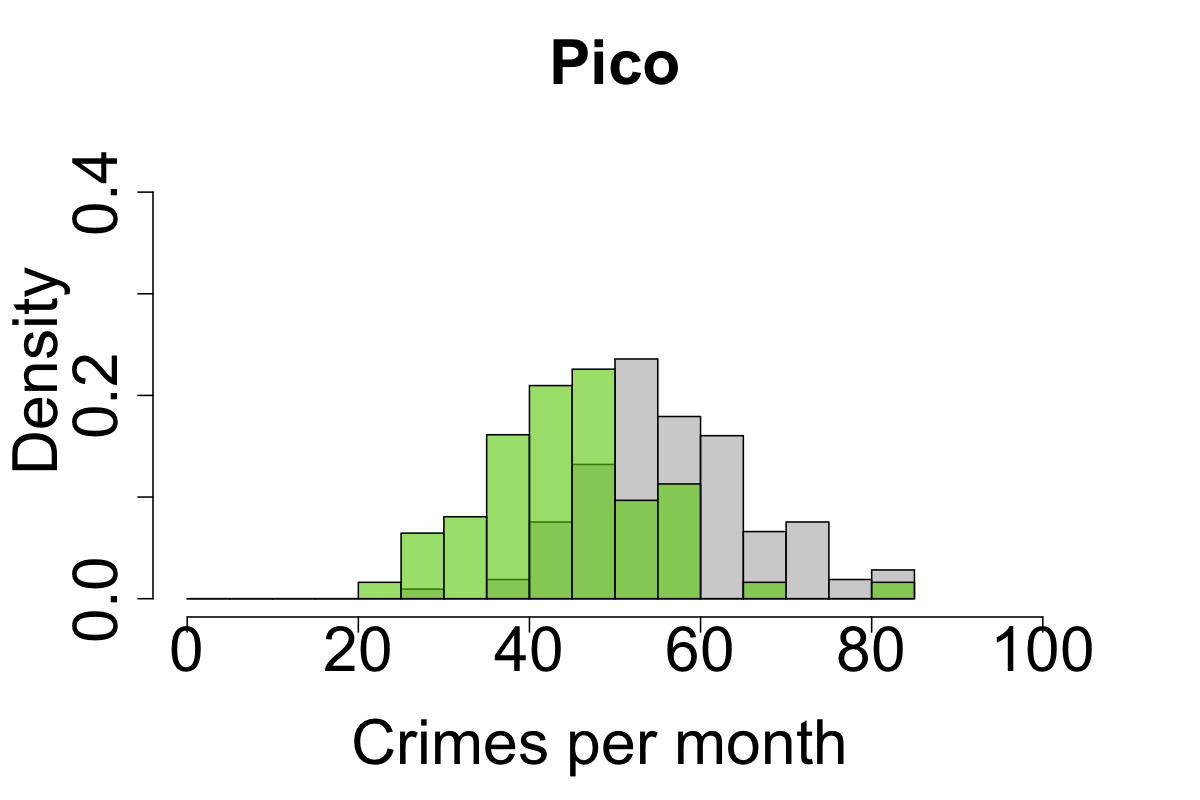}
\hspace{-0.7cm}
\includegraphics[width=0.27\textwidth]{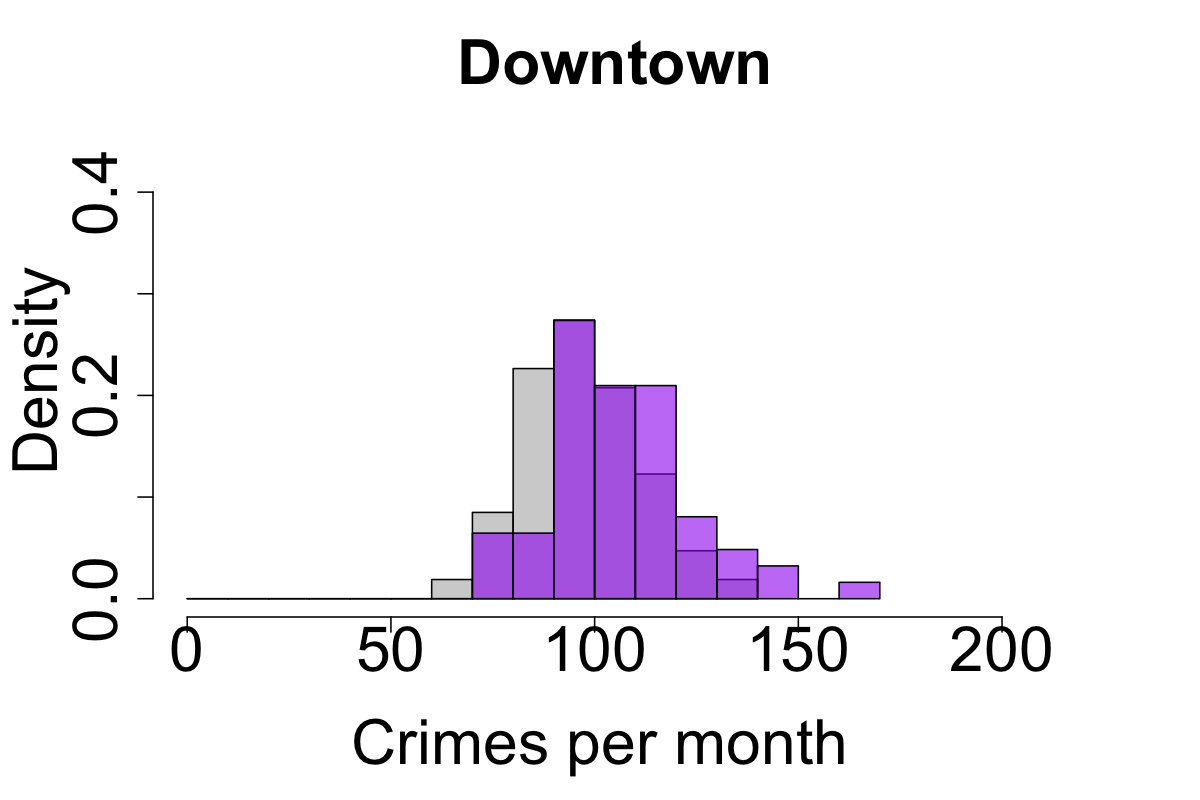}
\hspace{-0.7cm}
\includegraphics[width=0.27\textwidth]{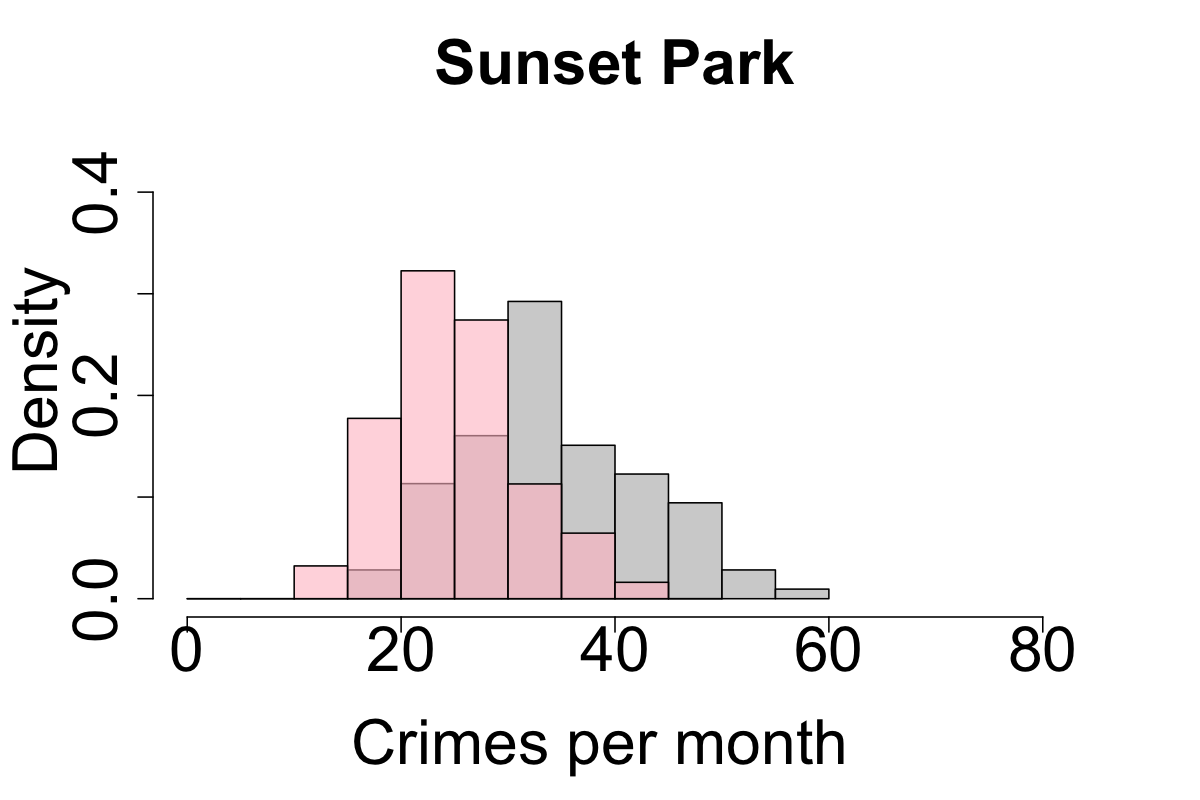}
\hspace{-0.7cm}
\includegraphics[width=0.27\textwidth]{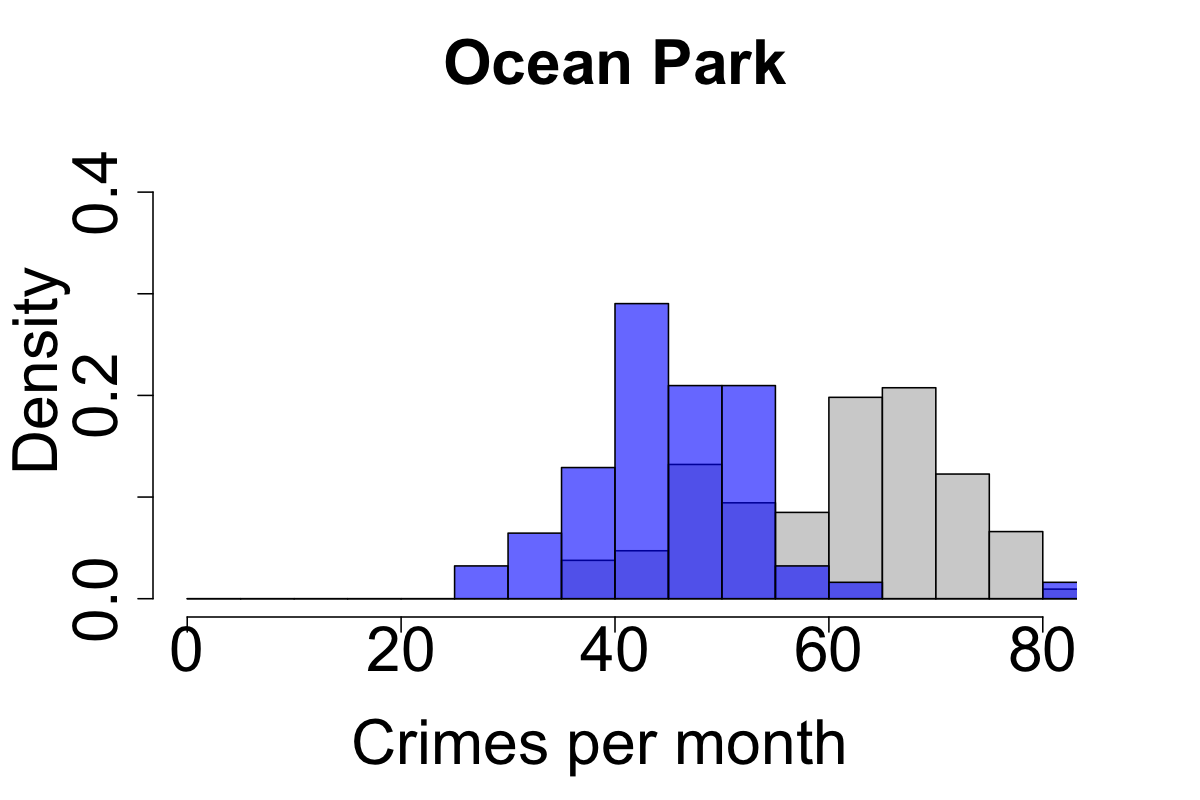}\\
\vspace{0.3cm}
\caption{Histograms of the number of Prop.\,47 (top) and non-Prop.\,47 (bottom) monthly crimes in the eight Santa Monica 
neighborhoods. The gray bars indicate crimes that occurred before November 2014;
 the colored ones those that occurred after November 2014. 
 Table\,\ref{tab:Welch8} provides summary statistics.}
\label{fig_crimePerMonth_neighborhood}
\end{figure}

\begin{table}[ht!]
    \centering
    \centering
    \caption*{\textbf{Prop.\,47 crimes; Monthly averages}}
    \vspace{0mm} 
    \begin{tabular}{|l|c|c|c| c|}
    \hline
    \cellcolor[gray]{0.7} \textbf{Neighborhood} & \cellcolor[gray]{0.7} \textbf{Before} $\{\mu_{\rm b}, \sigma_{\rm b} \}$ 
    & \cellcolor[gray]{0.7} \textbf{After} $\{\mu_{\rm a}, \sigma_{\rm a} \}$ & \cellcolor[gray]{0.7} $t $ \cellcolor[gray]{0.7}&
    \cellcolor[gray]{0.7} \textbf{Significant?}\\
     \hline
    \cellcolor[gray]{0.9} North of Montana  &   \{14.64, 5.71\}  &  \{16.57, 6.02\} & 2.04 &  {\color{red} yes, up +13.2$\%$}\\
    \hline
     \cellcolor[gray]{0.9} Wilshire/Montana  &   \{37.75, 9.16\} &   \{41.07, 9.63\} & 2.20  & {\color{red}  yes, up +8.8$\%$}\\
    \hline
    \cellcolor[gray]{0.9} Northeast Neighbors  &  \{8.84, 3.55\}  &  \{8.50, 3.44\} &  0.62  & no\\
     \hline
    \cellcolor[gray]{0.9} Mid City   &   \{21.08, 6.03\}  &   \{18.62, 5.59\} &  2.67 &  yes, down -11.7$\%$\\
      \hline
    \cellcolor[gray]{0.9} Pico &   \{30.64, 6.92\}  & \{33.61, 8.37\} &  2.36  &  {\color{red} yes, up +9.7$\%$}\\
        \hline
      \cellcolor[gray]{0.9} Downtown  &   \{86.48, 14.59\}  &  \{118.69, 29.31\} & 8.09 &  {\color{red} yes, up +37.2$\%$}\\
      \hline
      \cellcolor[gray]{0.9} Sunset Park  &   \{23.37, 6.92\}  &   \{22.88, 7.49\} &  0.42 & no\\
    \hline
    \cellcolor[gray]{0.9} Ocean Park  & \{35.07, 8.44\}  &  \{39.27, 9.32\}  & 2.92  & {\color{red}  yes, up  +12.0$\%$} \\
    \hline
    \end{tabular}
    \vspace{4mm}
    \centering
    \caption*{\textbf{ non-Prop.\,47 crimes; Monthly averages}}
    \vspace{0mm} 
    \begin{tabular}{|l|c|c|c| c|}
    \hline
    \cellcolor[gray]{0.7} \textbf{Neighborhood} & \cellcolor[gray]{0.7}  \textbf{Before} $\{\mu_{\rm b}, \sigma_{\rm b} \}$ 
    & \cellcolor[gray]{0.7} \textbf{After} $\{\mu_{\rm a}, \sigma_{\rm a} \}$ & \cellcolor[gray]{0.7} $t $ \cellcolor[gray]{0.7}&
    \cellcolor[gray]{0.7} \textbf{Significant?}\\
     \hline
    \cellcolor[gray]{0.9} North of Montana  &   \{17.52, 8.24\} &   \{14.82, 4.13\} &  2.82  & yes, down -15.4$\%$\\
    \hline
     \cellcolor[gray]{0.9} Wilshire/Montana  &   \{47.17, 9.34\} & \{40.67, 8.18\} &  4.71 & yes, down -13.8$\%$\\
    \hline
    \cellcolor[gray]{0.9} Northeast Neighbors  &    \{11.64, 4.24\}  &  \{8.45, 3.63\} & 5.17  & yes, down -27.4$\%$ \\
     \hline
    \cellcolor[gray]{0.9} Mid City   &   \{28.14, 6.86\} &   \{21.08, 5.48\} & 7.33  & yes, down -25.1$\%$\\
      \hline
    \cellcolor[gray]{0.9} Pico &   \{57.20, 10.26\}  &  \{44.72, 10.04\} & 7.71  & yes, down -21.8$\%$ \\
        \hline
      \cellcolor[gray]{0.9} Downtown  &    \{98.29, 14.15\} &  \{107.00, 17.70\} &  3.30  & {\color{red} yes, up +8.9$\%$} \\
      \hline
      \cellcolor[gray]{0.9} Sunset Park  &   \{34.92, 8.38\} &  \{25.50, 6.60\} & 8.05  &  yes, down -27.0$\%$\\
    \hline
    \cellcolor[gray]{0.9} Ocean Park  &   \{61.08, 10.81\} &   \{46.18, 8.59\} & 9.84  &  yes, down -24.4$\%$\\
    \hline
    \end{tabular}
    \vspace{0.2cm}
    \caption{The Welch's t-test applied to the histograms of the Santa Monica neighborhoods
    shown in Fig.\,\ref{fig_crimePerMonth_neighborhood} 
    for Prop.\,47 (top) and non-Prop.\,47 (bottom) crimes. The last column indicates
    whether changes are statistically significant and shows percent changes to the mean.
    The $\{\mu_{\rm b}, \sigma_{\rm b} \}$ quantities are monthly averages and standard deviations of crime before implementation of
    Prop.\,47 calculated over $N_{\rm b} = 106$ months. Their post-imple\-mentation counterparts
    are  $\{\mu_{\rm a}, \sigma_{\rm a} \}$ calculated over $N_{\rm a} = 62$ months.
    The Welch's t-test statistic, $t$, is compared to the Student's t-distribution reference value $t_{\rm s} = 1.66$.  
    See Sect.\,\ref{subsec:meanCrimes} for more details.}
         \label{tab:Welch8}
\end{table}

In five of the eight neighborhoods the number of reclassified crimes increased substantially 
after implementation of Prop.\,47, whereas in the other three, the increase was more modest or a small decrease was observed.
Changes to the occurrence of non-Prop.\,47 crimes after November 2014 are also neighborhood-dependent, but typically
involve significant decreases. More quantitatively, we find that the most impacted areas are 
the Downtown,  North of Montana, and Ocean Park neighborhoods which saw the greatest increase 
in the number of Prop.\,47 crimes per year with 37.5$\%$, 13.4$\%$,  and 12.2$\%$ 
increases after implementation of Prop.\,47, respectively.  Note that 
North of Montana has total crime counts that are much lower relative to the
Downtown and the Ocean Park neighborhoods as can be seen in Fig.\,\ref{fig_neighborhood}.
Non-Prop.\,47 crimes decreased in all areas, except for Downtown where crimes increase 
by an average of 8.9$\%$ per year.

Similarly to the city-wide analysis carried out in Sect.\,\ref{subsec:meanCrimes}, for each of the
eight neighborhoods we construct histograms of the average monthly crime rate before and after implementation of Prop.\,47 for both 
reclassified and non-reclassified crimes. Results for all neighborhoods are shown in 
Fig.\,\ref{fig_crimePerMonth_neighborhood} for 
both Prop.\,47 and non-Prop.\,47 crimes, where gray bars indicate crimes 
occurring prior to November 2014 and colored bars represent
those that occurred after November 2014. We use the same color scheme as in Fig.\,\ref{fig_neighborhood}.
In all districts, histograms for the Prop.\,47 crimes  shift to the right or remain unchanged, indicating 
an increase or stationarity, whereas outcomes for the non-reclassified crimes may shift to the left, indicating a decrease. 
The only exception is Downtown, where the non-Prop.\,47 crime distribution also shifts to the right, indicating an increase. 

In Table \ref{tab:Welch8} we quantify whether these shifts are statistically significant by performing a Welch's t-test
in all neighborhoods for both reclassified and non-reclassified crimes.
The increases of Prop.\,47 monthly crimes in Downtown,
North of Montana, Ocean Park, Pico, Wilshire/Montana neighborhoods are large and statistically significant.  
On the other hand, we observe statistically-significant decreases in non-reclassified crimes in all 
neighborhoods except Downtown, where we see a modest increase in crime per month. 
Overall we observe the largest effects of Prop.\,47 occur Downtown, where the average monthly incidence of 
reclassified crimes increased by $37.2\%$ after implementation of the new law.
This is to be expected as most of the reclassified crimes are crimes of opportunity and 
Downtown Santa Monica is rich in crime generators such as shopping, entertainment, and dining venues
that attract large numbers of residents and tourists, but also potential offenders due to the ample opportunities for crime
these settings offer \citep{BRA95}.
The current analysis reveals that while 
increased crime levels were observed for the reclassified crimes in most neighborhoods,  
the strongest effects were felt in areas that were already primed for a criminal surge. 

In Sect.\,\ref{sec:SItext3} we perform an STL decomposition in each of the
eight Santa Monica neighborhoods, similarly to what was done in Sect.\,\ref{subsec:std}
to analyze time dependent trends at the local level. Fig.\,\ref{fig:prop47_det3} in the SI shows that 
Downtown is marked by the highest increase in crime after implementation
of Prop.\,47 but also by the strongest decrease starting at the end of 2018.
The observed decrease may be an indicator of success for the concurrent
new police initiatives that added more approachable police officers to engage with the community, 
often on foot patrol and in highly frequented areas, and a series of measures to improve illumination and monitoring
of parking garages and other public places that are more prevalent Downtown.

\section{Impacts of the Metro Expo Line light rail extension}
 \label{subsec:train}
 
 \begin{figure}[t!]
\centering
\includegraphics[width=0.8\textwidth]{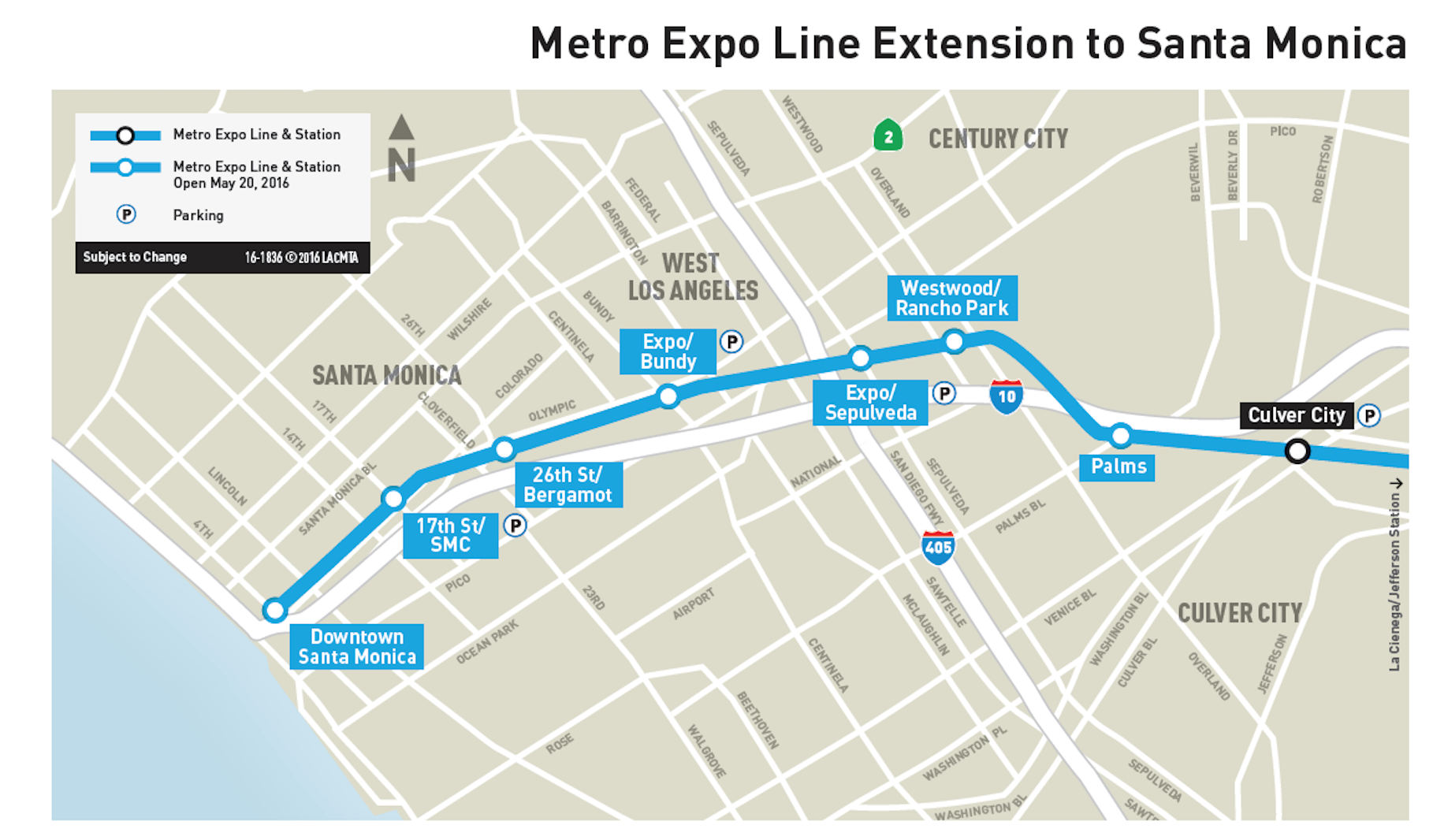}
\caption{Map of the Metro Expo Line light rail extension to Santa Monica. Operations began on May 20, 2016 at seven new stations,
of which four are located within Santa Monica city borders. They are Expo/Bundy,  26$^{th}$ Street/Bergamot,  17$^{th}$ Street/Santa Monica College, 
and Downtown Santa Monica.
The other three stations (Palms, Westwood/Rancho Park, Expo/Sepulveda) are within neighboring Culver City. Prior to May 20, 2016 
the Expo Line's terminus was at the Culver City train stop. Picture courtesy of the Los Angeles Metro.}
\label{fig_Expo}
\end{figure}

The Metro Expo Line extension was inaugurated on May 20, 2016, connecting 
Culver City to Santa Monica via light rail. As discussed earlier 
both the passage of Prop.\,47 and the opening of the Expo Line
have been inculpated for increases in crime \citep{CER16, NEW17, HAR18, RES19, SMN19, SMC19}. 
In this section we aim to better understand whether and how the extension of the light rail affected criminal activity around the four
new train stations located within Santa Monica municipal borders.

\begin{figure}[t!]
\caption*{\textbf{All crimes; Monthly averages before and after the Expo Line (May 2016)}}
\centering
\hspace{-1.33cm}
\includegraphics[width=0.34\linewidth]{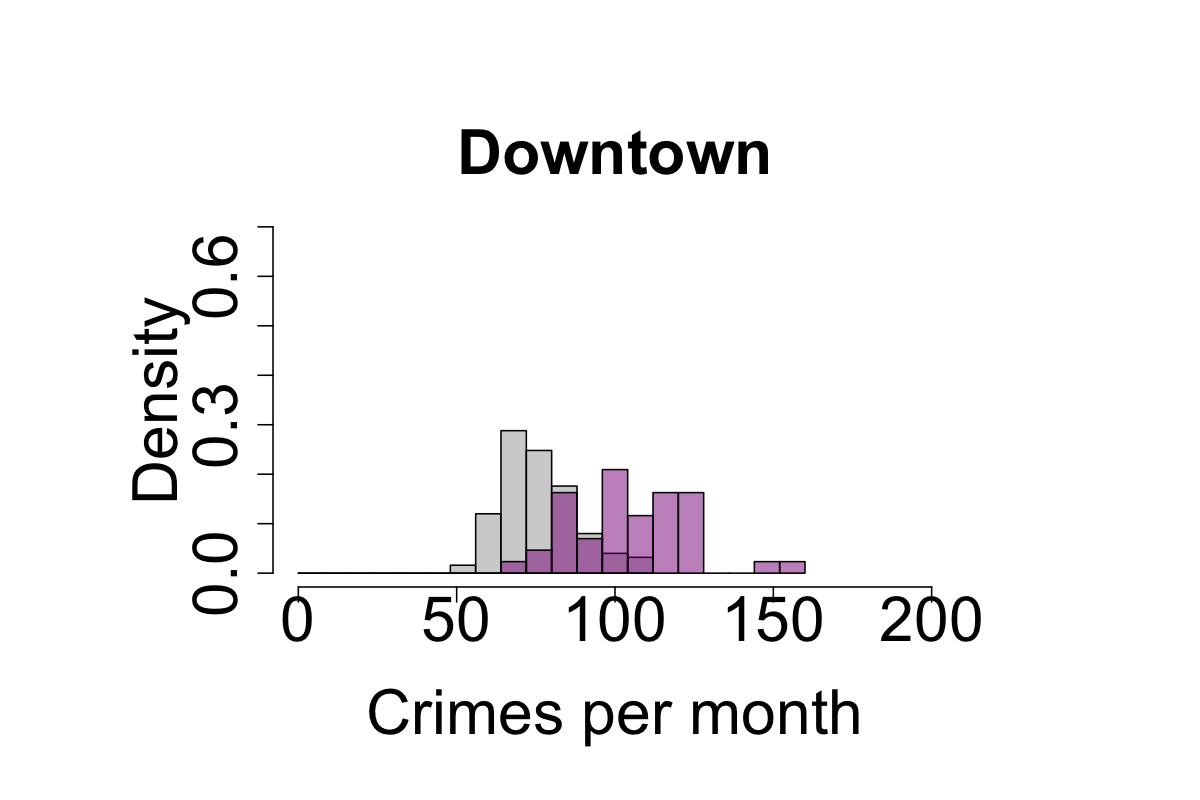}
\hspace{-1.85cm}
\includegraphics[width=0.34\linewidth]{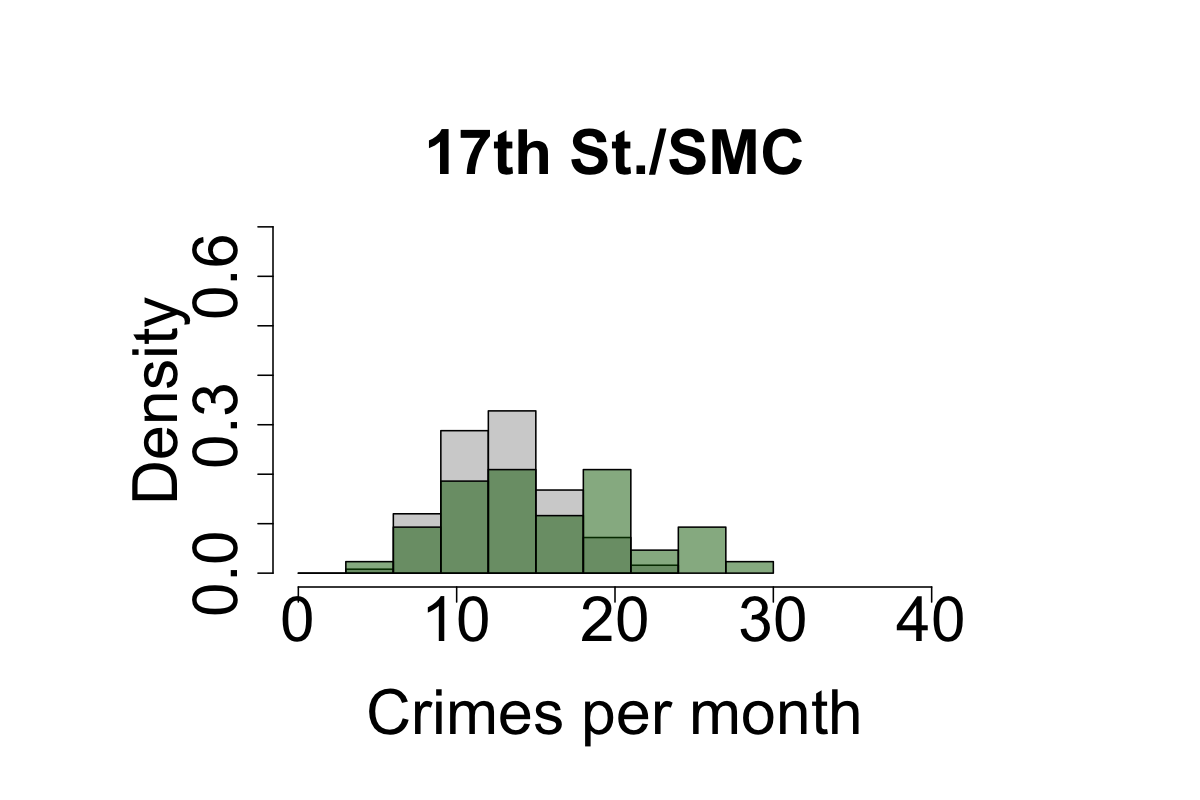}
\hspace{-1.85cm}
\includegraphics[width=0.34\linewidth]{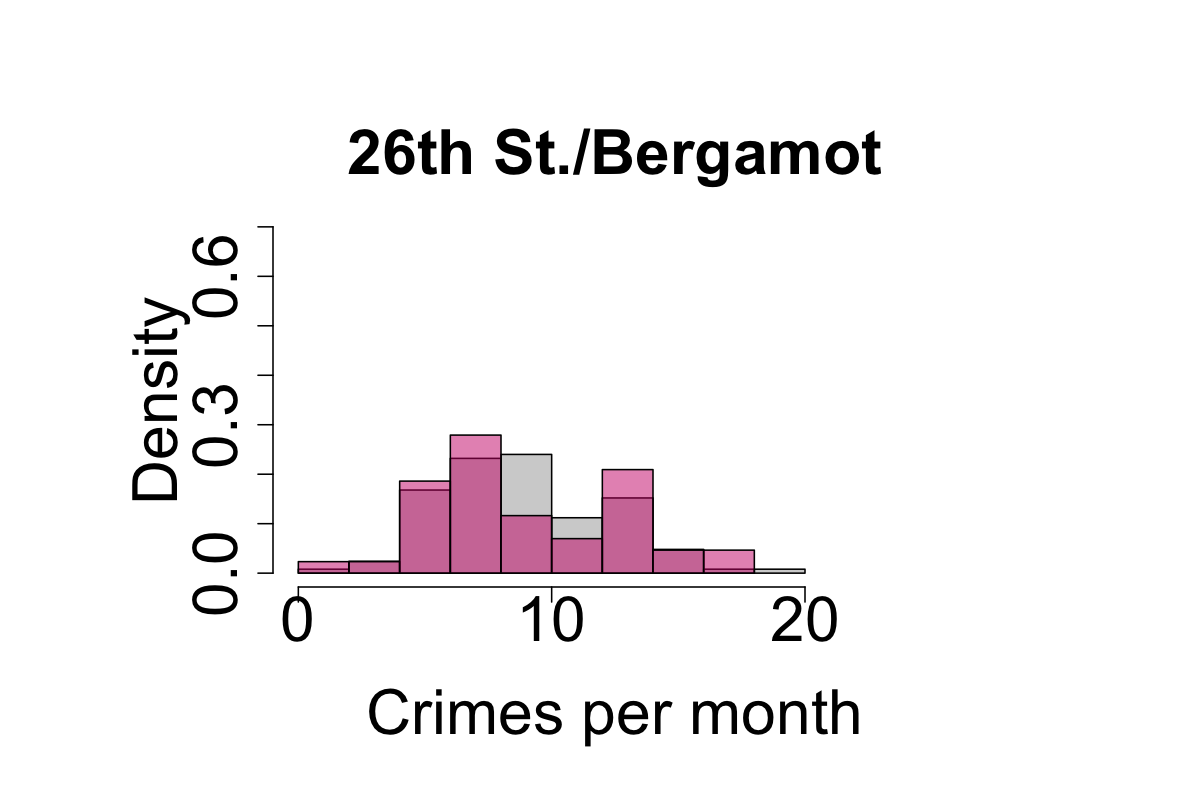}
\hspace{-1.75cm}
\includegraphics[width=0.35\linewidth]{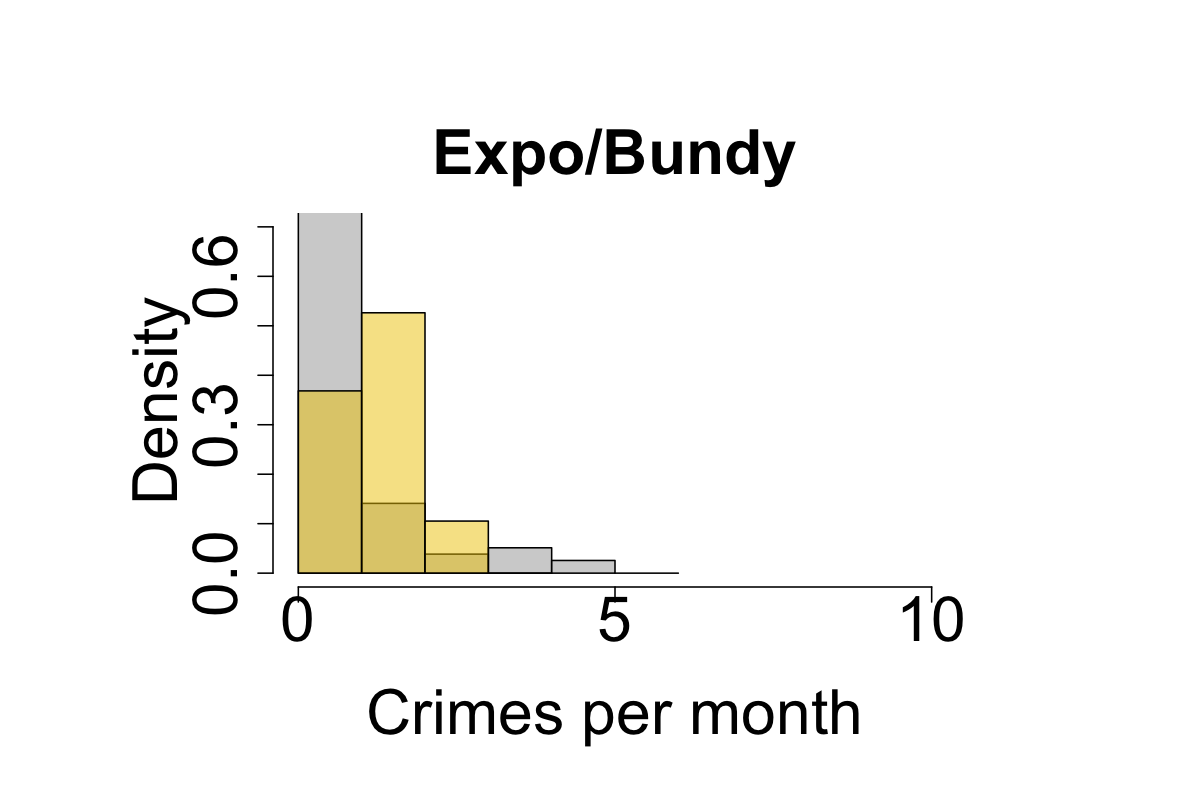}
\hspace{-2cm}
\caption{Histograms of monthly total crime counts in a 450 
meter radius area centered around the four new Expo Line train stations in Santa Monica.
The gray bars represent crime distributions prior to May 2016,  
the colored ones pertain to data after the same date.
Changes to the mean are statistically significant for the Downtown Santa Monica 
and the 17$^{th}$ Street/Santa Monica College 
stations;  they are not for 26$^{th}$ Street/Bergamot and Expo/Bundy.
Corresponding data is listed in Table \ref{tab:Welch4a}.}
\label{fig:expo_line4a}
\end{figure}
\begin{table}[t!]
    \centering
    \caption*{\textbf{All crimes; Monthly averages before and after the Expo Line (May 2016)}}
    \vspace{0mm} 
    \begin{tabular}{|l|c|c|c| c|}
    \hline
    \cellcolor[gray]{0.7} \textbf{Train Station} & \cellcolor[gray]{0.7}  \textbf{before train} 
    & \cellcolor[gray]{0.7} \textbf{after train}  & \cellcolor[gray]{0.7} $t , t_{\rm s}$ \cellcolor[gray]{0.7}&
    \cellcolor[gray]{0.7} \textbf{Significant?}\\
    \cellcolor[gray]{0.7}  & \cellcolor[gray]{0.7}  $\{\mu_{\rm b}, \sigma_{\rm b} \}$ 
    & \cellcolor[gray]{0.7}  $\{\mu_{\rm a}, \sigma_{\rm a} \}$ & \cellcolor[gray]{0.7} \cellcolor[gray]{0.7}&
    \cellcolor[gray]{0.7}\\
     \hline
     \hline
    \cellcolor[gray]{0.9} Downtown   & \{76.51, 12.17\} &  \{105.33, 18.84\} & 9.38, 1.68 & {\color{red} yes, up $+37.7\%$}\\
    \hline
     \cellcolor[gray]{0.9} $17^{th}$ St./SMC  & \{13.38, 3.55\} & \{16.07, 5.71\} & 2.91, 1.68& {\color{red} yes, up $+20.1\%$}\\
    \hline
    \cellcolor[gray]{0.9} 26$^{th}$ St./Bergamot  & \{9.50, 3.21\}  &  \{9.63, 3.77\} & 0.19, 1.67 & no\\
     \hline
    \cellcolor[gray]{0.9} Expo/Bundy   &  \{1.47, 0.98\}  &   \{1.74, 0.65\} & 1.41, 1.68 & no\\
      \hline
      \end{tabular}
    \vspace{0.2cm}
   \caption{The Welch's t-test applied to the total crime histograms for the Expo Line stations shown in Fig.\,\ref{fig:expo_line4a}.
   The last column indicates whether changes are statistically significant and shows percent changes to the mean.
   The $\{\mu_{\rm b}, \sigma_{\rm b} \}$ quantities are monthly  averages and standard deviations of crime before inauguration 
   of the Expo Line calculated over $N_{\rm b} = 125$ months.  Their post-inauguration counterparts 
   are $\{\mu_{\rm a}, \sigma_{\rm a} \}$ calculated over $N_{\rm a} = $ 43 months.
   The Welch's t-test statistic, $t$, is compared to the Student's t-distribution reference value $t_{\rm s}$.  
   See Sect.\,\ref{subsec:meanCrimes} for more details.}
      \label{tab:Welch4a}
\end{table}

The possibility of mass transit leading to rising criminal activity, both inside stations and in their immediate vicinity, 
is well studied. While many studies point to increases
in crime \citep{THR74, BRA91, BRA93, POI96, BLO00, IHL03}, others show that the establishment of mass transit 
does not necessarily lead to a decline in public safety \citep{LOU02, DEN06, SAN09}.
Train and bus routes are usually concentrated in areas with high human activity, offering
more opportunities for predatory crime. However, the impact of mass transit on crime is found to also depend on 
the overall demographic, socio-economic, and land-use contexts surrounding 
transit stops \citep{LEV96, LOU99, LOU02}. Thoughtful architectural, lighting, and environmental design of the station
themselves may help reduce the incidence of crime \citep{LAV96, FEL96, LOU02}. 

\begin{figure}[t!]
\caption*{\textbf{All crimes; Monthly averages (2006-2014; 2014-2016; 2016-2019)}}
\begin{subfigure}{.6\textwidth}
\hspace{-1cm}
\includegraphics[width=\linewidth]{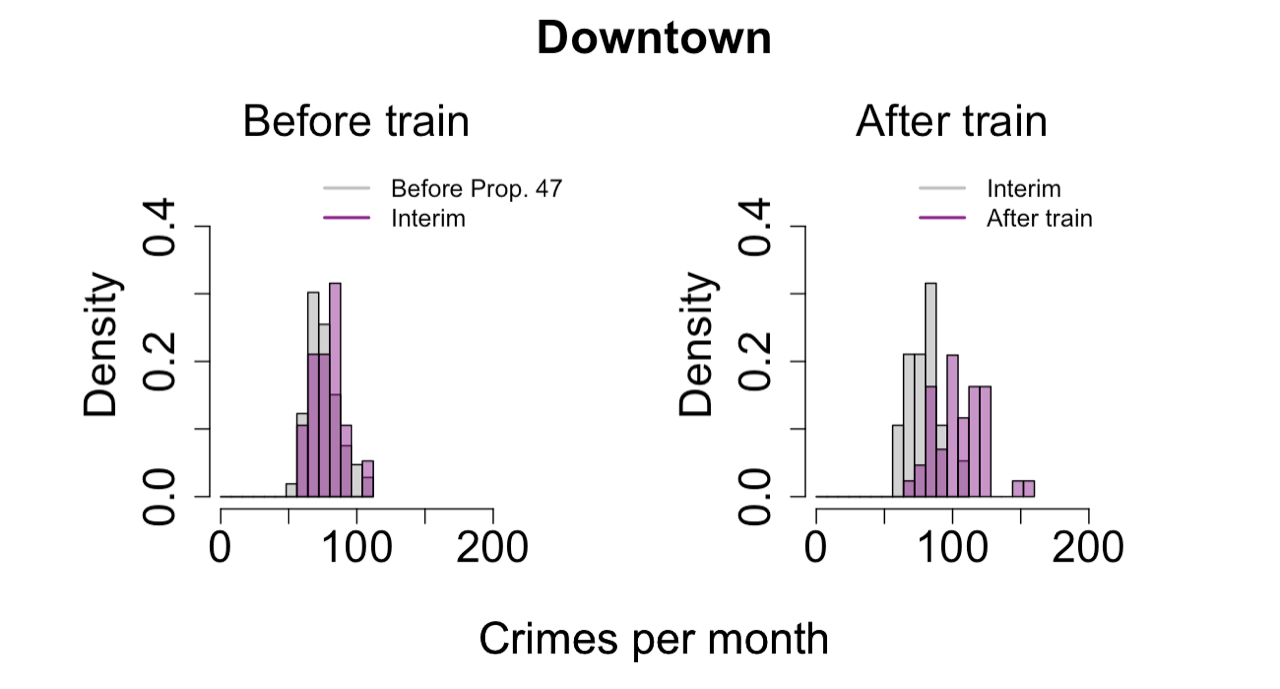}
\end{subfigure}
\begin{subfigure}{.6\textwidth}
\hspace{-3.7cm}
\centering
\includegraphics[width=\linewidth]{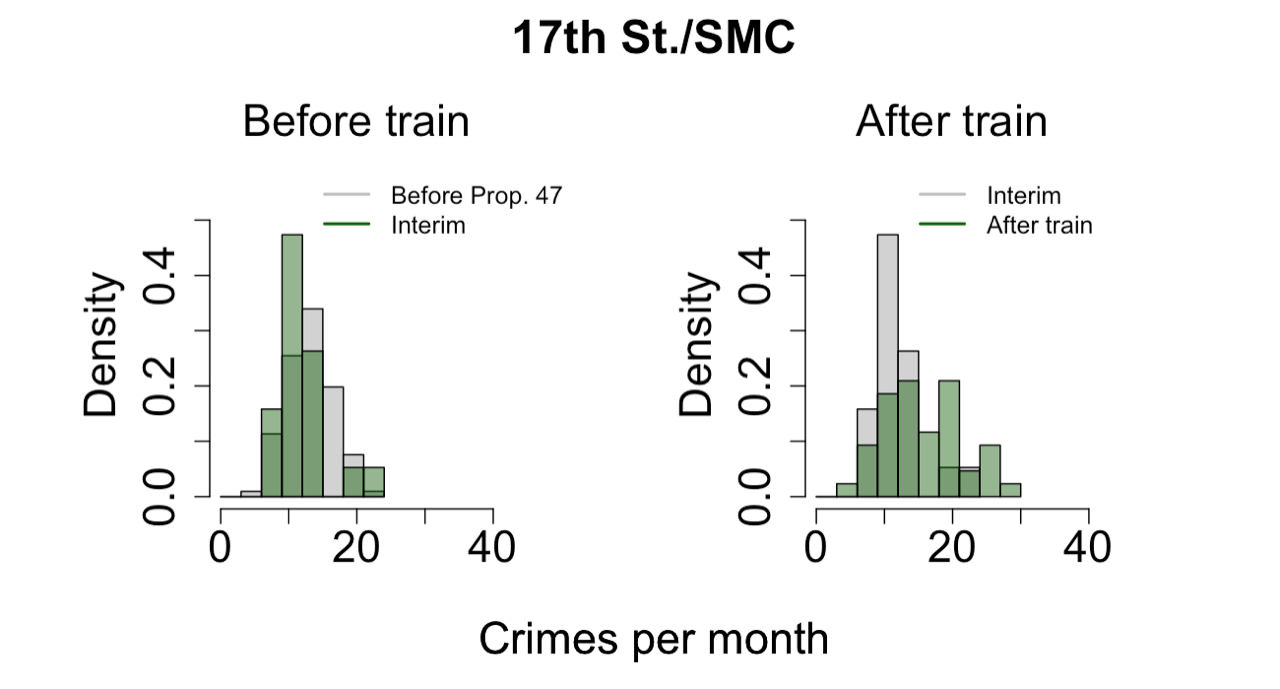}
\end{subfigure}
\vspace{0.5cm}
\newline
\begin{subfigure}{.6\textwidth}
\hspace{-1cm}
\centering
\includegraphics[width=\linewidth]{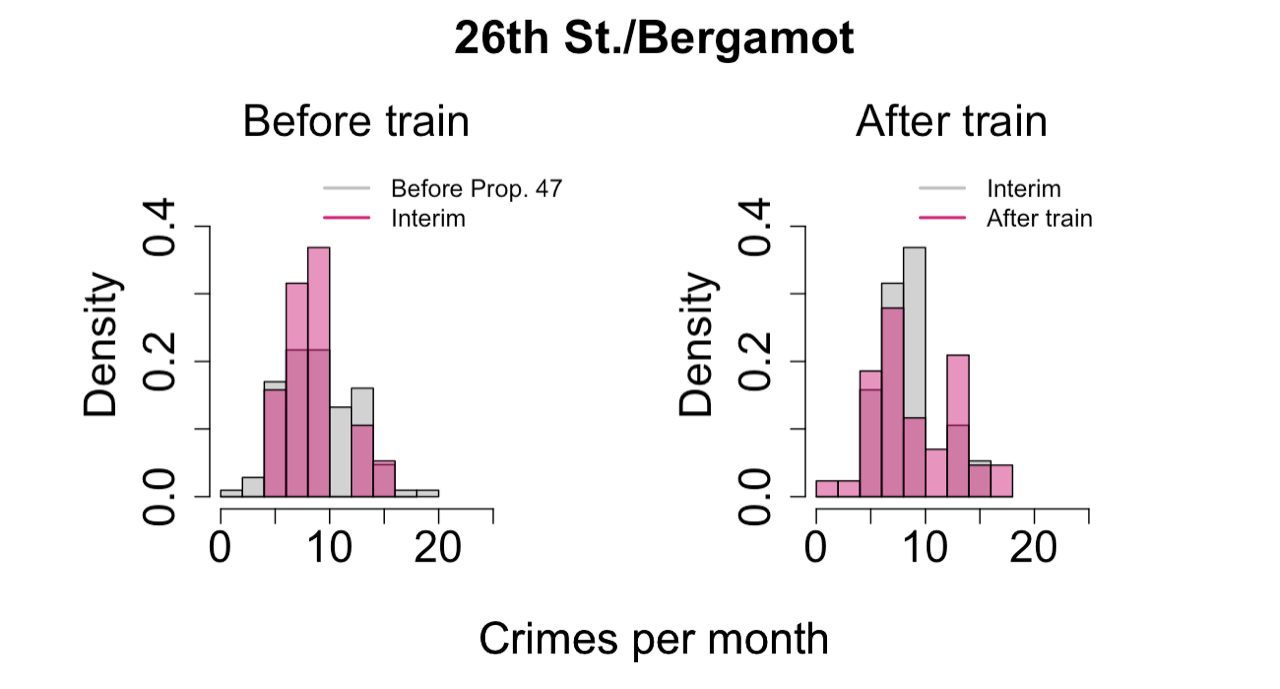}
\end{subfigure}
\begin{subfigure}{.6\textwidth}
\hspace{-3.7cm}
\centering
\includegraphics[width=\linewidth]{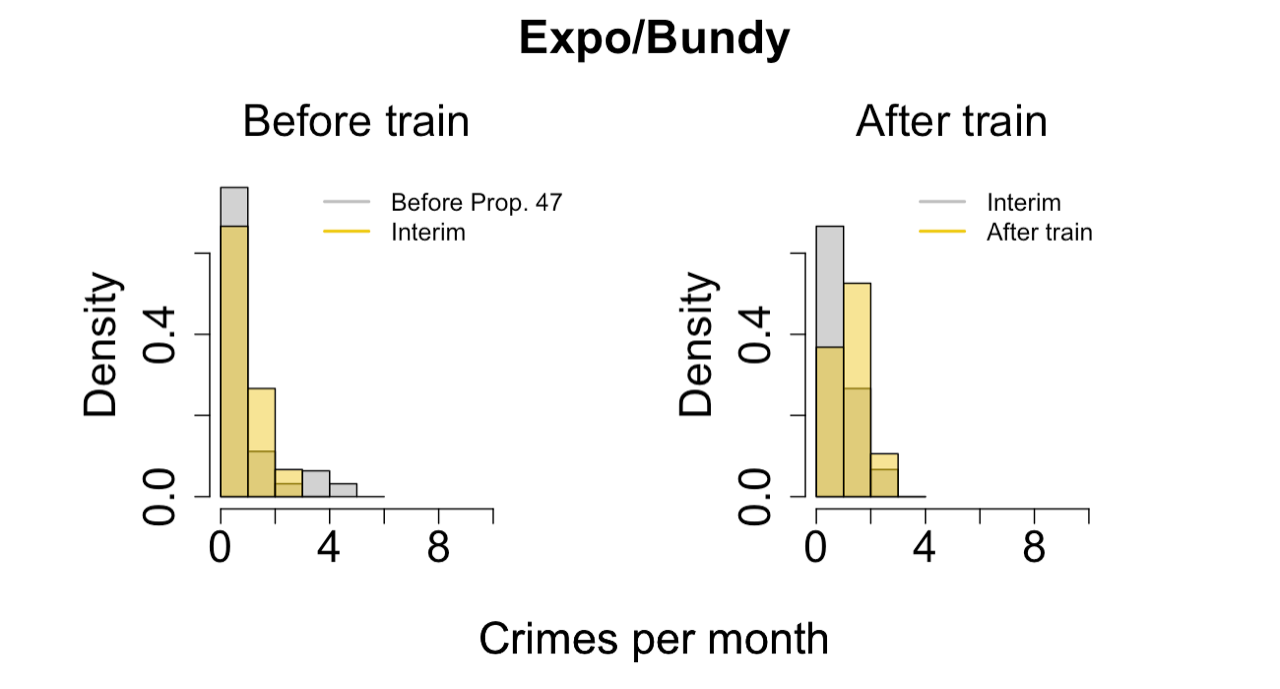}
\end{subfigure}
\caption{Histograms of monthly total crime counts in a 450 meter radius area centered around the
four new Expo Line stations in Santa Monica. In each panel, 
the left histograms are distributions prior to May 2016 
(before and after passage of Prop.\,47). Here,
the gray and colored bars represent crime densities before passage of 
Prop.\,47, and in the interim between passage of Prop.\,47 and before opening
of the Expo Line, respectively. The right panels are distributions after November 2014 
(before and after opening of the Expo Line). Here,
the gray and colored bars represent crime densities
in the interim between passage of Prop.\,47 and before opening
of the Expo Line, and after opening of the Expo Line, respectively.
Changes to the mean are not statistically significant prior to opening of the Expo Line; 
crimes increase significantly at the Downtown and 17$^{th}$ Street/Santa Monica College stops
after opening of the Expo Line. Corresponding data is listed in Table \ref{tab:Welch4b}.}
\label{fig:expo_line4b}
\end{figure}

\begin{table}[t!]
    \centering
    \caption*{\textbf{All crimes; Monthly averages (2006-2014; 2014-2016)}}
    \vspace{0mm} 
   { \color{black}
    \begin{tabular}{|l|c|c|c| c|}
    \hline
    \cellcolor[gray]{0.7} \textbf{Train Station} & \cellcolor[gray]{0.7}  \textbf{before Prop.\,47} 
    & \cellcolor[gray]{0.7} \textbf{after Prop.\,47}  & \cellcolor[gray]{0.7} $t , t_{\rm s}$ \cellcolor[gray]{0.7}&
    \cellcolor[gray]{0.7} \textbf{Significant?}\\
    \cellcolor[gray]{0.7}  & \cellcolor[gray]{0.7}  $\{\mu_{\rm b}, \sigma_{\rm b}, N_{\rm b} = 106 \}$ 
    & \cellcolor[gray]{0.7}  $\{\mu_{\rm a}, \sigma_{\rm a},  N_{\rm a} = 19\}$ & \cellcolor[gray]{0.7} \cellcolor[gray]{0.7}&
    \cellcolor[gray]{0.7}\\
     \hline
     \hline
    \cellcolor[gray]{0.9} Downtown   &  \{75.97, 11.96\}  &  \{79.53, 13.23\}  &  1.09, 1.71 & no \\
    \hline
     \cellcolor[gray]{0.9} $17^{th}$ St./SMC  &  \{13.52, 3.49\} &  \{12.58, 3.86\} & 0.99, 1.71 & no\\
    \hline
    \cellcolor[gray]{0.9} 26$^{th}$ St./Bergamot  &  \{9.57, 3.30\} &  \{9.16, 2.67\}  &  0.59, 1.70 & no\\
     \hline
    \cellcolor[gray]{0.9} Expo/Bundy  &   \{0.87, 0.93\}  & \{1.11, 0.60\} &1.32, 1.70 & N/A\\
      \hline
      \end{tabular}}
     \vspace{4mm}
    \caption*{\textbf{All crimes; Monthly averages (2014-2016; 2016-2019)}}
     \vspace{0mm} 
    \begin{tabular}{|l|c|c|c| c|}
    \hline
    \cellcolor[gray]{0.7} \textbf{Train Station} & \cellcolor[gray]{0.7}  \textbf{Before light rail} 
    & \cellcolor[gray]{0.7} \textbf{After light rail}  & \cellcolor[gray]{0.7} $t , t_{\rm s}$ \cellcolor[gray]{0.7}&
        \cellcolor[gray]{0.7} \textbf{Significant?}\\
        \cellcolor[gray]{0.7}  & \cellcolor[gray]{0.7}  $\{\mu_{\rm b}, \sigma_{\rm b}, N_{\rm b} = 19 \}$ 
    & \cellcolor[gray]{0.7}  $\{\mu_{\rm a}, \sigma_{\rm a},N_{\rm a} = 43 \}$ & \cellcolor[gray]{0.7} \cellcolor[gray]{0.7}&
    \cellcolor[gray]{0.7}\\
     \hline
\hline
    \cellcolor[gray]{0.9} Downtown &   \{79.53, 13.23\}&  \{105.33, 18.84\} & 6.17, 1.68 &  {\color{red} yes, up  $+32.4\%$}\\
        \hline
      \cellcolor[gray]{0.9}  $17^{th}$ St./SMC &   \{12.58, 3.86\} &  \{16.07, 5.71\} &  2.81, 1.68 &  {\color{red} yes, up $+27.7\%$ }\\
      \hline
      \cellcolor[gray]{0.9}  26$^{th}$ St./Bergamot   &  \{9.16, 2.67\} &  \{9.63, 3.77\} &   0.56, 1.68 & no\\
    \hline
    \cellcolor[gray]{0.9} Expo/Bundy  &   \{1.11, 0.60\}  &  \{0.77, 0.77\} &  1.86, 1.68 & N/A
\\
    \hline
    \end{tabular}
    \vspace{0.2cm}
    \caption{The Welch's t-test applied to the total crime histograms for the Expo Line stations shown in Fig.\,\ref{fig:expo_line4b}.
The last column indicates whether changes are statistically significant and shows percent changes to the mean.
    The $\{\mu_{\rm b}, \sigma_{\rm b} \}$ quantities are monthly  averages and standard deviations of crime before inauguration 
   of the Expo Line calculated over $N_{\rm b}$ months.  Their post-inauguration counterparts 
   are $\{\mu_{\rm a}, \sigma_{\rm a} \}$ calculated over $N_{\rm a}$ months.
    The Welch's t-test statistic, $t$, is compared to the Student's t-distribution reference value $t_{\rm s}$.  
    There is not sufficient data for meaningful conclusions at Expo/Bundy.
   See Sect.\,\ref{subsec:meanCrimes} for more details.}
    \label{tab:Welch4b}
\end{table}

To understand whether and how the extension of the Expo Line impacted crime rates in 
the vicinity of the four new train stations in Santa Monica we consider crime count distributions within a circle of radius 450 meters 
centered around the four new Expo Line stops. A detailed map 
is shown in Fig.\,\ref{fig_Expo}.  We first neglect passage of Prop.\,47 and consider the total crime distribution before and after
inauguration of the light rail extension on May 20, 2016.  
For simplicity, we categorize the entire month of May 2016 as falling before opening of the Expo Line
when creating monthly aggregates.  Results are shown in Fig.\,\ref{fig:expo_line4a}
and in Table \,\ref{tab:Welch4a}.  The crime distributions shift to the right in a statistically 
significantly manner after opening of the Expo Line at the Downtown Santa Monica and 
the 17$^{th}$ Street/Santa Monica College stops.  These shifts correspond to
 $37.7\%$ and $20.1\%$  increases in monthly crime rates, respectively. 
No statistically significant changes are observed around the 26$^{th}$ Street/Bergamot and the Expo/Bundy 
stations, which are located in areas with fewer crime attractors and human activity
compared to the other two, and where the average number of monthly crimes is also much lower
in comparison.  

\begin{figure}[t!]
\caption*{\textbf{Prop.\,47 crimes; Monthly averages (2006-2014; 2014-2016; 2016-2019)}}
\begin{subfigure}{.6\textwidth}
\hspace{-1cm}
\includegraphics[width=\linewidth]{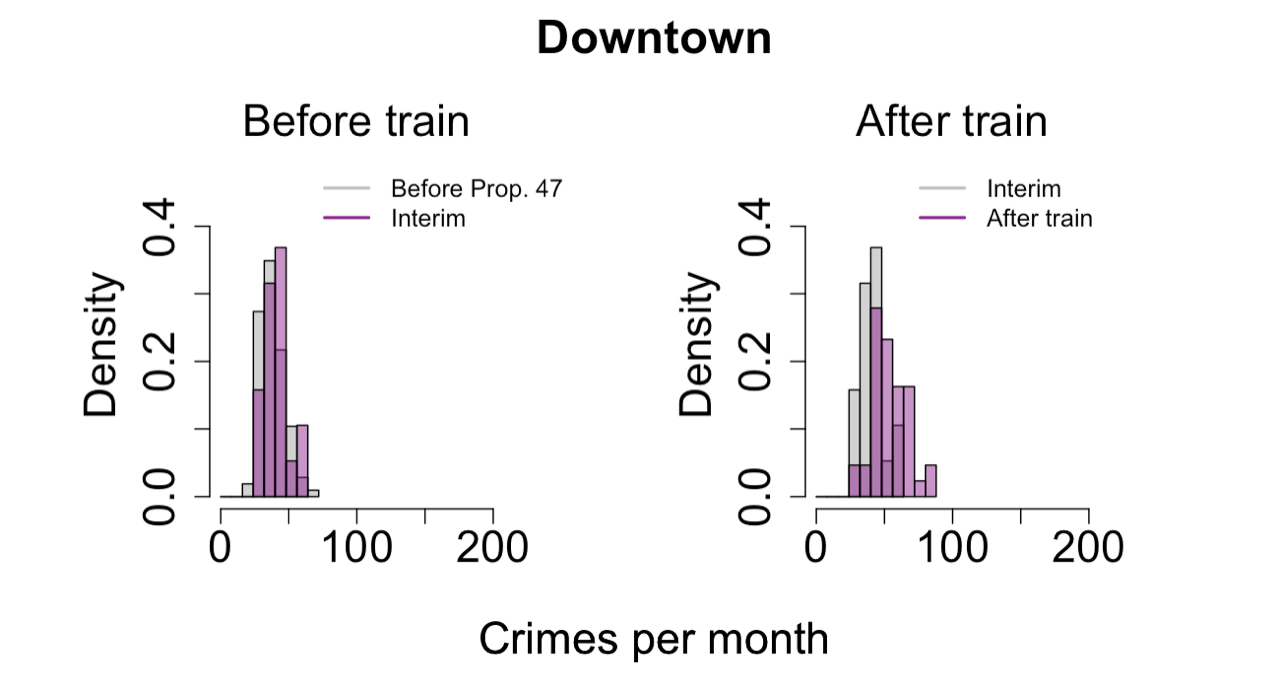}
\end{subfigure}
\begin{subfigure}{.6\textwidth}
\hspace{-3.7cm}
\centering
\includegraphics[width=\linewidth]{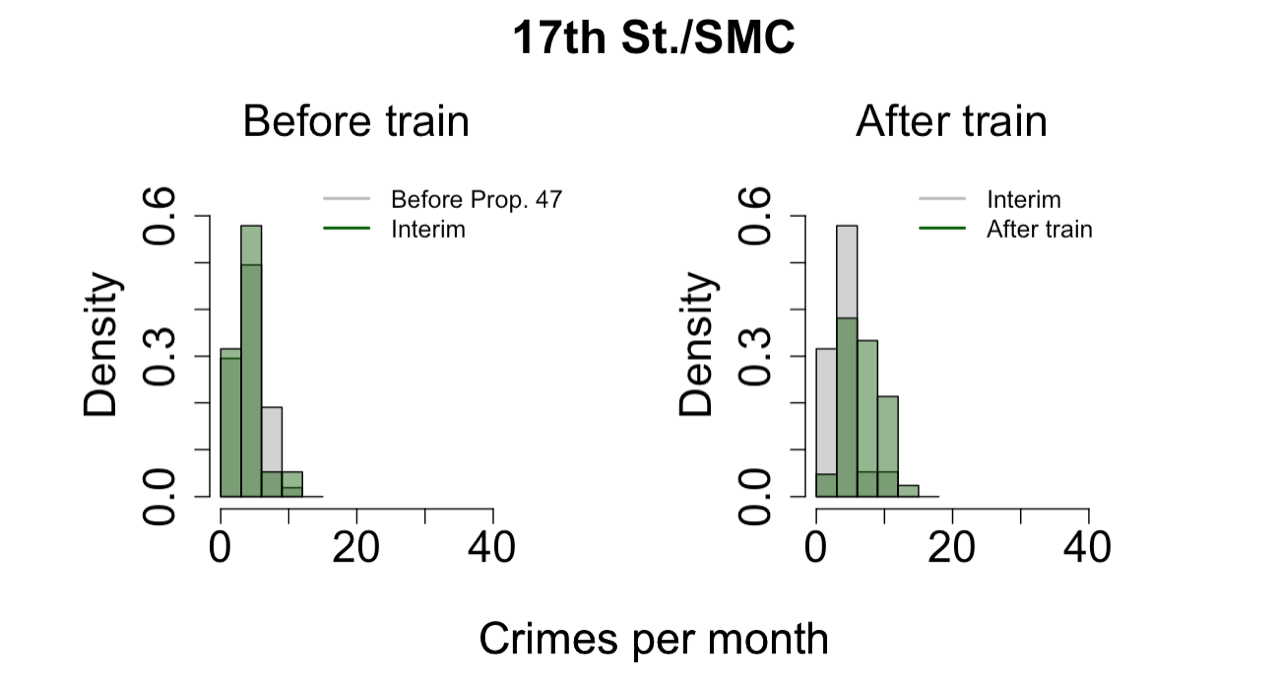}
\end{subfigure}
\vspace{0.5cm}
\newline
\begin{subfigure}{.6\textwidth}
\hspace{-1cm}
\centering
\includegraphics[width=\linewidth]{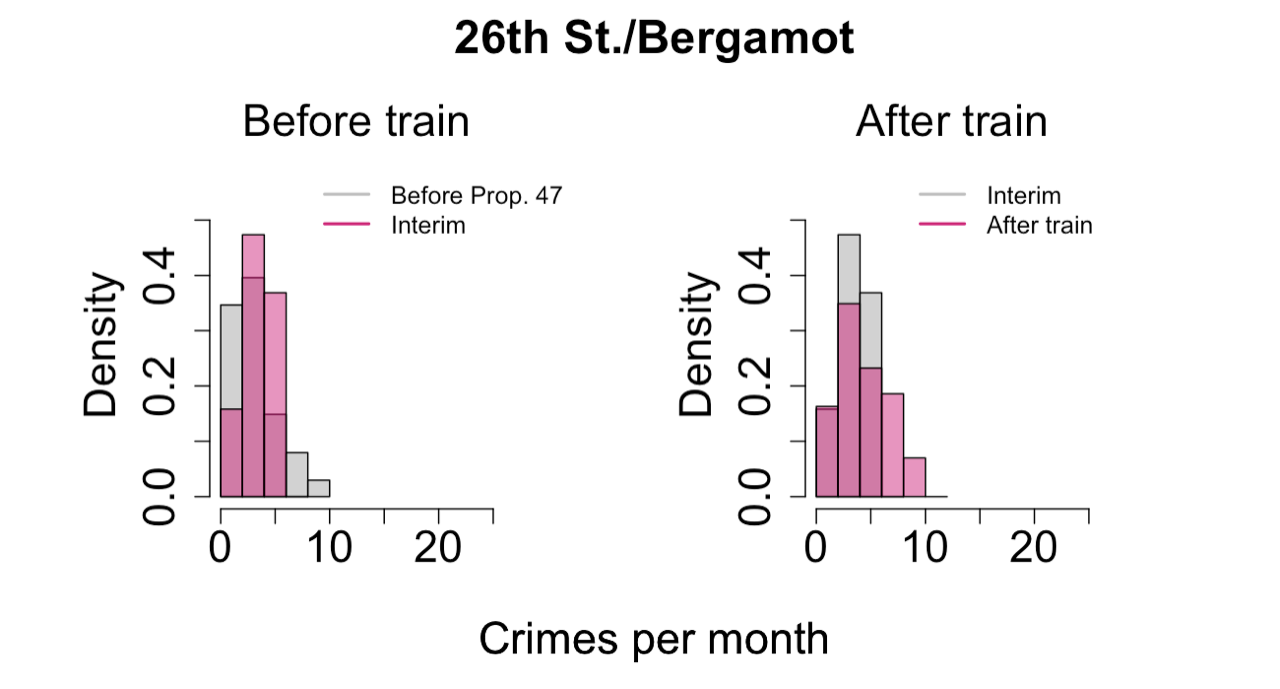}
\end{subfigure}
\begin{subfigure}{.6\textwidth}
\hspace{-3.7cm}
\centering
\includegraphics[width=\linewidth]{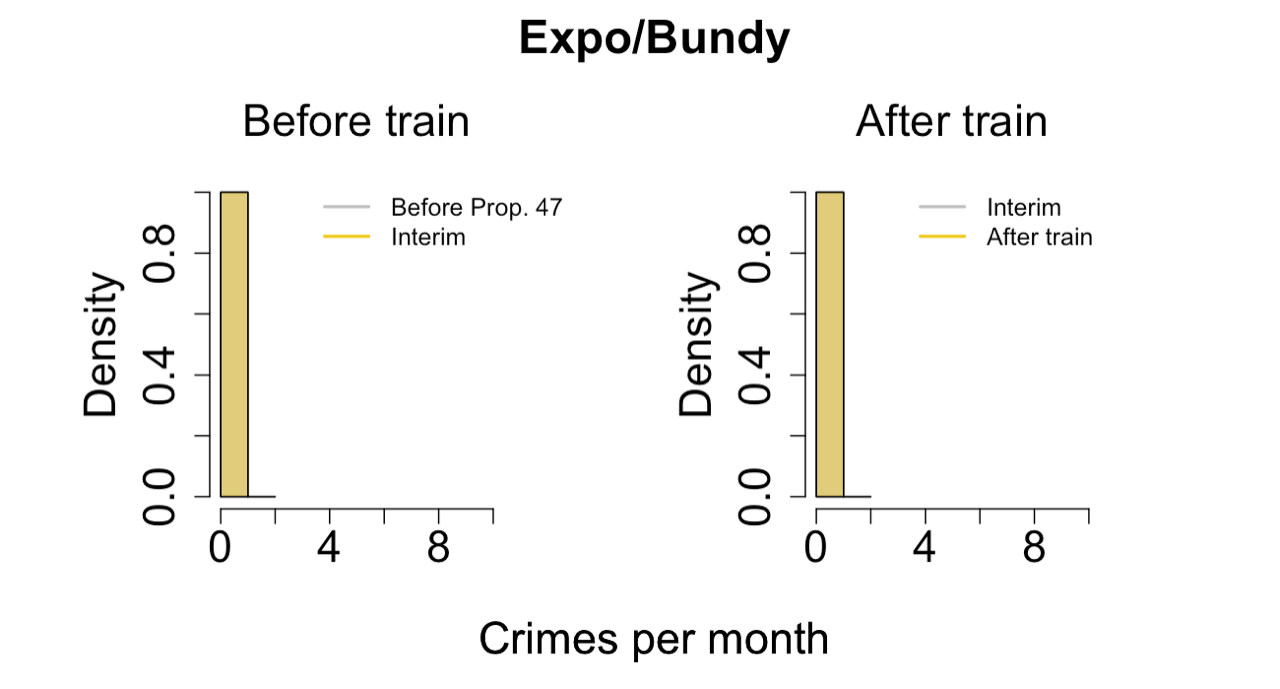}
\end{subfigure}
\caption{Histograms of monthly Prop.\,47 crime counts in a 450 meter radius centered around 
the four new Expo Line stations in Santa Monica.  In each panel, 
the left histograms are distributions prior to May 2016 
(before and after passage of Prop.\,47). Here,
the gray and colored bars represent crime densities before passage of 
Prop.\,47, and in the interim between passage of Prop.\,47 and before opening
of the Expo Line, respectively. The right panels are distributions after November 2014 
(before and after opening of the Expo Line). Here,
the gray and colored bars represent crime densities
in the interim between passage of Prop.\,47 and before opening
of the Expo Line, and after opening of the Expo Line, respectively.
Not enough events are recorded at Expo/Bundy for a meaningful analysis. 
Changes to the mean are not statistically significant prior to opening of the Expo Line; 
reclassified crimes increase in all stations for which there is sufficient data. 
Corresponding data is listed in Table \ref{tab:Welch4c}.}
\label{fig:expo_line4c}
\end{figure}

To separate the impacts of Prop.\,47 from the opening of the Expo Line
we further refine our data by binning it into three time periods:  January 2006--October 2014 (before implementation of Prop.\,47); 
November 2014--May 2016 (between the implementation of Prop.\,47 and the opening of the Expo Line);  
June 2016--December 2019 (after the opening of the Expo Line).
As a result of this stratification, only one or less crimes per month emerge at
the Expo/Bundy stop. Given the paucity of data, we do not perform any further statistical analysis on this station.
As seen in Fig.\,\ref{fig:expo_line4b} and in Table \ref{tab:Welch4b}, 
there is no significant increase in the number of overall monthly crimes near any of the other transit stops
immediately after passage of Prop.\,47 and prior to opening of the Expo Line (2014-2016). 
Total crimes instead increase dramatically after opening of the Expo Line at the Downtown 
and 17$^{th}$ Street/Santa Monica College stops 
by $32.4\%$ and 27.7\% respectively.
Upon restricting our analysis to the reclassified crimes, 
as shown in Fig.\,\ref{fig:expo_line4c} and in Table \ref{tab:Welch4c}, 
we find that Prop.\,47 crimes increased significantly not only
at the Downtown and 17$^{th}$ Street/Santa Monica College stops, 
but also at 26$^{th}$ Street/Bergamot 
by $30.6\%$, $38.6\%$, and 30.0$\%$ respectively. 
Finally, non-Prop.\,47 crimes increased at the Downtown and 17$^{th}$ 
Street/Santa Monica College stops by
$34.6\%$ and $20.6\%$, respectively, but did not affect 26$^{th}$ Street/Bergamot.

\begin{table}[ht!]
    \centering
    \caption*{\textbf{Prop.\,47 crimes; Monthly averages (2006-2014; 2014-2016)}}
    \vspace{0mm} 
    \begin{tabular}{|l|c|c|c| c|}
    \hline
    \cellcolor[gray]{0.7} \textbf{Train Station} & \cellcolor[gray]{0.7}  \textbf{before Prop.\,47} 
    & \cellcolor[gray]{0.7} \textbf{after Prop.\,47}  & \cellcolor[gray]{0.7} $t , t_{\rm s}$ \cellcolor[gray]{0.7}&
    \cellcolor[gray]{0.7} \textbf{Significant?}\\
    \cellcolor[gray]{0.7}  & \cellcolor[gray]{0.7}  $\{\mu_{\rm b}, \sigma_{\rm b}, N_{\rm b} = 106 \}$ 
    & \cellcolor[gray]{0.7}  $\{\mu_{\rm a}, \sigma_{\rm a},  N_{\rm a} = 19 \}$ & \cellcolor[gray]{0.7} \cellcolor[gray]{0.7}&
    \cellcolor[gray]{0.7}\\
     \hline
     \hline
    \cellcolor[gray]{0.9} Downtown   &  \{38.38, 8.93\}&  \{41.47, 8.95\} & 1.39, 1.71 & no \\
    \hline
     \cellcolor[gray]{0.9} $17^{th}$ St./SMC  & \{4.74, 2.00\} &  \{5.00, 2.52\}  &  0.44, 1.72  & no\\
    \hline
    \cellcolor[gray]{0.9} 26$^{th}$ St./Bergamot  &  \{3.48, 1.90\} & \{3.74, 1.56\}  &  0.65, 1.70 & no\\
     \hline
    \cellcolor[gray]{0.9} Expo/Bundy   &   \{0.12, 0.31\} & \{0.21, 0.36\}  & 1.00, 1.72 & N/A\\
      \hline
      \end{tabular}
     \vspace{4mm}
    \caption*{\textbf{Prop.\,47 crimes; Monthly averages (2014-2016; 2016-2019)}}
     \vspace{0mm} 
    \begin{tabular}{|l|c|c|c| c|}
    \hline
    \cellcolor[gray]{0.7} \textbf{Train Station} & \cellcolor[gray]{0.7}  \textbf{Before light rail} 
    & \cellcolor[gray]{0.7} \textbf{After light rail}  & \cellcolor[gray]{0.7} $t , t_{\rm s}$ \cellcolor[gray]{0.7}&
        \cellcolor[gray]{0.7} \textbf{Significant?}\\
        \cellcolor[gray]{0.7}  & \cellcolor[gray]{0.7}  $\{\mu_{\rm b}, \sigma_{\rm b}, N_{\rm b} = 19 \}$ 
    & \cellcolor[gray]{0.7}  $\{\mu_{\rm a}, \sigma_{\rm a}, N_{\rm a} = 43 \}$ & \cellcolor[gray]{0.7} \cellcolor[gray]{0.7}&
    \cellcolor[gray]{0.7}\\
     \hline
\hline
    \cellcolor[gray]{0.9} Downtown & \{41.47, 8.95\} & \{54.14, 12.23\}  & 4.57, 1.71 & {\color{red} yes, up  $+30.6\%$}\\
        \hline
      \cellcolor[gray]{0.9}  $17^{th}$ St./SMC &  \{5.00, 2.52\} &  \{6.93, 2.74\} &  2.76, 1.70  & {\color{red} yes, up $+38.6\%$}\\
      \hline
      \cellcolor[gray]{0.9}  26$^{th}$ St./Bergamot   &  \{3.74, 1.56\} &  \{4.86, 2.41\}  &  2.19, 1.68 & {\color{red} yes, up $+30.0\%$}\\
    \hline
    \cellcolor[gray]{0.9} Expo/Bundy  &   \{0.21, 0.36\} & \{0.23, 0.37\}  &  0.22, 1.70 & N/A\\
    \hline
    \end{tabular}
    \vspace{0.2cm}
    \caption{The Welch's t-test applied to the Prop.\,47 crime histograms for the Expo Line stations shown in Fig.\,\ref{fig:expo_line4c}. 
    The last column indicates whether changes are statistically significant and shows percent changes to the mean. 
    We consider two time frames, before and after passage of Prop.\,47, and before and after inauguration of the Expo Line.
    The respective before averages and standard deviations $\{\mu_{\rm b}, \sigma_{\rm b} \}$ calculated over $N_{\rm b}$ months, and 
    after averages and standard deviations $\{\mu_{\rm a}, \sigma_{\rm a} \}$ calculated over $N_{\rm a}$ months, are listed.  
     The Welch's t-test statistic, $t$,
     is compared to the Student's t-distribution reference value $t_{\rm s}$.  
     There is not sufficient data for meaningful conclusions at Expo/Bundy.
    See Sect.\,\ref{subsec:meanCrimes} for more details.}
    \label{tab:Welch4c}
\end{table}

\begin{figure}[t!]
\caption*{\textbf{non-Prop.\,47 crimes; Monthly averages (2006-2014; 2014-2016; 2016-2019)}}
\begin{subfigure}{.6\textwidth}
\hspace{-1cm}
\includegraphics[width=\linewidth]{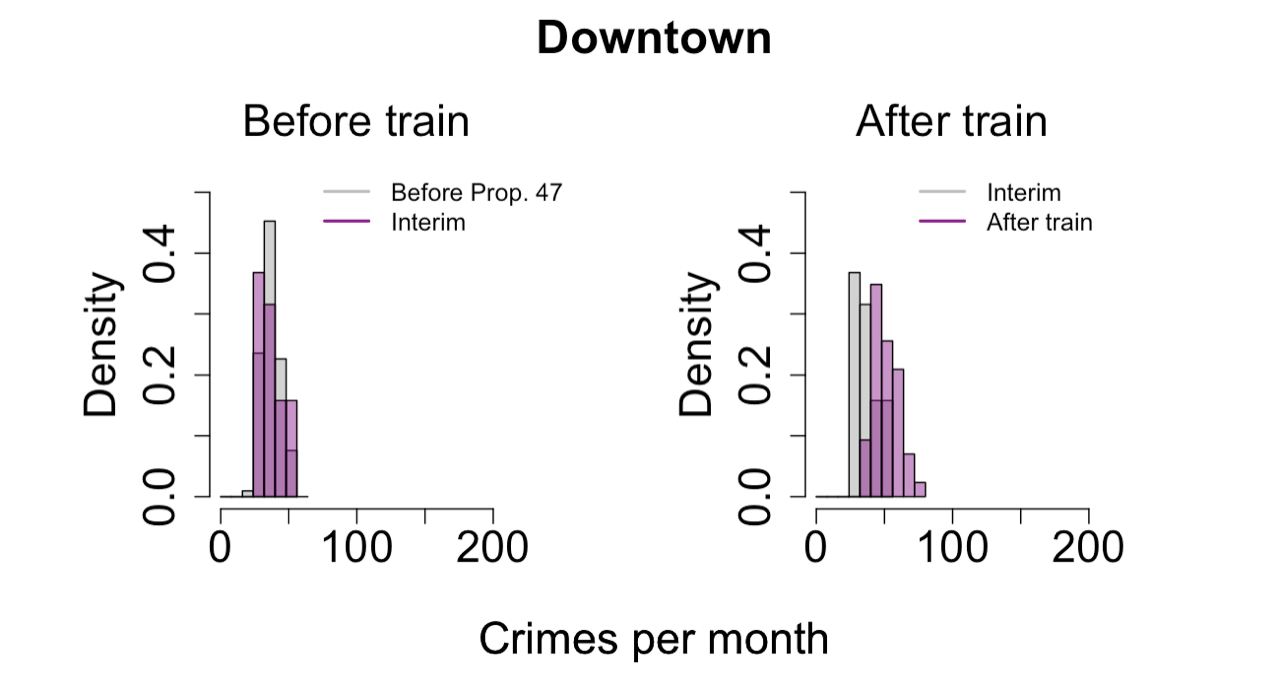}
\end{subfigure}
\begin{subfigure}{.6\textwidth}
\hspace{-3.7cm}
\centering
\includegraphics[width=\linewidth]{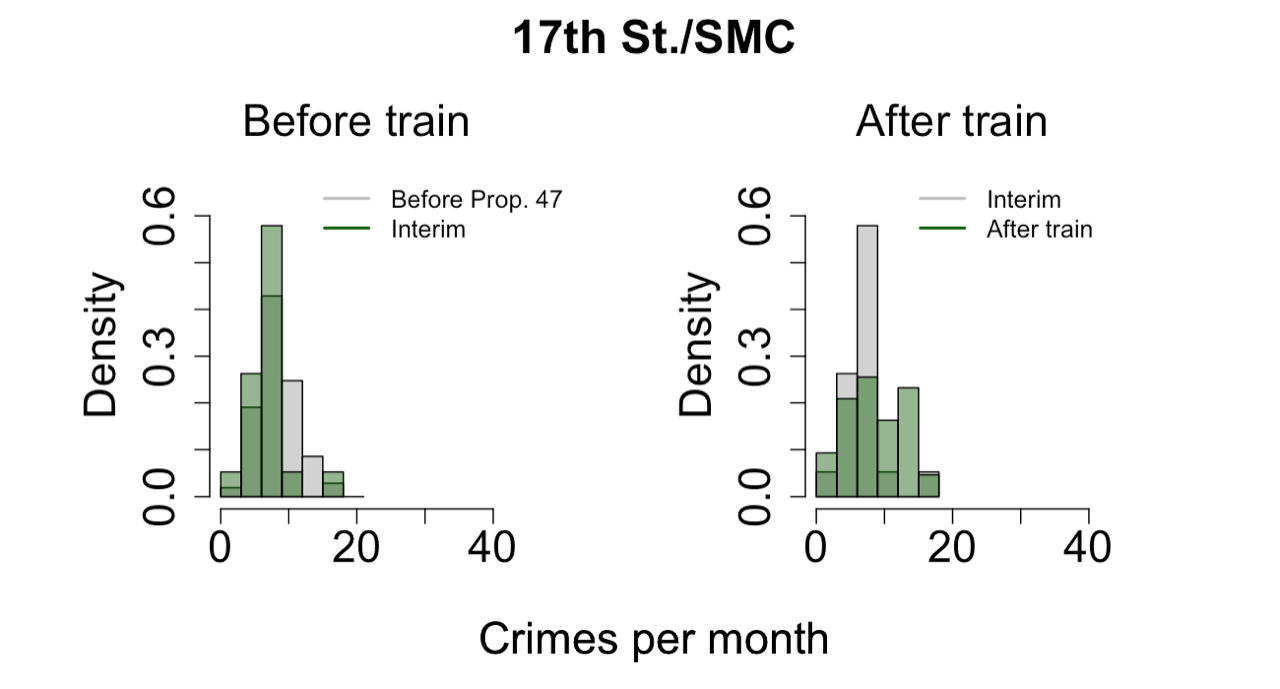}
\end{subfigure}
\vspace{0.5cm}
\newline
\begin{subfigure}{.6\textwidth}
\hspace{-1cm}
\centering
\includegraphics[width=\linewidth]{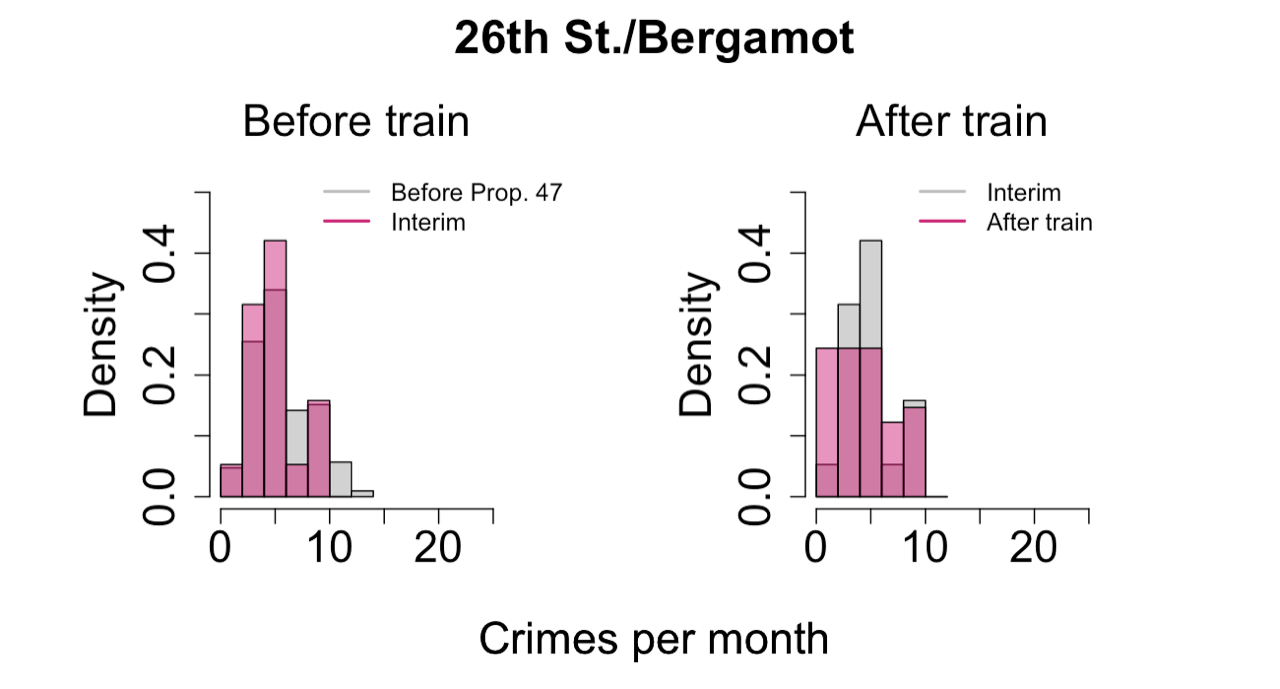}
\end{subfigure}
\begin{subfigure}{.6\textwidth}
\hspace{-3.7cm}
\centering
\includegraphics[width=\linewidth]{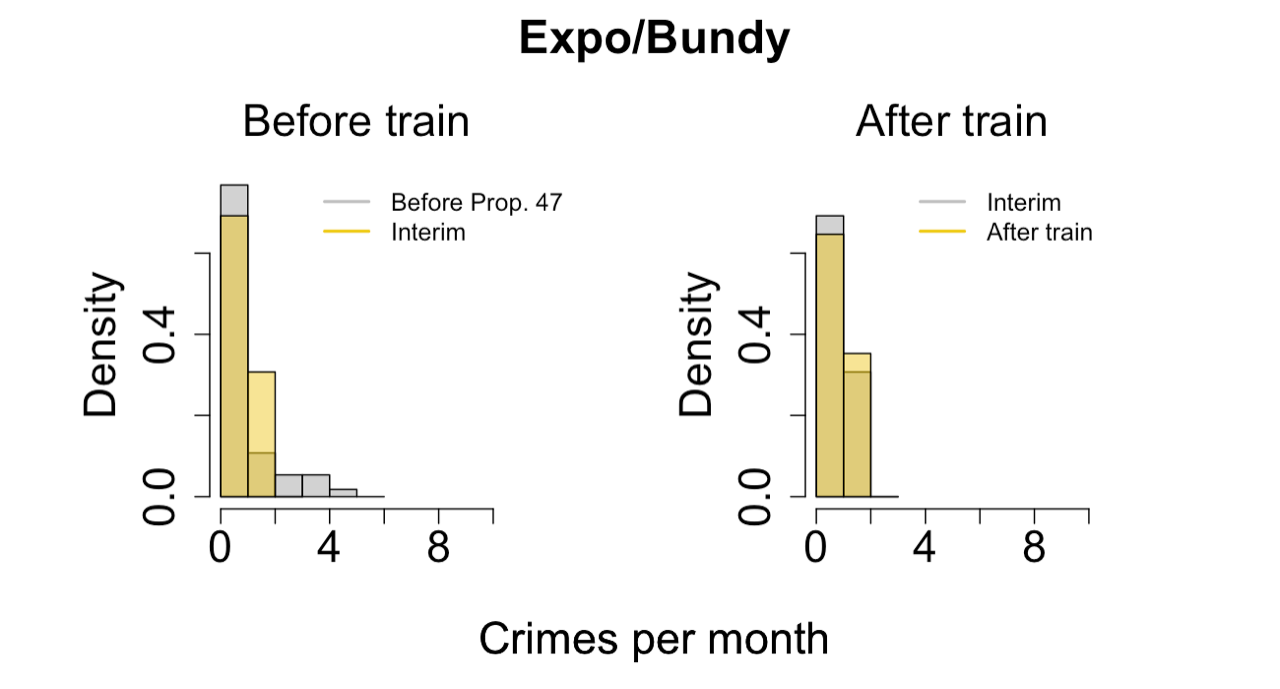}
\end{subfigure}
\caption{Histograms of monthly non-Prop.\,47 crime counts in a 450 meter radius centered around 
the four new Expo Line stations in Santa Monica.  In each panel, 
the left histograms represent distributions prior to May 2016 
(before and after passage of Prop.\,47). Here,
the gray and colored bars represent crime densities before passage of 
Prop.\,47, and in the interim between passage of Prop.\,47 and before opening
of the Expo Line, respectively. The right panels are distributions after November 2014 
(before and after opening of the Expo Line). Here,
the gray and colored bars represent crime densities
in the interim between passage of Prop.\,47 and before opening
of the Expo Line, and after opening of the Expo Line, respectively.
Not enough events
are recorded at Expo/Bundy for a meaningful analysis. 
The mean decreases or stays constant both prior and after opening of the Expo Line,
except for Downtown and 17$^{th}$ Street/Santa Monica College where non-Prop.\,47 
crimes increased after opening of the Expo Line.
Corresponding data is listed in Table \ref{tab:Welch4d}.
}
\label{fig:expo_line4d}
\end{figure}

\begin{table}[ht!]
    \centering
    \caption*{\textbf{non-Prop.\,47 crimes; Monthly averages (2006-2014; 2014-2016)}}
    \vspace{0mm} 
    \begin{tabular}{|l|c|c|c| c|}
    \hline
    \cellcolor[gray]{0.7} \textbf{Train Station} & \cellcolor[gray]{0.7}  \textbf{before Prop.\,47} 
    & \cellcolor[gray]{0.7} \textbf{after Prop.\,47}  & \cellcolor[gray]{0.7} $t , t_{\rm s}$ \cellcolor[gray]{0.7}&
    \cellcolor[gray]{0.7} \textbf{Significant?}\\
    \cellcolor[gray]{0.7}  & \cellcolor[gray]{0.7}  $\{\mu_{\rm b}, \sigma_{\rm b},  N_{\rm b} = 106\}$ 
    & \cellcolor[gray]{0.7}  $\{\mu_{\rm a}, \sigma_{\rm a},  N_{\rm a} = 19 \}$ & \cellcolor[gray]{0.7} \cellcolor[gray]{0.7}&
    \cellcolor[gray]{0.7}\\
     \hline
     \hline
    \cellcolor[gray]{0.9} Downtown   &  \{37.59, 6.7\} & \{38.05, 7.49\}  & 0.25, 1.71 & no \\
    \hline
     \cellcolor[gray]{0.9} $17^{th}$ St./SMC  &  \{8.78, 3.02\} &  \{7.58, 2.69\} &  1.80, 1.70 & yes, down  -13.7$\%$ \\
    \hline
    \cellcolor[gray]{0.9} 26$^{th}$ St./Bergamot  &  \{6.08, 2.67\} &  \{5.42, 2.29\}  &  1.13, 1.70  & no \\
     \hline
    \cellcolor[gray]{0.9} Expo/Bundy   &  \{0.76, 0.85\} &  \{0.89, 0.51\}  & 0.91, 1.70  & N/A\\
      \hline
      \end{tabular}
     \vspace{4mm}
    \caption*{\textbf{non-Prop.\,47 crimes; Monthly averages (2014-2016; 2016-2019)}}
     \vspace{0mm} 
    \begin{tabular}{|l|c|c|c| c|}
    \hline
    \cellcolor[gray]{0.7} \textbf{Train Station} & \cellcolor[gray]{0.7}  \textbf{Before light rail} 
    & \cellcolor[gray]{0.7} \textbf{After light rail}  & \cellcolor[gray]{0.7} $t , t_{\rm s}$ \cellcolor[gray]{0.7}&
        \cellcolor[gray]{0.7} \textbf{Significant?}\\
        \cellcolor[gray]{0.7}  & \cellcolor[gray]{0.7}  $\{\mu_{\rm b}, \sigma_{\rm b}, N_{\rm b} = 19 \}$ 
    & \cellcolor[gray]{0.7}  $\{\mu_{\rm a}, \sigma_{\rm a}, N_{\rm a} = 43 \}$ & \cellcolor[gray]{0.7} \cellcolor[gray]{0.7}&
    \cellcolor[gray]{0.7}\\
     \hline
\hline
    \cellcolor[gray]{0.9} Downtown & \{38.05, 7.49\} & \{51.19, 9.4\} & 5.87, 1.68 & {\color{red} yes, up  +34.6$\%$}\\
        \hline
      \cellcolor[gray]{0.9}  $17^{th}$ St./SMC &   \{7.58, 2.69\}&  \{9.14, 3.78\}  &  1.85, 1.68 & {\color{red} yes, up  +20.6$\%$ }\\
      \hline
      \cellcolor[gray]{0.9}  26$^{th}$ St./Bergamot   &  \{5.42, 2.29\} &  \{4.77, 2.69\}  &  1.00, 1.70 & no\\
    \hline
    \cellcolor[gray]{0.9} Expo/Bundy  & \{0.89, 0.51\} &  \{0.53, 0.60\}  &  2.42, 1.70  & N/A \\ 
    \hline
    \end{tabular}
    \vspace{0.2cm}
    \caption{The Welch's t-test applied to the non-Prop.\,47 crime histograms for the Expo Line stations shown in Fig.\,\ref{fig:expo_line4d}. 
    The last column indicates whether changes are statistically significant and shows percent changes to the mean. 
     We consider two time frames, before and after passage of Prop.\,47, and before and after inauguration of the Expo Line.
    The respective before averages and standard deviations $\{\mu_{\rm b}, \sigma_{\rm b} \}$ calculated over $N_{\rm b}$ months, and 
    after averages and standard deviations $\{\mu_{\rm a}, \sigma_{\rm a} \}$ calculated over $N_{\rm a}$ months, are listed.  
    The Welch's t-test statistic, $t$, is compared to the Student's t-distribution reference value $t_{\rm s}$.  
    There is not sufficient data for meaningful conclusions at Expo/Bundy.
    See Sect.\,\ref{subsec:meanCrimes} for more details.}
    \label{tab:Welch4d}
\end{table}

Our analysis indicates that the arrival of light rail was accompanied by a surge in
the occurrence of crime in proximity of the Downtown Santa Monica stop, where
all types of crime, reclassified and non-reclassified, increased by large percentages.
Light rail was also accompanied by an increase in Prop.\,47 crimes near two of the other three stations
($17^{th}$ Street/Santa Monica College and $26^{th}$ Street/Bergamot), while
non-Prop.\,47 crimes remained stationary or decreased in all three.
Tables  \ref{tab:Welch4c} and \ref{tab:Welch4d} reveal that there were no statistically significant increases in crime after passage of Prop.\,47 
but before opening of the Expo Line (2006-2016) at any of the four new train stations, including Downtown. 
This may be due to ongoing construction of the Expo Line, which began in 2011 and was still active when Prop.\,47 
was implemented so that any associated effects could only emerge after the train began its operations in May 2016. 
Finally, the interval between November 2014--May 2016 comprises only 18 months
and includes one summer and two winters. This period is thus marked by an 
unbalanced seasonality, which may be reflected in the corresponding monthly averages.

\section{Conclusions}
\label{subsec:concl}

Using a publicly-available database compiled and maintained by the 
Santa Monica Police Department, we investigated whether passage of Proposition
47 in the state of California had any impact on criminal activity.
We specifically focused on crimes that were directly affected by legislative change
and were reclassified from felonies to misdemeanors. Our analysis 
shows that overall the monthly count of these crimes (larceny, fraud, possession of narcotics, forgery,
receiving/possessing stolen property) increased by about 15$\%$ after
implementation of Proposition 47 in November 2014.
By contrast, the non-reclassified crime count 
decreased by 13$\%$ after the new legislation became effective. 
We used a Welch's t-test to verify that the reclassified crime distribution
shift from less crime before November 2014 to more crime
after the same date was statistically significant, as well 
as signal decomposition to isolate the main crime trends from 
seasonal effects. Using change-point analysis, we identified a discontinuity in 
crime trend at the end of 2014. A segmented regression analysis on the monthly time series
led us to identify November 2014 as the main breakpoint.  
We also identified a secondary discontinuity in the crime trend and 
a secondary break-point in the monthly time series, both occurring towards the end of 2018, 
indicating a decreases in crime concurrent with the new SMPD policing efforts. 
While these changes are too recent to reverse the overall crime surge observed
after the 2014 implementation of Prop.\,47, and while it is unclear what the longer term implications of the
new initiatives will be, our results show that community partnerships, a responsive and outward facing 
police force, and targeted measures, may help ameliorate crime.

We also considered the impact of Prop.\,47 on
the eight neighborhoods that comprise the city and verified that
the largest monthly change in crime (+37.2$\%$) occurred Downtown, an area with many opportunistic
crime attractors, such as nightlife, tourists, dining venues,  
and shopping centers.  Finally, we examined the effects
of the opening of the Expo Line on monthly crime rates within 450 meters from four new 
transit stations.  We find that total crime counts increase significantly at the Downtown Santa Monica
and 17$^{th}$ Street/Santa Monica College stops. 
Prop.\,47 and non-Prop.\,47 percent increases were comparable at
the Downtown Santa Monica stop (+30.6$\%$ and +34.6$\%$ respectively); at the 17$^{th}$ Street/Santa Monica College
stop instead, the increase for Prop.\,47 crimes (+38.6$\%$) was much higher than for non-Prop.\,47 ones (+20.6$\%$).

Several observations are in order. The city of Santa Monica does not report the monetary value of
relevant crimes. Partitioning of the data into reclassified vs.\,non-reclassified offenses
is thus based on our best estimate of which crimes would, on average, 
fall under the 950 USD threshold, one of the conditions specified
by Prop.\,47 for reclassification. For example, the initiative applies to grand theft auto but 
only for vehicles worth less than 950 USD. Since the typical 
value of stolen cars in Santa Monica surpasses this threshold, 
we do not include grand theft auto in the list of Prop.\,47 crimes.
It is clear that exceptions may exist and that our partitioning may have introduced errors; however, due to the 
large data sample, we expect these not to be systematic and not to have significantly affected our results. 
Other biases could arise from unreported victimizations differentially 
affecting the reclassified vs.\,non-reclassified crime categories. For instance,
the US Department of Justice estimates that the average annual incidence of unreported larceny was
41$\%$ nationwide over the 2006--2010 period; for motor vehicle theft the same figure was 
17$\%$ \citep{LAN12}. Similarly, reporting rates could change over time
in response to changing perceptions of the effectiveness of reporting crimes.
We also observe that in the immediate vicinity of the four new train stations in Santa Monica
no increases in reclassified or non-reclassified crimes were observed after passage of Prop.\,47 but before
opening of light rail. This may be due to ongoing construction of the new crime attractor,
which delayed the effects of Prop.\,47. Indeed, crimes increased dramatically at the busiest transit stations
after opening of the Expo Line, especially in the Downtown area and for the reclassified crimes.  
Finally, the period between passage of Prop.\,47 in November 2014 and the opening of the 
Expo Line in May 2016 is only 18 months. Compounding effects 
of the two events may have led to the observed increases in crime;  
disentangling their overlap may require more discriminants than the
data analyzed here.

Possible extensions of this work would involve analyzing crime occurrence in 
neighboring cities that share similar socio-economic backgrounds with Santa Monica,
such as Culver City (population 39,000), Pasadena (population 138,000) or Glendale (population 203,000);
all within Los Angeles County, although not directly adjacent to the Pacific Ocean
and less touristic.  Similarly, it would be interesting to study the incidence of crime
on the other three new Expo Line stations operating in Culver City 
(Palms, Westwood/Rancho Park, Expo/Sepulveda) to compare and contrast results
between the two municipalities.   
A longer term monitoring of crime in Santa Monica would also be desirable. 
On one hand, it would allow us to determine whether the decrease in the number of monthly crimes
observed in late 2018 persists or stabilizes over time. On the other, it would allow us
to better discern the effects of the Expo Line, since the incidence of crime may temporarily increase
around newly opened stations and settle back to their original levels
once novelty effects subside \citep{POI96}. This may not be possible due to the
COVID-19 pandemic that severely impacted the global economy. 
Suspension of all non-essential activities and stay-at-home orders within the city of Santa Monica, as
well as reduction to service and ridership of the Expo Line, do not allow for a
meaningful, continuous, long term analysis. Although conducted over short time frames,
both after passage of Prop.\,47 and the inauguration of the Expo Line, our results do suggest that
neighborhood characteristics may influence crime rates, since areas with more opportunities for crime appear 
to have been affected more by the light rail extension and by the new law. Similarly,
our results show that community-based policing, a stronger police presence and targeted interventions
may help reduce crime. 

Finally, although we find a rise in reclassified crimes that coincides with passage of Prop.\,47, 
our research does not provide a definite causative explanation for it.  While it is possible that the new law 
directly motivated offenders to commit more reclassified crimes, there may also be 
other relevant explanations contributing to the rise. Among them, the increased attention of police, heightened public awareness, 
more reporting, all of which may have been influenced by media coverage. The observed rise may also be 
a loose manifestation of the well known Hawthorne effect, whereby individuals modify their behavior 
as a result of being part of an experiment or study
\citep{ROE39, ADA84}. In this case, reporting behavior could have been affected
by awareness of the changes brought by Prop.\,47. Similarly, the decrease in the number of
reported reclassified crimes observed in late 2018 may be due to shifts in policing, but also due to 
fading of Prop.\,47 awareness, or habituation.  
We hope that these and other considerations relevant to public utility, respect for human rights, and 
existence of socioeconomic disparities,  
will be used  in combination with our results to assess the overall effect of Prop.\,47.

\section{Acknowledgements}
We thank the Santa Monica Police Department, Tricia Crane, Michele Modglin and Jude Higdon-Topaz
for valuable information and for the generous time spent discussing our work with us.  We thank the AMS Mathematics
Research Community where this work was initiated. We also thank Hwayeon Ryu for contributing to the beginning stages of this study. 
Finally, we acknowledge support from the National Science Foundation
under grants DMS 1440386 (CV), DMS 1703761 (JC) and from the Army 
Research Office under grant ARO W911NF-16-1-0165
(MRD).

\appendix
\section{Supplementary Information}
\label{sec:SItext}
\subsection{Seasonal and Trend decomposition using Loess (STL)}
\label{sec:SItext1}

In our work the  decomposition is performed via the `stl' function in the R software environment \citep{R_stats}
using an iterative process that uses the monthly time series $Y(t)$ as input data and user-specified initial trend and seasonality estimates. The latter
are typically null sets. In each iteration the data is cleared of the current trend estimate
and broken into cycle-subseries, one for each of the data points within a period. 
In our case, since we have monthly data with a periodicity of $n_{\rm p} =12$ 
months, twelve cycle-subseries arise, one for each month.
A new seasonal series is obtained through a combination
of loess (locally estimated scatterplot smoothing) polynomial regression with given weights, and moving averages
performed on each of the cycle-subseries.  The loess regression ensures that the obtained seasonality 
is defined for all times, not just when data points are available; the moving average procedures
guarantee the mean is nearly zero. The input data is then cleared of the newly derived seasonal effect and
a temporary trend is obtained, once more using a loess polynomial regression. 
The freshly derived trend and seasonal estimates are then used as inputs for the next iteration. 
Any residual elements are included in the remainder and used to 
compute robustness weights, to reduce the
influence of transient, aberrant behavior in the data on the trend and seasonal components. 
The procedure is run until a pre-set convergence is reached; typically two loops suffice.

 \begin{figure}[t]
     \centering
     \includegraphics[width=0.7\textwidth]{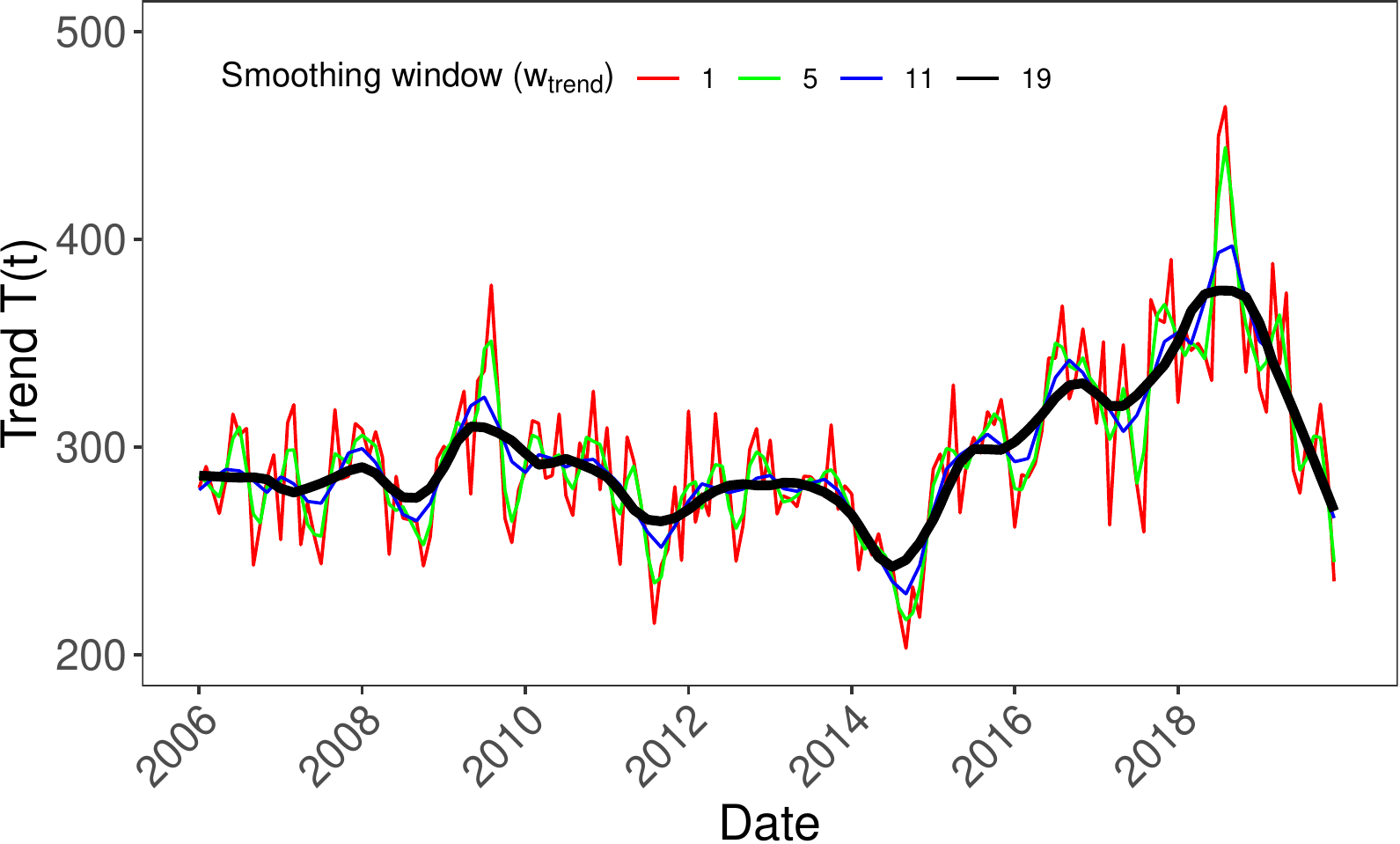}
    \caption{Trends $T(t)$ evaluated on Prop.\,47 crimes using  decomposition using loess from the R statistical package with 
     varying window lengths $w_{\rm trend}$.  Note that as $w_{\rm trend}$ increases the corresponding $T(t)$ becomes smoother. 
     Unless otherwise stated, we set $w_{\rm trend} = 19$ months (black curve) as
     determined by Eq.\,(\ref{eqn:t_window}) throughout this work.}
     \label{fig:prop47_changeWL}
 \end{figure}

Two window lengths must be specified to apply the loess regression in the `stl' function:
$w_{\rm{season}}$ (the s.window argument in R) and $w_{\rm{trend}}$ (the t.window argument in R). 
Since we do not assume seasonal patterns to have significantly evolved over
the thirteen year time span under investigation, we use the entire data to perform the loess
seasonal smoothing analysis and set $\rm{s.window} = \rm{periodic}$. Effectively, 
the monthly cycle-subseries are smoothed using weighted averages over all the pertaining monthly data.  Note that if seasonality were expected 
to change over the 2006-2019 arc the analysis would have to be performed using a more restricted window, 
so that older seasonal patterns do not affect more recent ones.  Finally, $w_{\rm {trend}}$ is assumed to be an odd integer and 
is set following standard procedures \citep{R_stats} as

\begin{equation}
\label{eqn:t_window}
\displaystyle{\rm{t.window} = \text{Nextodd}\left(\text{Ceiling}\left(\dfrac{1.5 \,n_{\rm p}}{\displaystyle{1 - 1.5 \, / \rm{s.window}}}\right)\right). }
\end{equation}

\noindent
Here, NextOdd$(.)$ is the smallest odd integer greater than, or equal to its argument, and
Ceiling$(.)$ is the smallest integer greater than or equal to its argument. 
Eq.\,(\ref{eqn:t_window}) yields values for $w_{\rm{trend}}$ that are known to prevent overlaps 
between the trend and seasonal components. 

The value of $w_{\rm {trend}}$ plays a fundamental role in the decomposition process: 
as this parameter increases more points are used in the smoothing process,
and sharper trends may be identified. Increases to $w_{\rm {trend}}$ however may also 
minimize, eliminate, or shift peaks and valleys. 
Our data set leads to $w_{\rm trend}= 19$ months. Unless otherwise noted,
we use this value as the default; 
when more resolution is necessary, for example to investigate trends around November 2014,
smaller window lengths are used. Fig.\,\ref{fig:prop47_changeWL} shows various smoothed trend curves for different
$w_{\rm trend}$: note how the minimum located towards the end of 2014 shifts in time and depth as $w_{\rm trend}$ 
is modified.

\subsection{Change-point analysis}
\label{sec:SItext2}      

\noindent
Our change-point analyses are performed using the R package `mosum' \citep{R_mosum} 
which uses a moving sum to average over subsets of the data; the size of the subset 
$G$ is termed bandwidth.  For a given data point, the prior and after averages over the bandwidth 
are evaluated together with the respective variances; the first and last $G$ points are discarded. 
A mosum statistic is then constructed as the difference between the
after and prior averages divided by an ad-hoc standard deviation, which 
may be chosen as the root of the average, minimum,  or maximum of the after and prior variances.  
We select the average.
The mosum statistic is then compared with a threshold derived from an asymptotic distribution that depends on the bandwidth, 
the size of the entire time series, and a significance level chosen by the user.
If the mosum statistic exceeds this critical threshold then the null hypothesis,
of no change-points, is rejected in favor of the alternative one; 
the corresponding data point is now considered a change-point estimator.

Other criteria must be met in order to identify true change-points from the estimators
above. These criteria are imposed so that spurious peaks are discarded  
and multiple estimates pertaining to the same underlying true change-point are avoided.
The $\eta$-criterion is used to set the minimum distance between possible change-points at $\eta G$, 
whereas the $\epsilon$-criterion imposes that not only the putative change-point but an entire
neighborhood of minimum width $\epsilon G$, with $\epsilon < 1$, centered about it must surpass the threshold. 
Confidence intervals for the change-point locations are evaluated using bootstrap methods as illustrated
in \citep{R_mosum}. 

\subsection{STL decomposition in the eight Santa Monica neighorhoods}
\label{sec:SItext3}

\noindent
We plot here the monthly crime time series $Y(t)$ for each of the eight neighborhoods in the city of Santa Monica
for both Prop.\,47 and non-Prop.\,47 crimes. As done in Sect.\,\ref{subsec:meanCrimes} we also evaluate and present
the respective trend $T(t)$, seasonality $S(t)$ and remainder $R(t)$ components through an  
STL decomposition. The greatest trend increase for Prop.\,47 crimes
is observed in Downtown, together with a sharp decrease starting in late 2018.

 \begin{figure}[t!]
 \centering
 	\begin{subfigure}{0.95\linewidth}
	 \centering
 	\caption*{\textbf{North of Montana}}
	\includegraphics[width=\linewidth]{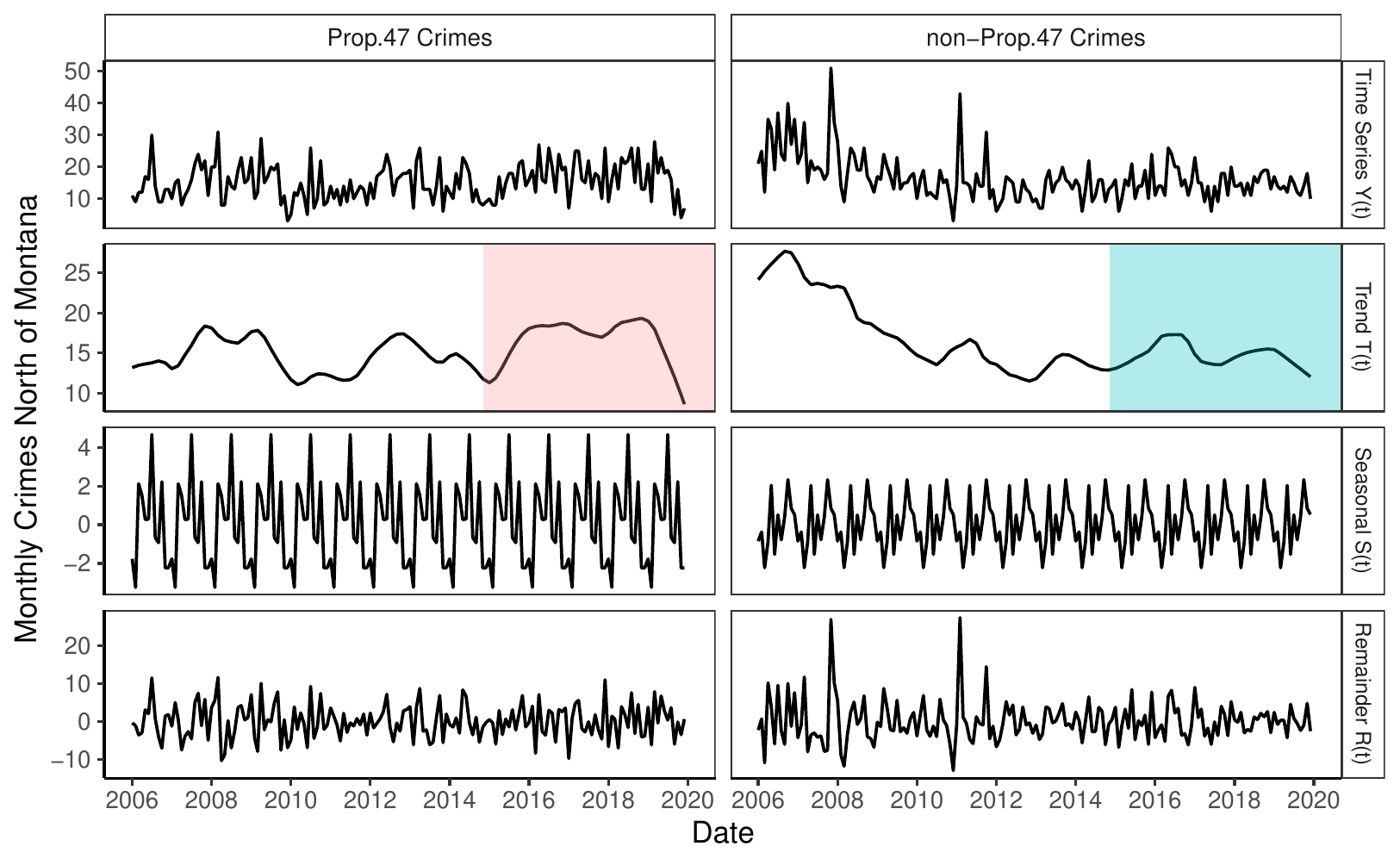}
	\end{subfigure}
\begin{subfigure}{0.95\textwidth}
 \centering
 	\caption*{\textbf{\quad Wilshire/Montana}}
	\includegraphics[width=\linewidth]{Fig_north_montana_decomp.pdf}
	\end{subfigure}
\caption{STL decomposition for 2006-2019 monthly crimes in the North of Montana and Wilshire/Montana neighborhoods.
	Prop.\,47 (non-Prop.\,47) crimes are to the left (right).}
\end{figure}
\begin{figure}[t!]
 \centering
	\begin{subfigure}{0.95\linewidth}
	 \centering
	\caption*{\textbf{Northeast Neighbors}}
 	\includegraphics[width=\textwidth]{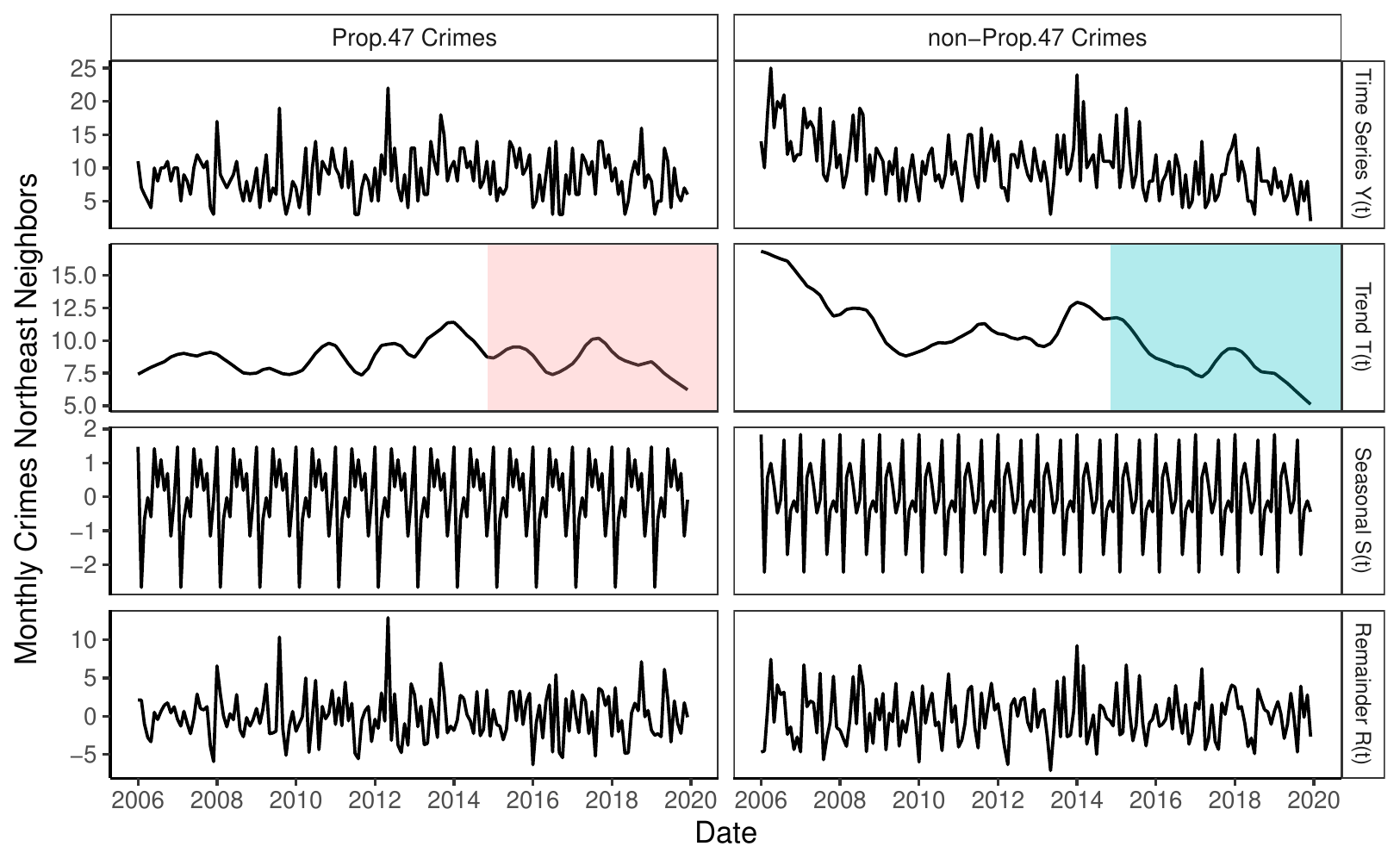}
	\end{subfigure}
\begin{subfigure}{0.95\textwidth}
 \centering
  	\caption*{\textbf{\quad Mid City}}
     	\includegraphics[width=\textwidth]{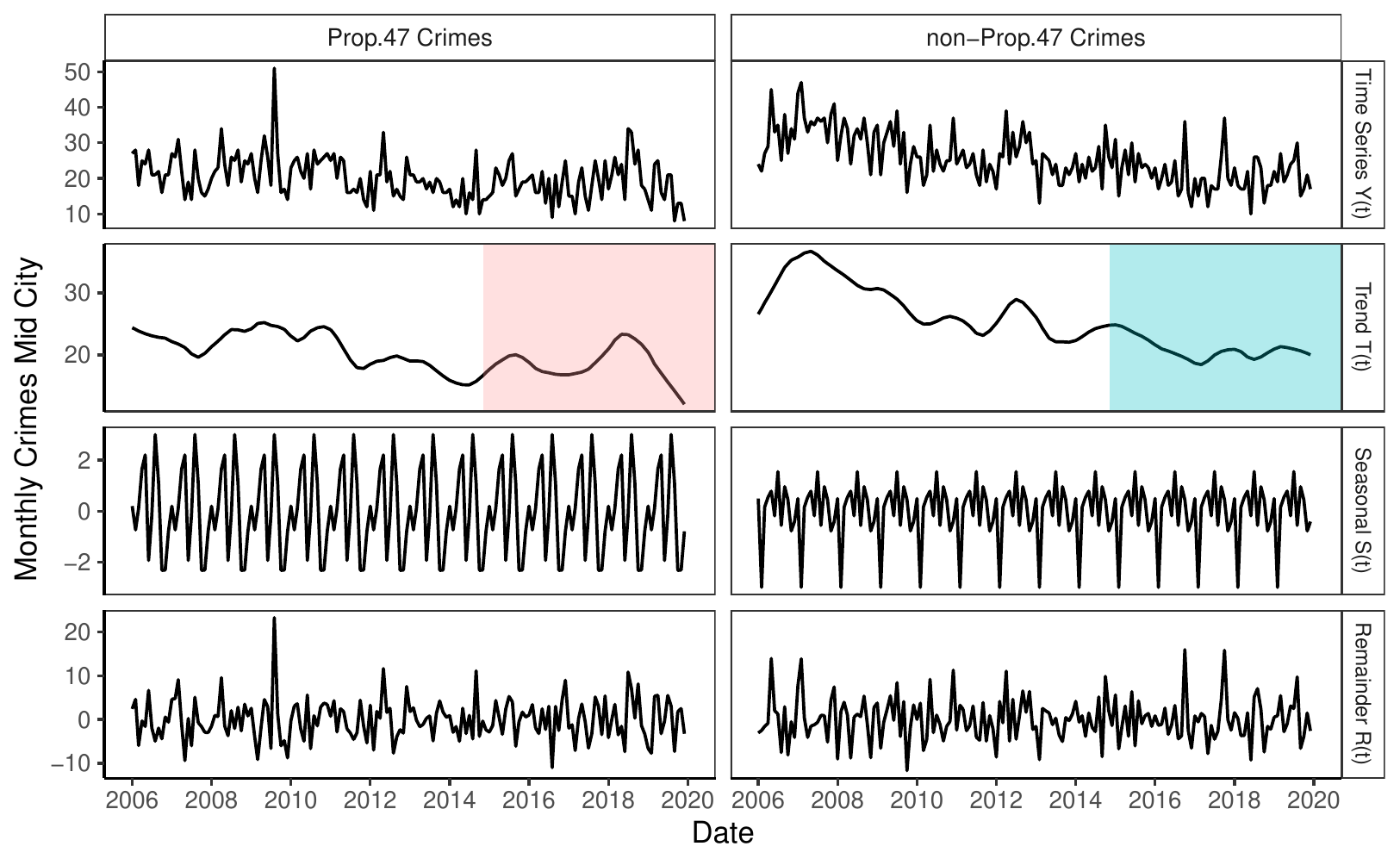}
	\end{subfigure}
\caption{STL decomposition for 2006-2019 monthly crimes in the Northeast Neighbors and Mid City neighborhoods.
	Prop.\,47 (non-Prop.\,47) crimes are to the left (right).}
 \label{fig:prop47_det2}
 \end{figure}
 \begin{figure}[t!]
  \centering
	\begin{subfigure}{0.95\linewidth}
	 \centering
     	\caption*{\textbf{Pico}}
     	\includegraphics[width=\textwidth]{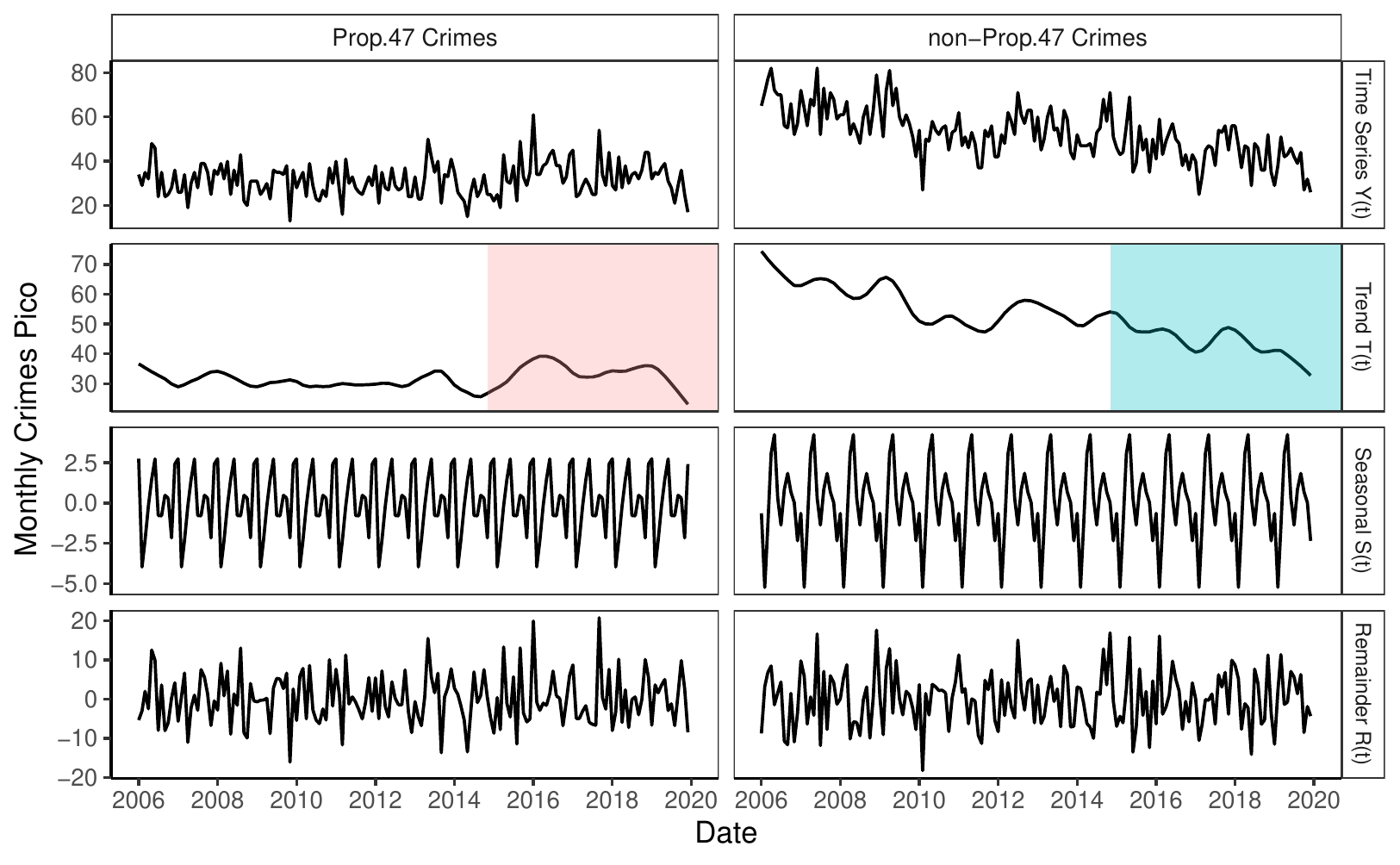}
     	\end{subfigure}
\begin{subfigure}{0.95\textwidth}
 \centering
      \caption*{\textbf{\quad Downtown}}
     \includegraphics[width=\textwidth]{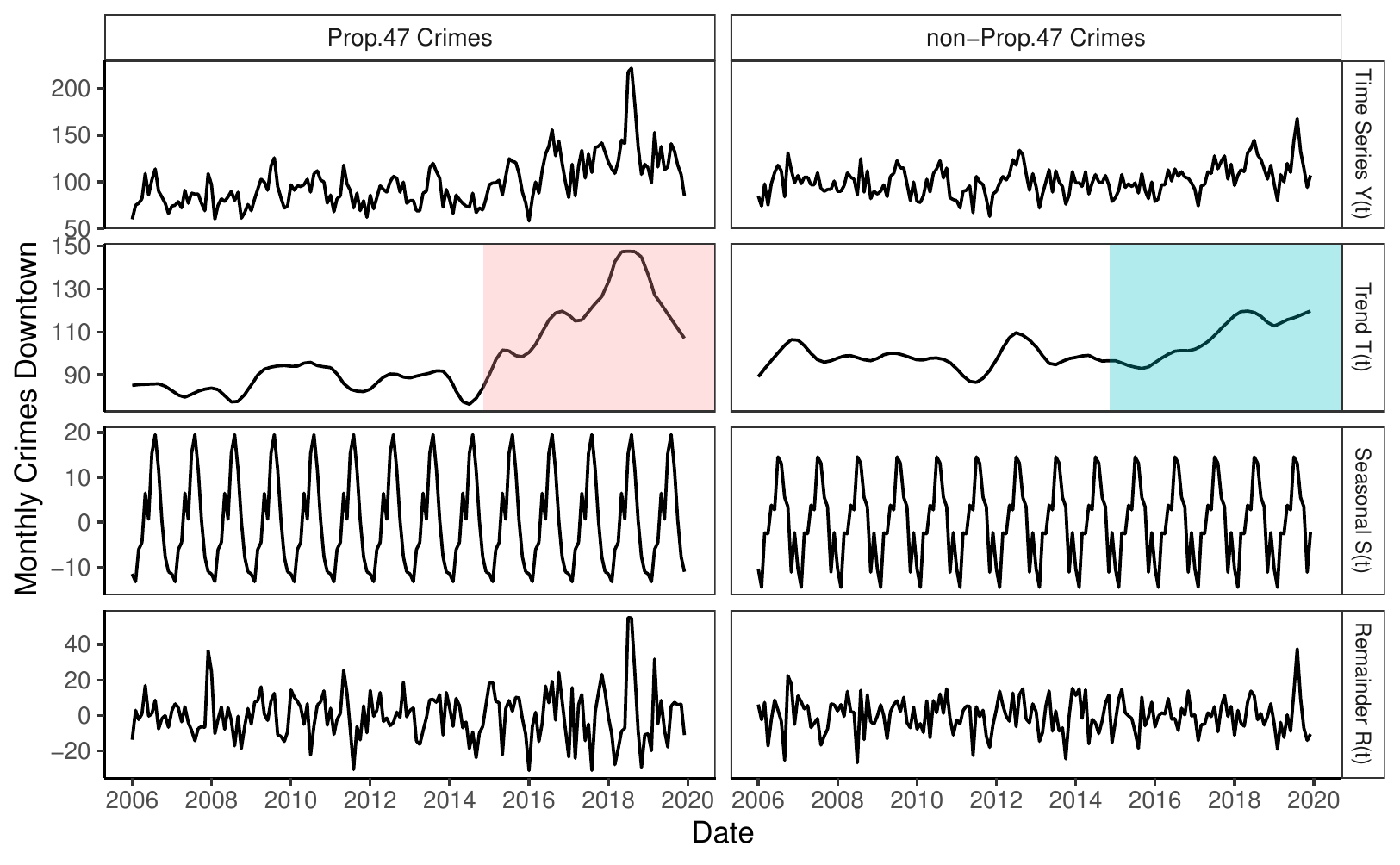}
     \end{subfigure}
\caption{STL decomposition for monthly crimes from 2006 to 2019 in the Downtown and Pico neighborhoods.
	Prop.\,47 (non-Prop.\,47) crimes are to the left (right).}
     \label{fig:prop47_det3}
 \end{figure}
\begin{figure}[t!]
 \centering
     	\begin{subfigure}{0.95\linewidth}
	 \centering
     	\caption*{\textbf{Sunset Park}}
         \includegraphics[width=\textwidth]{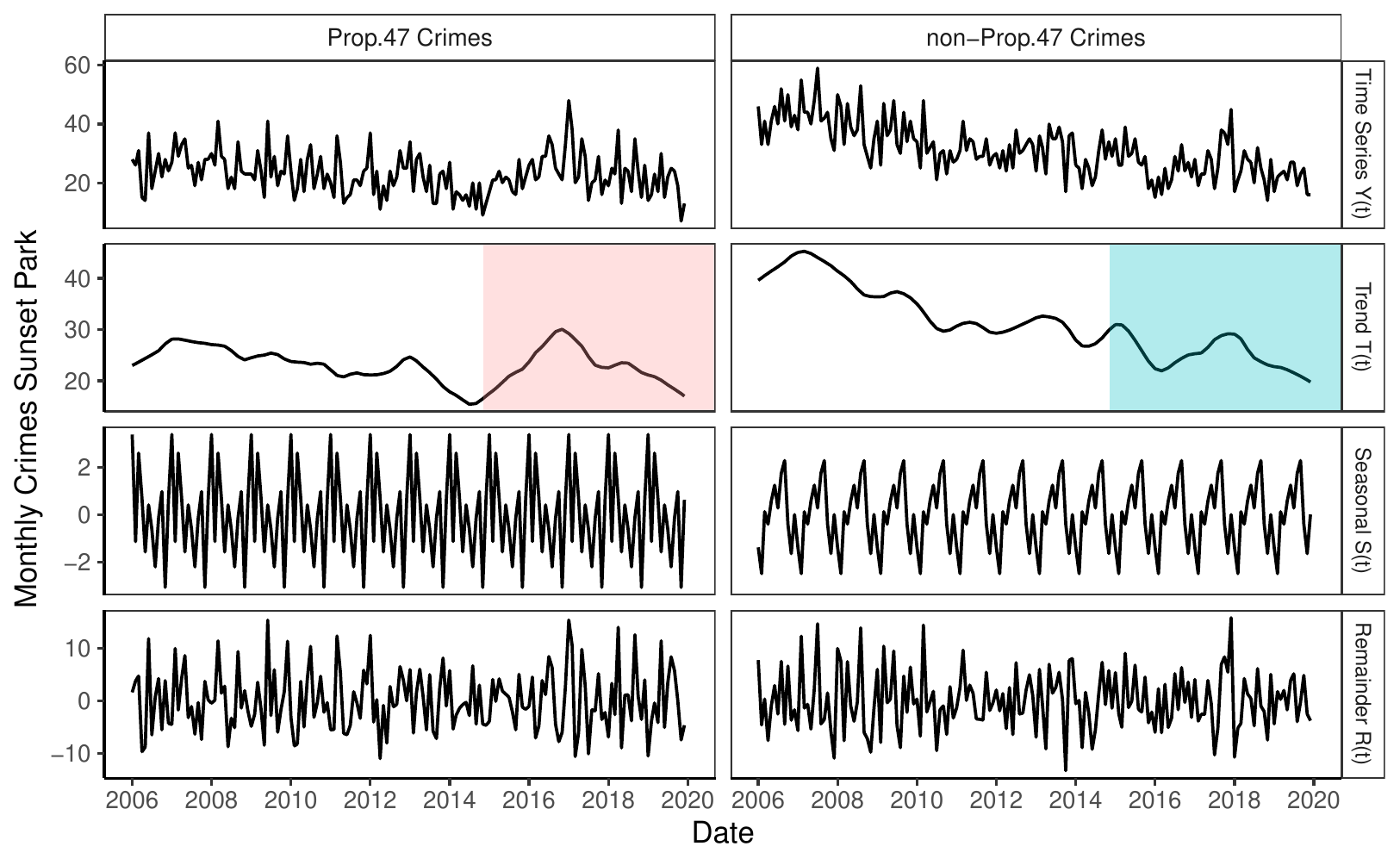}
         \end{subfigure}
\begin{subfigure}{0.95\textwidth}
 \centering
        	   \caption*{\textbf{\quad Ocean Park}}
        	     \includegraphics[width=\textwidth]{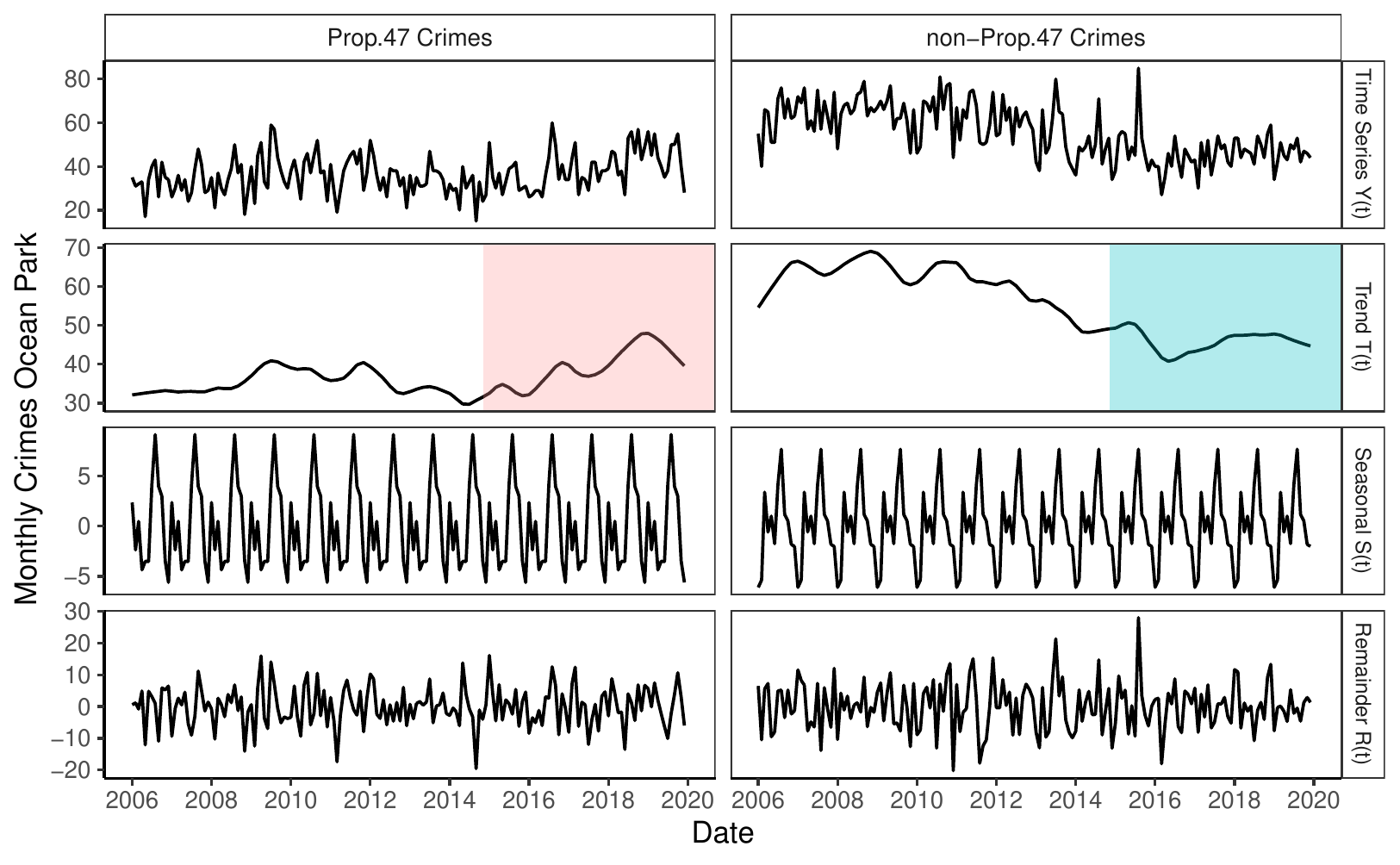}
             \end{subfigure}
               \caption{STL decomposition for monthly crimes from 2006 to 2019 in the Sunset Park and Ocean Park neighborhoods.
	Prop.\,47 (non-Prop.\,47) crimes are to the left (right).}
     \label{fig:prop47_det4}
 \end{figure}

\newpage
\bibliographystyle{apalike}
\bibliography{references.bib}

\end{document}